\documentclass[a4paper,11pt]{article}
\pdfoutput=1
\usepackage{jcappub}
\usepackage{afterpage,booktabs}
\usepackage{graphicx,subfig}
\usepackage{amssymb,amsmath,bm}
\usepackage{mathtools}
\usepackage[usenames,dvipsnames]{xcolor}
\usepackage[capitalise]{cleveref}
\usepackage{expl3,xstring}
\usepackage{dcolumn}
\usepackage{bm}
\usepackage{color}
\usepackage{breqn}
\usepackage{multirow}
\usepackage[normalem]{ulem}
\usepackage{tabu}
\usepackage{xcolor}
\newcommand\crule[3][black]{\textcolor{#1}{\rule{#2}{#3}}}

\definecolor{earlyde}{RGB}{255, 192, 203}
\definecolor{unphysical}{RGB}{106,90,205}
\definecolor{nomatterdom}{RGB}{218,112,214}
\definecolor{viable}{RGB}{0,0,128}
\definecolor{star}{RGB}{255,165,0}

\newcommand{\hiclass}{\texttt{Hi-Class}}
\newcommand{\HiCOLA}{\texttt{Hi-COLA}}
\newcommand{\nbody}{{\it N}-body}
\newcommand{\fml}{\texttt{FML}}
\newcommand{\lcdm}{$\Lambda$CDM}

\newcommand{\tK}{\tilde{K}}
\newcommand{\tX}{\tilde{X}}
\newcommand{\tphi}{\tilde{\phi}}
\newcommand{\tgthree}{\tilde{G}_3}
\newcommand{\tgfour}{\tilde{G}_4}

\newcommand{\dS}{\textrm{dS}}
\newcommand{\kds}{k_{1\mathrm{dS}}}
\newcommand{\kkds}{k_{2\mathrm{dS}}}
\newcommand{\gds}{g_{31\mathrm{dS}}}
\newcommand{\ggds}{g_{32\mathrm{dS}}}

\title{\boldmath \texttt{Hi-COLA}: Fast, approximate simulations of structure formation in Horndeski gravity}

\author{Bill S. Wright$^1$,}
\author{Ashim Sen Gupta$^1$,}
\author{Tessa Baker$^1$,}
\author{Georgios Valogiannis$^2$,}
\author{Bartolomeo Fiorini$^1$,}
\author{and The LSST Dark Energy Science Collaboration}

\affiliation{$^1$Astronomy Unit, Queen Mary University of London, Mile End Road, London E1 4NS, UK}
\affiliation{$^2$Department of Physics, Harvard University, Cambridge, MA 02138, USA}

\emailAdd{a.sengupta@qmul.ac.uk}
\emailAdd{t.baker@qmul.ac.uk}
\emailAdd{gvalogiannis@g.harvard.edu}
\emailAdd{b.fiorini@qmul.ac.uk}

\date{\today}

\abstract{
We introduce \texttt{Hi-COLA}, a code designed to run fast, approximate \textit{N}-body simulations of non-linear structure formation in reduced Horndeski gravity. Given an input Lagrangian, \texttt{Hi-COLA} dynamically constructs the appropriate field equations and consistently solves for the cosmological background, linear growth, and screened fifth force of that theory. Hence \texttt{Hi-COLA} is a general, adaptable, and useful tool that allows the mildly non-linear regime of many Horndeski theories to be investigated for the first time, at low computational cost. In this work, we first describe the screening approximations and simulation setup of \texttt{Hi-COLA} for theories with Vainshtein screening. We validate the code against traditional \textit{N}-body simulations for cubic Galileon gravity, finding $2.5\%$ agreement up to $k_{\rm max}=1.2~h/{\rm Mpc}$. To demonstrate the flexibility of \texttt{Hi-COLA}, we additionally run the first simulations of an extended shift-symmetric gravity theory. We use the consistency and modularity of \texttt{Hi-COLA} to dissect how the modified background, linear growth, and screened fifth force all contribute to departures from $\Lambda$CDM in the non-linear matter power spectrum. \HiCOLA{} can be found at \url{https://github.com/Hi-COLACode/Hi-COLA}.
}

\begin{document}
\maketitle


\section{Introduction}
\label{sec:intro}

\subsection{Motivations}
Testing gravity on cosmological scales is a key aim of the upcoming generation of large-scale structure surveys, such as LSST \cite{LSST}, Euclid \cite{Euclid}, and DESI \cite{DESI}. These surveys will give unprecedented access to information about the clustering of matter on non-linear scales. They will thus provide an ideal arena for constraining theories of gravity beyond General Relativity (GR). Such theories generically result in fifth-force enhancements of gravity and feature screening mechanisms in order to mimic GR within our Solar System. Notably, many of these deviations from GR lie beyond well-established tests of gravity in the linear regime \cite{Zuntz:2011aq,Baker:2012zs,Simpson:2012ra,Leonard:2015hha}.

However, to make use of this new data, we will need fast and accurate modelling of non-linear structure formation. Currently, such modelling is only available for a small number of individual modified gravity (MG) theories, largely nDGP gravity \cite{DGP}, $f(R)$ gravity \cite{fofr1,fofr2}, cubic Galileon gravity \cite{Nicolis:2008in, Deffayet:2009wt, Deffayet:2009mn}, and interacting dark energy \cite{Baldi:2008ay, Baldi:2011qi}. Indeed, nDGP and $f(R)$ gravity have become the standard workhorse theories for tests of gravity beyond the linear regime (e.g. \cite{Thomas:2015dfa, Achitouv:2015yha, Alam:2020jdv, Fiorini:2021dzs}); but this is principally because there exist well-established codes for them\footnote{These include tools based on the halo model \cite{Zhao:2013dza, Mead:2016zqy, Bose:2020wch}, traditional \nbody{} codes \cite{Khoury:2009tk, Schmidt:2009sg, Schmidt:2009sv, Li:2011vk, Puchwein:2013lza, Llinares:2013jza, Ruan:2021wup, Hernandez-Aguayo:2021kuh}, approximate methods \cite{PINOCCHIO2, Valogiannis:2016ane, Winther2017, Moretti:2019bob}, and emulators \cite{Winther:2019mus, Ramachandra:2020lue, Arnold:2021xtm}. See \cite{Winther:2015wla} for a comparison between some of the traditional \nbody{} codes.}, and not due to strong theoretical motivations\footnote{For example, it is long established that neither is successful as a dark energy theory. nDGP is (by definition) the non-accelerating branch of the original DGP theory \cite{DGP}, and hence still requires a standard cosmological constant (whilst the accelerating branch, sDGP, fails to fit CMB and large-scale structure data \cite{Fang:2008kc}). Chameleon $f(R)$ gravity cannot both successfully screen and accelerate without the presence of an effective cosmological constant \cite{Wang:2012kj}.}. 
A single tool capable of modelling non-linear structure formation across a much broader section of modified gravity theory space would remove these limitations. Indeed for many theories, it would enable them to be subjected to non-linear constraints for the first time. The creation of such a tool is the motivation for this work.

Horndeski gravity \cite{Horndeski1, Horndeski2, Horndeski3} is a class of theories that subsumes a substantial swathe of modified gravity theory space.
Any constraints placed on the general Horndeski framework can be rapidly translated into statements about individual members of the Horndeski class. As such, constraining Horndeski gravity has emerged over the past ten years as an efficient and relatively agnostic approach for performing tests of gravity with new data \cite{Bellini:2015xja,Alonso:2016suf,Kreisch:2017uet,Arai:2017hxj,SpurioMancini:2018apc,Reischke:2018ooh,Noller:2018wyv,SpurioMancini:2019rxy,Arai:2019zul,Brando:2020ouk, Baker:2020apq,Traykova:2021hbr}.
At the level of linear perturbation theory, the popular parameterisation developed by \cite{Bellini:2014fua}, sometimes known as the effective field theory of dark energy (EFTofDE), has been successfully constrained by a variety of observations, including the binary neutron star merger GW170817 \cite{Lombriser:2015sxa,Bettoni:2016mij,Creminelli:2017sry,Sakstein:2017xjx,Ezquiaga:2017ekz,Baker:2017hug}. However, obtaining constraints on Horndeski gravity from the non-linear structure formation that will be observed by upcoming galaxy surveys would require parameterisations to be extended to non-linear perturbations. Several attempts to model the mildly non-linear regime within the general framework of EFTofDE have been made \cite{Bellini:2015wfa,Bellini:2015oua,BoseKoyama2016,Cusin:2017mzw,Cusin:2017wjg}, but these do not address the deeply non-linear regime that we focus on in this paper. The framework of \cite{Lombriser:2016zfz} has been proposed to model the deeply non-linear regime for a general modified gravity parameterisation and a spherically symmetric top-hat mass configuration, and we discuss its implementation in an \nbody{} code \cite{Hassani:2020rxd} more below. Interestingly, the relativistic effects of EFTofDE that appear at large, linear scales have been implemented in \nbody{} simulations in \cite{Brando:2021jga}. In what follows, we refer to the subset of Horndeski theories with luminal gravitational wave speed as \textit{reduced} Horndeski gravity.

\subsection{The \HiCOLA{} code} \label{ssec:intro_code}
The standard method for predicting the observables of non-linear structure formation is to run \nbody{} simulations \cite{TrentiHut2008, DehnenRead2011, Kuhlenetal2012, Baldi2012}. 
Typically, in order to compute the screened fifth forces that arise in MG theories, the non-linear Klein--Gordon equation must be solved for any new field(s) present in the theory. For a general framework like Horndeski gravity, this adds a considerable degree of complexity as well as a significant increase to the computational requirements of the simulation compared to \lcdm{}.
For these reasons, in this paper we take an alternative approach. We show that the screened fifth force in general Horndeski gravity can be estimated using an approach involving a coupling and a screening coefficient. Our approach essentially equates to solving a linearised Klein--Gordon equation that has been corrected to account for the impact of screening. This technique was first described in \cite{2015PhRvD..91l3507W} and was implemented for nDGP and $f(R)$ gravity in the {\tt MG-PICOLA} code (\cite{Winther2017}, see also \cite{Valogiannis:2016ane}), where it led to percent-level agreement with traditional \nbody{} codes up to $k\sim3h/{\rm Mpc}$.
Another method for running \nbody{} simulations in parameterised MG without solving the full Klein--Gordon equation was investigated in \cite{Hassani:2020rxd}. We note there are some similarities in the derivations of the screened-fifth-force expressions given in \cite{Hassani:2020rxd} and that of this work, significantly that they both consider the Vainshtein mechanism in spherical symmetry. However, the screened-fifth-force expression used in \cite{Hassani:2020rxd} is ultimately a phenomenological parameterisation, whereas ours is intrinsically connected to the reduced Horndeski action. 

In this work, we present a fast, approximate simulation code called \HiCOLA{}\footnote{Can be found at \url{https://github.com/Hi-COLACode/Hi-COLA}, along with documentation to guide those who wish to use \HiCOLA{}.} (Horndeski-in-COLA) that implements our approach for computing screened fifth forces. \HiCOLA{} uses the COmoving Lagrangian Acceleration (COLA) simulation method \cite{OriginalCOLA}, which relies on 2$^{\rm nd}$ order Lagrangian perturbation theory (2LPT) to reduce the number of simulation timesteps required to reproduce accurate large-scale clustering. Thus we can trade computational speed for accuracy at non-linear scales without sacrificing accuracy at large, linear scales. \HiCOLA{} has two main components. The first is a Python module that takes a Horndeski Lagrangian as input and outputs intermediate quantities necessary for estimating the screened fifth force. 
The second is an extension of the COLA solver within the publicly available code \fml{}\footnote{Information about \fml{} can be found at \url{https://fml.wintherscoming.no/} and the COLA solver is specifically available at \url{https://github.com/HAWinther/FML/tree/master/FML/COLASolver}.} that uses the intermediate quantities from our Python module to compute the screened fifth force throughout the simulation. Whilst \HiCOLA{} is not itself a `full' \nbody{} simulation due to its use of COLA, we note that other simulation codes could be adapted to read in the output of our Python module to compute the screened fifth force.

While \HiCOLA{} allows us to run fast, approximate simulations of non-linear structure for any reduced Horndeski Lagrangian with a Vainshtein screening mechanism, we must first validate the code against well-studied individual theories. For this reason, we first use \HiCOLA{} to produce simulations for cubic Galileon gravity, with which we explore the parameter space as well as validate our code by comparing against existing traditional \nbody{} simulations. However, we also investigate a relatively new extended shift-symmetric (ESS) theory studied in \cite{Traykova:2021hbr}, creating the first simulations of non-linear structure formation for this theory. We intend to use \HiCOLA{} to study a more diverse array of modified gravity theories in future work.
\vspace{3mm}

The paper is structured as follows.
In \S\ref{sec:Horndeski_cosmo}, we review the cosmological behaviour of Horndeski gravity and introduce the cubic Galileon and ESS theories that we focus on in later sections.
In \S\ref{sec:screened_fifth_force}, we introduce the coupling-plus-screening coefficient approach for estimating a screened fifth force and derive the relevant expressions in terms of background Horndeski quantities.
In \S\ref{sec:COLA_implemenation}, we briefly introduce the COLA simulation method, before discussing the implementation of the Horndeski screened fifth force in \HiCOLA{}.
In \S\ref{sec:cuGalsims}, we present non-linear power spectra obtained from \HiCOLA{} for cubic Galileon gravity, and discuss their features.
We then do the same for the new ESS models introduced in \S\ref{sec:ESSsims}.
We conclude in \S\ref{sec:conc}. Additional details regarding validation of the code and in-depth analysis of its outputs, are presented in a set of appendices.

For the busy reader, we suggest those who are principally interested in simulations focus on \S\ref{ssec:Horndeski_background}, \S\ref{ssec:Horndeski_perts}, \S\ref{sec:screened_fifth_force}, and \S\ref{sec:COLA_implemenation} to understand our simulation approach, Appendix~\ref{app:valid} to understand our validation process, and \S\ref{sec:cuGalsims} and \S\ref{sec:ESSsims} for demonstrations of our code. For those principally interested in MG theories, we suggest focusing on \S\ref{sec:Horndeski_cosmo} and \S\ref{ssec:Vainshtein_spherical_symmetry} to understand the theory behind our screened-fifth-force computation, then \S\ref{ssec:cuGal_viability}, \S\ref{ssec:cuGal_sim_res}, \S\ref{ssec:ESS_viability}, and \S\ref{ssec:ESS_sim_res} for our results.


\section{Horndeski cosmology} \label{sec:Horndeski_cosmo}

In this section we describe in detail the cosmology of the Horndeski family of gravity theories, including their background and perturbation equations. This is the fundamental theoretical framework encoded in \HiCOLA{}. We also introduce the two example Horndeski members for which we will display results in \S\ref{sec:cuGalsims} and \S\ref{sec:ESSsims}. We describe our procedure for choosing initial conditions and free parameters for these theories in \S\ref{ssec:cuGal_theory}.

\subsection{Action and background evolution equations}
\label{ssec:Horndeski_background}

Horndeski gravity is a generalised class of the simplest modified gravity theories called scalar-tensor theories. These theories feature four dimensions and second order equations of motion, but also an additional scalar field coupled to matter via the metric tensor. The theories contained within Horndeski gravity feature a variety of screening mechanisms that reduce gravity to GR in certain environments, to maintain consistency with solar system constraints \cite{Brax2013}. In this paper, where explicit theories are considered, we will focus on members of the Horndeski class that possess the Vainshtein screening mechanism \cite{Vainshtein:1972sx}; we leave a detailed study of chameleon-screened theories \cite{Khoury:2003aq, Khoury:2003rn} to a future work.

One of the most prominent constraints on Horndeski gravity comes from the near-simultaneous observation of the gravitational wave event GW170817 and its gamma-ray counterpart GRB170817A \cite{LIGOScientific:2017ync, LIGOScientific:2017vwq}. This enabled a stringent constraint to be placed on the speed of propagation of gravitational waves relative to the speed of light, bounding the fractional difference to be smaller than approximately $10^{-15}$. This in turn strongly constrains several terms in the original Horndeski action \cite{Lombriser:2015sxa,Bettoni:2016mij,Creminelli:2017sry,Sakstein:2017xjx,Ezquiaga:2017ekz,Baker:2017hug} (though see subtleties discussed in \cite{deRham:2018red,deRham:2021fpu,Baker:2022rhh,Oikonomou:2020sij,Clifton:2020xhc}\footnote{In brief, when treated as a low-energy effective field theory, Horndeski gravity may still permit non-luminal low-frequency gravitational waves, whilst high-frequency gravitational waves (such as those detected by LIGO-Virgo) return to luminality. For simplicity we do not consider this possibility in the present work.}).

We refer to the surviving, viable part of the Horndeski action as `reduced Horndeski'. The reduced Horndeski action is
\begin{equation}
\label{eq:Horn_action}
S = \int d^4 x \sqrt{-g}\left( \tgfour(\tphi) R + \tK(\tphi, \tX) - \tgthree(\tphi, \tX) \Box \tphi  - M_{\rm P}^2 \Lambda + \mathcal{L}_m \right)\,,
\end{equation}
where $\mathcal{L}_m$ is the matter Lagrangian, $M_{\rm P}$ is the Planck mass, and $\tX$ represents the kinetic term of the scalar field $\tphi$:
\begin{equation}
\tX = -\frac{\partial_{\mu}\tphi \partial^{\mu} \tphi}{2}\,.
\end{equation}
This action describes the subset of the full Horndeski class in which gravitational waves travel at the speed of light.
A member of the reduced Horndeski class is specified via $\tK, \tgthree, \tgfour$, which are free functions of the scalar field and its derivative. With respect to the original Horndeski Lagrangian, the quintic `${\cal L}_5$' term has been eliminated, and $\tgfour$ reduced to a function of $\tphi$ only.

$\tgfour$ acts as a conformal coupling to the Einstein--Hilbert term (with GR recovered in the limit $\tgfour = M_{\rm P}^2/2$), $\tK$ represents a generalised kinetic term, and $\tgthree$ is a scalar self-interaction term that gives rise to the Vainshtein screening mechanism (see \S\ref{ssec:Vainshtein_spherical_symmetry}). Note that we have also included a cosmological constant term $\Lambda$, though its value in Eq.~(\ref{eq:Horn_action}) should not generally be taken to match that of the \lcdm{} model.

The tildes on quantities in Eq.~(\ref{eq:Horn_action}) indicate that these are generally dimensionful. From here on it will be convenient to work with dimensionless quantities, which we denote without tildes.
See Appendix~\ref{app:mass} for an overview of how the dimensionless analogues are defined. In this paper we will make standard choices for all the mass scales and timescales involved in this process, such that the Horndeski scalar field $\phi$ is dynamically relevant on cosmological scales. We denote time derivatives with respect to $x=\ln(a)$ by a prime.

From the extremisation of Eq.~(\ref{eq:Horn_action}) we obtain a set of modified Einstein equations and an equation of motion for the scalar field \cite{Kimura:2011dc}. Evaluating these on a spatially flat Friedman-Robertson-Walker metric, the first Friedman equation can be written as a closure equation:
\begin{align}
\label{eq:closure}
1&=\Omega_{\rm m}(z) + \Omega_{\rm r}(z) +\Omega_\Lambda(z) +\Omega_\phi(z)\,,
\end{align}
where $\Omega_{\rm m}$, $\Omega_{\rm r}$, and $\Omega_\Lambda$ all have their standard meanings (hereafter we suppress their arguments), and the energy density in the Horndeski scalar field is
\begin{align}\label{eq:EOM_OmPhi_nomassratios}
\Omega_{\phi} = (\Omega_{\rm r} + \Omega_{\rm m} + \Omega_{\Lambda})\bigg(\frac{1}{2G_4} -1 \bigg) + \frac{1}{3G_4} \bigg[ \frac{X K_X}{E^2}  - \frac{K}{2E^2}  + 3 X \phi^{\prime} G_{3X} - \frac{ X G_{3\phi}}{E^2}  -  3\phi^{\prime} G_{4\phi} \bigg]\,.
\end{align}
As is conventional, we denote derivatives with respect to $\phi$ and $X$ via subscripts, e.g. $\partial_X \partial_{\phi} G_3 \equiv G_{3\phi X}$.
The second Friedman equation gives an evolution equation for the dimensionless Hubble function $E=H/H_0$: 
\begin{align} \label{eq:EOM_E}
\frac{E^{\prime}}{E} = &\frac{1}{4G_4} \Bigg[ \frac{K}{E^2} - 2X \left(\frac{G_{3\phi}}{E^2} + G_{3X} \left\{ \frac{E^{\prime}}{E}\phi^{\prime}  + \phi^{\prime\prime} \right\}  \right)
 + 2G_{4\phi}\left( \left\{\frac{E^{\prime}}{E}+2\right\}\phi^{\prime} + \phi^{\prime\prime}  \right)  +\frac{4XG_{4\phi\phi}}{E^2} \Bigg] \nonumber\\
 &- \frac{1}{2}\bigg(\frac{\Omega_r - 3\Omega_{\Lambda}}{2G_4}+3\bigg)\,.
\end{align}
Finally, the evolution equation for the scalar field can be written as
\begin{align}
\phi^{\prime\prime} = - \frac{B}{A} - \frac{E^{\prime}}{E}\phi^{\prime}\,,
\end{align}
where
\begin{align}
A &= K_X - G_{3\phi} + 2X G_{3\phi X}  + E^2 \bigg[6\phi^{\prime}\bigg(  G_{3X} + X G_{3XX} \bigg) + {\phi^{\prime}}^2\bigg( K_{XX} - 2G_{3\phi X}\bigg)\bigg]\,,\\
B &= \frac{E^{\prime}}{E}B_1 + B_2\,,\\
B_1 &= 6 X  G_{3X} - 6 G_{4\phi}\,,\\
B_2 &= 3\phi^{\prime}\bigg[ K_X - 2 G_{3\phi} + 2X G_{3\phi X} \bigg]+ {\phi^{\prime}}^2\bigg[ K_{X\phi} - 2G_{3\phi \phi} \bigg] - \frac{K_{\phi}}{E^2} - 12G_{4\phi} \nonumber\\
&\ \ \ \ + 18X G_{3X} + \frac{2X G_{3\phi \phi} }{E^2} ~.
\end{align}
By appropriate substitution, Eqs.~\eqref{eq:EOM_OmPhi_nomassratios} and \eqref{eq:EOM_E} can be rearranged as a system of two second-order ODEs for $\phi$ and $E$ (the resulting expressions are too cumbersome to display here). Together with standard evolution equations for $\Omega_{\rm m}$, $\Omega_{\rm r}$, and $\Omega_\Lambda$, we can solve these as a function of redshift to determine the background cosmology of a given Horndeski model. $\Omega_\phi$ is then constructed from $\phi$ and $E$ via Eq.~\eqref{eq:EOM_OmPhi_nomassratios}. One can verify that the closure relation Eq.~\eqref{eq:closure} is satisfied to numerical accuracy during this procedure (as it must be).

\subsection{Horndeski perturbations}
\label{ssec:Horndeski_perts}

Having solved for the background quantities in a general reduced Horndeski theory, our next step will be to study the corresponding evolution of matter density perturbations. On large cosmological scales linear perturbation theory applies, and such modifications to the growth of structure have been computed in many gravity theories \cite{Linder:2005in,Pogosian:2007sw, 2011PhRvD..84l3524B,Berg:2012kn,Baker:2013hia,Perenon:2019dpc, Brando:2020ouk}. However, in this work we are interested in pushing beyond this to the (mildly) non-linear regime. In this section we borrow results from the helpful computations of \cite{Kimura:2011dc}.

We work in the Newtonian gauge, expressing the perturbed metric as
\begin{eqnarray}
ds^2=-(1+2\Phi)dt^2+a^2(1-2\Psi)d\mathbf{x}\,.
\label{eq:lineel}
\end{eqnarray}
We also perturb the Horndeski scalar field as $\phi\to\phi(t)+\delta\phi(t, \mathbf{x})$ and the matter density field as $\rho_{\rm m} \to\rho_{\rm m}(t)[1+\delta_m(t,\mathbf{x})]$. It will be convenient to use the dimensionless variable
\begin{eqnarray}
Q=H\frac{\delta\phi}{\dot\phi}\equiv \frac{\delta\phi}{\phi^\prime}\,,
\end{eqnarray}
where in the last inequality we have indicated the conversion to $x = \ln(a)$ we will employ later.

On scales substantially below the cosmological horizon we may apply the quasi-static approximation (QSA) \cite{Silvestri:2013ne,Noller:2013wca,Sawicki:2015zya,Pace:2020qpj}, in which we ignore time derivatives in the field equations, but keep spatial derivatives. This approximation holds provided the sound speed of the scalar perturbations does not become small, i.e. $c_s^2\sim 1$. Since we are working with a general Horndeski framework, this condition cannot be guaranteed to hold for all models, and must be checked on a case-by-case basis\footnote{We plan to automate these checks in a future version of our \HiCOLA{} code. The QSA holds for the specific models presented in our results sections (\ref{sec:cuGalsims} \& \ref{sec:ESSsims}) within relevant regimes.}.

Under the QSA, and following \cite{Kimura:2011dc}, we assume that $\Phi$, $\Psi$, and $Q$ are small, but their spatial derivatives may not be and must be kept. In this work we choose to focus on theories that screen via the Vainshtein mechanism. Hence in addition to the term $\sim \partial^2\Psi$ in the Poisson equation, we also keep terms like $(\partial^2\Psi)^2$ and $(\partial^2\Psi)^3$, and equivalents for $\Phi$ and $Q$. We will assume that terms without derivatives like $\Psi^2$ remain second-order in smallness and are discarded. We note that this would not necessarily be true for theories possessing chameleon, dilaton or k-mouflage mechanisms (for example); we leave the inclusion of these terms for future work. The spherical collapse solutions of \cite{Lombriser:2016zfz} give good indication that the current work should generalise to screening involving nonlinearity of zero- or first-order derivatives of potentials.

As in \S\ref{ssec:Horndeski_background}, by projecting out components of the perturbed gravitational field equations, we arrive at two gravitational equations\footnote{Note that there are a further two components available; however these introduce additional variables in the perturbed stress-energy sector, so we do not use them here.} and an equation of motion for the perturbed scalar field. From the traceless spatial part of the gravitational field equations, and assuming matter sources of anisotropic stress are negligible, we obtain
\begin{eqnarray}
\nabla^2\left(2G_4\left[\Psi-\Phi\right]-A_1 Q\right)&=0\,. \label{trlseq}
\end{eqnarray}
The time-time $(00)$ component of the gravitational field equations further yields
\begin{eqnarray}
2G_4\nabla^2\Psi&=&\frac{a^2}{2}\rho_{\rm m}\delta_m-A_2 \nabla^2 Q\,, \label{00eq}
\end{eqnarray}
where the coefficients are (some of these will be needed further below)
\begin{eqnarray}
\Theta&=&-\dot\phi XG_{3X}+2HG_4+\dot\phi G_{4\phi}\,,
\\
A_0&=&\frac{\dot\Theta}{H^2}+\frac{\Theta}{H}
-2G_4-4\frac{\dot{G_4}}{H}-\frac{{\cal E}+{\cal P}}{2H^2}\,,
\\
A_1&=&\frac{2\dot{G_4}}{H}\,,
\\
A_2&=& 2G_4-\frac{\Theta}{H}\,,
\\
B_0&=&\frac{X}{H}\dot\phi G_{3X}\,,
\end{eqnarray}
and where ${\cal E}$ and ${\cal P}$ are the effective energy density and pressure of the Horndeski field in the Friedmann equations, defined in \cite{Kimura:2011dc}. (For the reduced Horndeski case presented here, $A_2$ is trivially related to $A_1$ and $B_0$; however, we maintain the notation of \cite{Kimura:2011dc} for ease of comparison.) For clarity, the objects above are all `background' quantities, that is, they are evaluated using the homogeneous solutions for $\dot\phi$ and $a$.
The perturbed equation of motion for $Q$ can be written as
\begin{eqnarray}
&&A_0\nabla^2Q
-A_1\nabla^2\Psi
-A_2\nabla^2\Phi+\frac{B_0}{a^2H^2}{\cal Q}^{(2)}= 0,\label{eq:seom}
\end{eqnarray}
where
${\cal Q}^{(2)}:=\left(\nabla^2Q\right)^2-\left(\partial_i\partial_j Q\right)^2$.

For a given matter density distribution, Eqs.~(\ref{trlseq}), (\ref{00eq}), and~(\ref{eq:seom}) are sufficient to solve for the gravitational potentials and scalar field on mildly non-linear scales. Note how the importance of the new scalar field terms in equations (\ref{trlseq}) and (\ref{00eq}) are controlled by the background objects $A_0, A_1$, etc. This is our first indication that a full solution of the modified cosmological expansion history will be relevant for an accurate description of non-linear scales. Meanwhile, the Horndeski Lagrangian function $G_4$ clearly plays the role of an effective Planck mass in Eqs.~(\ref{trlseq}) and (\ref{00eq}).

In general, Eqs.~(\ref{trlseq}), (\ref{00eq}), and~(\ref{eq:seom}) cannot be solved analytically, except in configurations with a high degree of symmetry. In \S\ref{ssec:Vainshtein_spherical_symmetry} we will solve them in a spherically symmetric system, to compute an approximate screening coefficient for our \HiCOLA{} code.

\subsection{Cubic Galileon gravity} 
\label{ssec:cuGal_theory}

Later in this paper, we will use cubic Galileon gravity \cite{Nicolis:2008in, Deffayet:2009wt, Deffayet:2009mn} as a test case with which to validate our general Horndeski simulation code. Although cubic Galileon gravity has been strongly constrained observationally \cite{Renk:2017rzu}, it is one of the simplest Horndeski theories that displays the full phenomenology of Vainshtein screening. Furthermore, it has been extensively studied in \cite{Barreira:2013xea, Frusciante:2019puu, Peirone:2017vcq} and implemented into \nbody{} simulations in \cite{Wyman:2013jaa, Barreira:2013eea,Zhang:2020qkd}. We emphasise that the theory we refer to hereafter as `cubic Galileon' is really the original cubic Galileon model plus a cosmological constant in some cases; this is similar to what is done to restore viability to DGP.

The cubic Galileon theory is defined by the following choices of dimensionless Lagrangian functions (recall these are the dimensionless versions of the functions appearing in \eqref{eq:Horn_action}):
\begin{equation} \label{eq:CuGal}
        K(\phi,X) = k_1 X\,, \qquad\qquad
        G_3(\phi,X) = g_{31} X\,, \qquad\qquad
        G_4(\phi) = \frac{1}{2}\,,
\end{equation}
where $X$ is defined in \eqref{eq:dimless_X} and $k_1, g_{31} \in \mathbb{R}$ are the free model parameters of the theory.
Within this parameter space, we wish to identify regions that produce a cosmological expansion history reasonably close to that of \lcdm{}.
One technique for finding such solutions is to demand that the cubic Galileon universe possesses a de Sitter (dS) phase at some future time. This naturally leads us to models in which the universe at $z=0$ is evolving towards a dS state, without having reached it fully -- much like the present-day \lcdm{} model \cite{DeFelice:2010nf,DeFelice:2010pv,Nesseris:2010pc} (see also \cite{Appleby:2020dko,Linder:2020xey,Appleby:2020njl} for related ideas).

A dS phase is characterised by total domination of the dark energy sector; in our case this is the sum of both cosmological constant and scalar field contributions, i.e. $\Omega_{\textrm{DE}} \coloneqq \Omega_{\Lambda} + \Omega_{\phi} = 1$ and $\Omega_{\textrm{DE}}^{\prime} = 0$. In order to satisfy these conditions for an extended period, a dS phase then requires\footnote{Strictly speaking this is sufficient but not necessary. See \cite{Appleby:2020dko, Linder:2020xey, Appleby:2020njl} for a more general treatment of imposing a dS phase on Horndeski models. The general condition reduces to the one in this work for the cubic Galileon and ESS model.} $E^{\prime}(a_{\textrm{dS}}) = \phi^{\prime\prime}(a_{\textrm{dS}}) = 0$ (where $a_{\textrm{dS}}$ denotes the function for the scale factor in the dS phase). Imposing these dS conditions upon the equations of motion from \S\ref{ssec:Horndeski_background} and solving, we quickly find the following relations:
\begin{align}\label{eq:CuGal_dS}
    \begin{split}
        k_1 &= - \frac{6}{ {\phi^{\prime}_{\textrm{dS} }}^2} \left( 1 - \frac{\Omega_{\Lambda 0}}{E^2_{\textrm{dS}}} \right)\,, \\
        g_{31} &= \frac{2}{ E^2_{\textrm{dS}} {\phi^{\prime}_{\textrm{dS}}}^3} \left( 1 - \frac{\Omega_{\Lambda 0}}{E^2_{\textrm{dS}}} \right)\,,
    \end{split}
\end{align}
where $E_{\rm dS} = E(a_{\mathrm dS})$, $\phi_{\rm dS} = \phi(a_{\mathrm dS})$ and $\Omega_{\Lambda0}=\Omega_\Lambda(z=0)$. For later convenience, we define the quantity in rounded brackets above as
\begin{equation}
\label{eq:alpha_def}
    \alpha =  1 - \frac{\Omega_{\Lambda 0}}{E^2_{\textrm{dS}}}\,.
\end{equation}

At first glance, this appears to simply shift the need to specify the model parameters $k_1$ and $g_{31}$, to now instead specifying the values for the Hubble function and scalar field derivative in the dS phase, i.e. $E_{\rm dS}$ and $\phi^{\prime}_{\rm dS}$. However, we can take advantage of the rescaling symmetry of cubic Galileon gravity \cite{DeFelice:2010nf,DeFelice:2010pv,Nesseris:2010pc,Barreira:2013jma} to perform a change of variables; rescaling symmetry implies the equations of motion are invariant under a transformation of the kind $k_i \rightarrow C^{n(i)} k_i$ provided $\phi \rightarrow C^{-1} \phi$, where $C$ and $n$ are scalars and $k_i$ is a generic model parameter. The value of $n$ depends on which parameter(s) are being rescaled, hence $n=n(i)$. $E$ can similarly be rescaled along with counterpart rescalings for the model parameters while leaving the equations of motion invariant. We define
\begin{align}
\label{eq:Uy_def}
        U &= \frac{E}{E_{\dS}}\,, &
        y &= \frac{\phi^{\prime}}{\phi^{\prime}_{\dS}}.
\end{align}
Substituting $U$ and $y$ into the equations of motion then motivates the following rescaling of the cubic Galileon model parameters:
\begin{align}
\label{eq:k1g31_rescale}
        k_{1 \dS} &= k_1 {\phi^{\prime}_{\dS}}^2\,, &
        g_{31 \dS} &= g_{31} E^2_{\dS} {\phi^{\prime}_{\dS}}^3.
\end{align}
Now we can rewrite our entire system of equations from \S\ref{ssec:Horndeski_background} and \S\ref{ssec:Horndeski_perts} in terms of the variables $\{U,y\}$ with parameters $\{k_{1 \dS}, g_{31 \dS}\}$ without changing any of the physics involved. Making use of Eqs.~\eqref{eq:CuGal_dS}-\eqref{eq:Uy_def}, we see that the model parameters for a cubic Galileon theory consistent with a future dS phase are now given by $k_{1 \dS} = -6 \alpha$, $g_{31 \dS} = 2 \alpha$. For a given choice of $E_{\dS}$ and $\Omega_{\Lambda 0}$, $\alpha$ is known via Eq.~\eqref{eq:alpha_def} and hence the model parameters are determined. In what follows, rather than fix $\Omega_{\Lambda 0}$ directly, we will make use of the ratio $f_{\phi}$, defined by
\begin{equation}
\label{eq:fphi}
    f_{\phi} = \frac{\Omega_{\phi 0}}{\Omega_{\phi 0} + \Omega_{\Lambda 0}}.
\end{equation}
$f_{\phi}$ determines what fraction of the energy density of the dark energy sector is due to the cosmological constant $\Lambda$, and how much is due to the scalar field energy density. When $f_{\phi}=1$, the value of dark energy at $a=1$ is solely due to the scalar field. In this scenario $\alpha = 1$, which then implies the model parameters $\{k_{1 \dS}, g_{31 \dS}\}$ are completely specified.

The determination of initial conditions for our variables, $U(a=1)=U_0\equiv E_{\dS}^{-1}$ and $y(a=1) \equiv y_0$, remains. However, one of these can be fixed by demanding the closure equation \eqref{eq:closure} is satisfied at $z=0$. We will generally choose this to be $y_0$
\footnote{A potential downside of this choice is that the closure equation may have multiple solutions for $y_0$; then some care is required to select the correct one.}.
Therefore, in the end, there remain two free parameters for a cubic Galileon model which possesses a future de Sitter phase: $U_0\equiv E_{\dS}^{-1}$ and $f_{\phi}$.

However, we will note here that there is an alternative way to apply the shift symmetry properties of cubic Galileon gravity.
We could also choose to normalise the scalar field by its present-day value, defining
$v = {\phi}/{\phi_0}$. The variables used in the system of equations are now $\{v,E\}$, which by definition have their initial conditions at $z=0$ both fixed to 1.
Then, analogously to the de Sitter rescaling above, one can define new model parameters $k_{1T}$ and $g_{31T}$ (subscript T for `today'), which are related to the original model parameters by $k_{1T} = k_1 \left( \phi^{\prime}_0\right)^2, \;g_{31T} = g_{31}\left(\phi^{\prime}_0\right)^3$. Now one of $k_{1T}$ or $g_{31T}$ can be fixed using the closure equation, leaving the other as a free parameter along with $f_{\phi}$ as previously.

We stress that under correct rescaling, the solutions of the cubic Galileon equations remain identical\footnote{Note that one can transform between the de Sitter and `today' rescalings as $k_{1 \dS} = \frac{k_{1 T}}{y^2_0}, g_{31 \dS} = \frac{g_{31 T} E^2_{\dS}}{y^3_0}$.}. In fact this property will be shared by any \textit{shift-symmetric} Horndeski Lagrangian -- this implies theories for which the Lagrangian functions only depend on $X$. Hence we are free to choose any such rescaling which facilitates our exploration of the parameter space of the theory. This becomes increasingly important as we explore theories with higher-dimensional parameter spaces, such as the one in the next subsection. One of our priorities in this work will be to find models that display suitably \lcdm{}-like expansion histories; for this purpose, we generally find the de Sitter choice of rescaling the most beneficial.

\subsection{Extended shift-symmetric (ESS) gravity} \label{ssec:ESS_theory}

As indicated in \S\ref{sec:intro}, the aim of this work is to build a fully flexible simulation code that can handle any theory within the reduced Horndeski family. To start putting this generality through its paces, we will further explore a less well-studied theory than cubic Galileon gravity: the extended shift-symmetric (ESS) model (also studied in \cite{Traykova:2021hbr}) defined by
\begin{align}\label{eq:ESS}
    K &= k_1 X + k_2 X^2\,,\\
    G_3 &= g_{31} X + g_{32} X^2\,,\\
    G_4 &= \frac{1}{2}\,.
\end{align}
Comparing with Eqs.~\eqref{eq:CuGal}, we see that ESS gravity is a natural extension of cubic Galileon gravity.
In addition to the model parameters shared with cubic Galileon gravity, the ESS model has two extra parameters, $k_2, g_{32} \in \mathbb{R}$.

As we did with cubic Galileon gravity, we will demand that our ESS model possesses a de Sitter phase at future time (defined by $E^{\prime}_{\dS} = \phi^{\prime\prime}_{\dS}=0$). For cubic Galileon gravity this yielded two constraint equations, reducing the number of degrees of freedom by two. This is also what occurs for the ESS model: the dS constraint manifests as two constraint equations that relate one pair of model parameters to the other pair. These constraint equations are
\begin{align}\label{eq:ESS_dS}
\begin{split}
    \kkds  &= -2 \kds - 12(1- \Omega_{\Lambda})\,,\\
    \ggds &= \frac{1}{2} - \frac{1}{2} \left( \Omega_{\Lambda} + \frac{\kds}{6} + \frac{\kkds}{4} + \gds \right)\,.
\end{split}
\end{align}
Note that, thanks to shift symmetry, we have again rescaled our variables and model parameters to their dS versions (see Eqs.~(\ref{eq:Uy_def})-(\ref{eq:k1g31_rescale})). With these constraints in hand, and again using the closure equation to remove $y_0$ as a degree of freedom, we are left with four free parameters for the ESS model: $U_0\equiv E_{\dS}^{-1}$, $f_{\phi}$, and any \textit{two} of $\{k_{1 \dS}, k_{2 \dS}, g_{31 \dS}, g_{32 \dS}\}$.

In \S\ref{ssec:cuGal_theory}, for the cubic Galileon case without a cosmological constant ($f_{\phi}=\alpha=1$), we found $\kds = -6$ and $\gds = 2$. Plugging these values into Eqs.~\eqref{eq:ESS_dS} we obtain $\kkds = \ggds = 0$. That is, we recover the cubic Galileon model of \S\ref{ssec:cuGal_theory} as a consistent limit of the ESS model. Choosing a model connected to cubic Galileon gravity gives us guidance on finding viable cosmological solutions in the enlarged parameter space of ESS -- we will discuss this further in \S\ref{ssec:ESS_viability}. 

We note that both theories introduced in this section possess shift symmetry. We stress that shift symmetry is not in any way required for a model to be used with our \HiCOLA{} code; it is merely a convenient device. In future works we will study theories beyond the shift-symmetric ones presented here.


\section{Estimating screened fifth forces} \label{sec:screened_fifth_force}

Our method for estimating screened fifth forces is derived by considering the Vainshtein mechanism in spherical symmetry, which we do in \S\ref{ssec:Vainshtein_spherical_symmetry}. Treating the calculation as spherically symmetric introduces a well-documented error, which we discuss in \S\ref{ssec:force_interpolation}.
Our two-part solution to this error involves: 1) interpolation between the spherically symmetric expression and the linear fifth force at large scales, and 2) smoothing of the density field that enters the spherically symmetric expression. We discuss these two additions in \S\ref{ssec:force_interpolation} \& \S\ref{ssec:dens_smoothing} respectively.

\subsection{Vainshtein mechanism in spherical symmetry} \label{ssec:Vainshtein_spherical_symmetry}

Mimicking the derivation in \cite{Kimura:2011dc} (see also \cite{Lombriser:2016zfz,Frusciante:2020zfs,Song:2021msd}), we consider a spherically symmetric overdensity on a cosmological background, with metric potentials defined by Eq.~(\ref{eq:lineel}). The Horndeski perturbation equations from \S\ref{ssec:Horndeski_perts} can be integrated once with respect to a comoving radial coordinate $r$, becoming (note we suppress arguments)
\begin{eqnarray}
\frac{1}{r}\frac{\partial\Psi}{\partial r}-\frac{1}{r}\frac{\partial\Phi}{\partial r}-\frac{\alpha_1}{r} \frac{\partial Q}{\partial r}
&=&0\,, \label{eq1}\\
\frac{1}{r}\frac{\partial\Psi}{\partial r}+\frac{\alpha_2}{r}\frac{\partial Q}{\partial r}
&=&\frac{1}{16\pi G_4}\frac{\delta M_r}{r^3}\,, \label{eq2}\\
\frac{\alpha_0}{r}\frac{\partial Q}{\partial r}-\frac{\alpha_1}{r}\frac{\partial\Psi}{\partial r}-\frac{\alpha_2}{r}\frac{\partial\Phi}{\partial r}
&=&-\frac{2\beta_0}{H^2}
\left(\frac{1}{r}\frac{\partial Q}{\partial r}\right)^2\,, \label{eq3}
\end{eqnarray}
where primes denote derivatives with respect to $r$, and we define the enclosed mass perturbation
\begin{eqnarray}
\delta M_r(t, r)=4\pi\bar{\rho}_{\rm m}(t)\int_0^r\delta_m(t, \hat{r})\, \hat{r}^2 d{\hat r}\,,
\label{eq:mpert}
\end{eqnarray}
and dimensionless coefficients
\begin{eqnarray}
\alpha_i(t)=\frac{A_i}{2 G_4}\,,
\quad
\beta_0(t)=\frac{B_0}{2 G_4}\,. \label{eq:beta0def}
\end{eqnarray}
For clarity, we explicitly indicate in Eqs.~(\ref{eq:mpert})-(\ref{eq:beta0def}) that the density distribution and modified gravity coefficients appearing here will change with time steps in our simulations. 

Re-arranging Eqs.~(\ref{eq1})-(\ref{eq3}) to eliminate $\partial\Psi/\partial r$ and $\partial\Phi/\partial r$, we quickly arrive at a quadratic equation for $r^{-1}\partial Q/\partial r$. This has the following solution\footnote{Note the positive root ensures $r^{-1}\partial Q/\partial r \rightarrow 0$ for vanishing $\delta M_r$.}:
\begin{eqnarray}
\frac{1}{r}\frac{\partial Q}{\partial r}=\frac{H^2}{{\cal B}}\left(\sqrt{1+\frac{2{\cal B}{\cal C}\mu_r}{H^2r^3}} -1 \right)\,,
\label{eq:qsol}
\end{eqnarray}
where
\begin{eqnarray}
{\cal B}=\frac{4\beta_0}{\alpha_0+2\alpha_1\alpha_2+\alpha_2^2}\,, \quad
{\cal C}=\frac{\alpha_1+\alpha_2}{\alpha_0+2\alpha_1\alpha_2+\alpha_2^2}\,, \quad
\mu_r=\frac{\delta M_r}{16\pi G_4}\,.
\label{eq:BCmu}
\end{eqnarray}
From Eq.~(\ref{eq:qsol}) we can extract the Vainshtein radius:
\begin{eqnarray}
 r_*^3=2{\cal B}{\cal C}\frac{\mu_r}{H^2}\,.
 \label{eq:Vainsh}
\end{eqnarray}
We introduce the useful dimensionless quantity $\chi$ (where $\Omega_{\rm m0}=\Omega_{\rm m}[z=0]$):
\begin{eqnarray}
 \chi = \left(\frac{r_*}{r}\right)^3 =\left(\frac{H_0}{H}\right)^2\frac{{\cal B}{\cal C}\, \Omega_{\rm m0}}{16 \pi G_{\rm N} G_4 a^3}\delta_{\rm AV}\,.
\label{eq:chiexpunsmoothed}
\end{eqnarray}
To reach the second equality above we have used the definition of $\mu_r$ from Eq.~(\ref{eq:BCmu}), and also expressed the mean matter density via $\bar{\rho}_m=\frac{3 H^2_0 \Omega_{\rm m0}}{8 \pi G_{\rm N} a^3}$. $\delta_{\rm AV}$ is a radially-averaged matter density perturbation:
\begin{eqnarray}
\delta_{AV}(r)&=&\frac{3}{r^3}\int_0^r \,\tilde{r}^2 \delta_m(\hat{r})\;d\hat{r}\,.
\end{eqnarray}
Note that in the limit of a top-hat perturbation, $\delta_{\rm AV}(r\leq R) = \delta_m(r)$. In \S\ref{ssec:dens_smoothing} we will see that the density field of our \HiCOLA{} simulations must be smoothed for practical reasons.
This smoothing also has a side-effect of bringing the non-spherically symmetric density field in the simulation closer to the spherically symmetry $\delta_{\rm AV}$. For this reason, in our simulations we replace $\delta_{\rm AV}$ in Eq.~(\ref{eq:chiexpunsmoothed}) with the smoothed density field; our validation results in Appendix \ref{app:valid} will justify that this is acceptable. We discuss the use of the spherically symmetric approximation further in \S\ref{ssec:force_interpolation}.

Returning to Eqs.~(\ref{eq1}) and (\ref{eq2}), we can eliminate $\partial\Psi/\partial r$ and substitute in our solutions for $\partial Q/\partial r$ and $r_*$. To make the resulting expressions as familiar as possible we define $G_{G_4} = \left(16\pi G_4\right)^{-1}$. This quantity is the contribution to the effective Newton's constant coming from  conformal coupling alone, i.e. the first term in the Lagrangian of Eq.~(\ref{eq:Horn_action}). If this part of the Lagrangian is unmodified from GR, then $G_4 = M_{\rm P}^2/2\equiv (16\pi G_{\rm N})^{-1}$, and hence $G_{G_4} = G_{\rm N}$. The result of the substitution is
\begin{eqnarray}
\frac{\partial\Phi}{\partial r}&\simeq&\frac{G_{G_4}\delta M}{r^2}
\left[1-2(\alpha_1+\alpha_2 ){\cal C}\left(\frac{r^3}{r_*^3}\right)\left(
\sqrt{1+\frac{r_*^3}{r^3}} -1
\right)\right]\,,
\label{nat}
\end{eqnarray}
or equivalently, when re-written as an expression for the modified total force in terms of $\chi$ (using Eq.~(\ref{eq:chiexpunsmoothed})):
\begin{eqnarray}
F_{\rm tot} &=&F_{\rm N}\, \frac{G_{G_4}}{G_{\rm N}}
\left[1-2(\alpha_1+\alpha_2 ){\cal C}\,\chi^{-1}\left(
\sqrt{1+\chi} -1
\right)\right]\,,
\label{eq:MGforce}
\end{eqnarray}
where $F_N$ represents the standard Newtonian force law used in simulations (see \cite{Adamek:2016zes, Adamek:2017uiq, Hassani:2019lmy, Adamek:2020jmr} for work on fully relativistic \nbody{} codes).
There are two key components appearing in this expression that are purely time-dependent (we remind the reader $E=H/H_0$):
\begin{align}
 \beta &= -(\alpha_1+\alpha_2){\cal C}\,, \label{eq:coupling}
 \\
 \frac{\chi}{\delta_{\rm m}} &= \frac{{\cal B}{\cal C} \,\Omega_{\rm m0}}{E a^3}~\frac{G_{G_4}}{G_{\rm N}}\,. \label{eq:chioverdelta}
\end{align}
Hereafter we refer to $\beta$ as the coupling. Rather than $\chi/\delta_{\rm m}$ itself, we will work in terms of the screening coefficient $S$ defined by
\begin{align}
S=\frac{2}{\chi}\left(\sqrt{1+\chi} -1\right)\,, \label{eq:screen_coeff}
\end{align}
which is density-dependent (and therefore implicitly scale-dependent), as well as time-dependent. On large scales where density perturbations are small, $\chi\ll 1$ and hence the screening coefficient $S\rightarrow 1$. On small scales $\chi$ is large, and hence $S\rightarrow 0$ like $\chi^{-\frac{1}{2}}$; the fifth force is screened away and the force law returns to that of GR, except for the potential factor of $G_{G_4}/G_{\rm N}$ in Eq.~(\ref{eq:MGforce})\footnote{This residual, non-screened factor has been tightly constrained by solar system tests of gravity, e.g. \cite{Burrage:2020jkj}.}. 

Putting this all together, our modified force law including the screened fifth force can be written in the helpful schematic form:
\begin{eqnarray}
F_{\rm tot} &=&F_{\rm N} \frac{G_{G_4}}{G_{\rm N}}
\left[1+\beta\, S\right].
\label{eq:MGforce_simple}
\end{eqnarray}
We reiterate here, for later use, the salient features of this expression: $\beta$ is a purely time-dependent coupling that sets the maximum strength for the fifth-force term. It is determined entirely by the solution of the background equations for $H$ and the scalar field. $S$ is a screening coefficient that smoothly takes the fifth force from full-strength to fully-screened as determined by the density field. It is also determined by the background solutions through the quantity $\chi/\delta_{\rm m}$, and the smoothed density field in the simulation. Plots of $\beta$ and $S$ for cubic Galileon gravity and ESS gravity will appear in \S\ref{sec:cuGalsims} and \S\ref{sec:ESSsims}.

As mentioned in \S\ref{ssec:intro_code}, this is schematically the same as the approach described in \cite{2015PhRvD..91l3507W}. The similarity can be seen by comparing Eq.~(\ref{eq:MGforce_simple}) to Eq.~(45) in \cite{2015PhRvD..91l3507W}. A key difference is that our expression for the screened fifth force is directly connected to the functions that appear in the Jordan frame reduced Horndeski action. We can use this to compute screened fifth forces for any Vainshtein screened theory contained within that action via the \HiCOLA{} front-end module described in \S\ref{ssec:HiCOLA_precomp}.
Note also that \cite{Fiorini:2022srj} found that modified gravity COLA simulations using the screening approach described in \cite{2015PhRvD..91l3507W} reproduced inaccurate results for some 3-point statistics in cases where the effect of screening is strong on quasi-non-linear scales, so it is possible \HiCOLA{} would do the same, although we restrict our study to 2-point statistics in this work.

We note in passing that the appearance of a square root in Eq.~(\ref{eq:MGforce_simple}) is potentially dangerous in voids, where the underdensities could result in a negative $\chi$ (Eq.~(\ref{eq:chiexpunsmoothed})). This issue is known to occur in Galileon models for some parameter values; its physical interpretation is poorly understood. Existing simulations simply set the scalar field to a constant value when such pathologies occur. We do not encounter this problem in the models used in \S\ref{sec:cuGalsims} and \S\ref{sec:ESSsims}; a pre-check for likely pathological solutions can be implemented in future work.

Finally, we remind the reader that Eq.~(\ref{eq:MGforce_simple}) is specialised to the Vainshtein mechanism due to assumptions made in \S\ref{ssec:Horndeski_perts}. We note that the linearised limit of Eq.~(\ref{eq:MGforce_simple}) depends on redshift only; this implies that the linear growth rate will remain scale independent, see also \S\ref{ssec:HiCOLA_2LPT} . Scale-independence of the linear growth rate will {\bf not} be maintained under other screening mechanisms.

\subsection{Linear-screened force interpolation} \label{ssec:force_interpolation}

Expressions for the screened fifth force derived in the manner above are known to underestimate the linear modification to gravity at large scales when used in \nbody{} simulations.
This behaviour is explained clearly in Appendix~A of \cite{Khoury:2009tk} in the context of DGP, which, like the theories we consider here, features a Vainshtein screening mechanism. 
Essentially, when we assume spherical symmetry the $\partial_i\partial_j\phi$ term that is present in Eq.~(\ref{eq:seom}) becomes proportional to the $\nabla^2\phi$ term and is thus absorbed into it. While this simplifies the system of equations and allows us to derive our expressions above, it does mean our solution is approximate compared to the exact solution where spherical symmetry is not assumed and the $\partial_i\partial_j\phi$ term remains.
The key consequence is that for high densities $\delta_{\rm m}\gg1$, our approximation underestimates the far-field effect of $\phi$. This is because, for the exact solution, the fifth force $\nabla\phi$ decreases more slowly than the $1/r^2$ that we see in Eq.~(\ref{nat}) for our approximate solution.
This topic is discussed further in \cite{HU2009230, PhysRevD.81.063005,Schmidt:2009sg,Hui:2009kc,Scoccimarro:2009eu,Chan:2009ew}.

Fortunately, it is easy to compute the correct far-field fifth force due to a dense, highly-screened source using linear theory. However, this linear solution of course includes no screening, so will overestimate the fifth force on small scales.

To summarise, we have a solution for the fifth force from linear theory that is accurate at large scales but not at small scales due to the lack of screening. We also have a solution for the fifth force derived in \S\ref{sec:screened_fifth_force} that is accurate at small scales due to the inclusion of screening, but inaccurate at large scales due to the assumption of spherical symmetry. Thus to circumvent the issues with each, we can interpolate at large scales between the linear solution for the total force $F_{\rm lin}$ and the approximate solution for the total force $F_{\rm approx}$ computed by Eq.~(\ref{eq:MGforce_simple}) to ensure we get the correct behaviour at all scales.

One way to do this is to use a low-pass filter such as
\begin{align}
f &= \exp\left(-\frac{1}{2}\frac{k^2}{k_{\rm filter}^2}\right)\,,
\\
F_{\rm filtered} &= f F_{\rm lin} + (1-f) F_{\rm approx}\,, \label{eq:force_interp}
\end{align}
which has a single parameter $k_{\rm filter}$ to calibrate. This ensures that at large scales, $f\rightarrow1$ and $F_{\rm filtered}\rightarrow F_{\rm lin}$; whereas at small scales $f\rightarrow0$ and $F_{\rm filtered}\rightarrow F_{\rm approx}$.
Other interpolation filters, including those where the width and midpoint can be varied independently, could be used instead.

We discuss how to calibrate the value of $k_{\rm filter}$ in Appendix~\ref{sapp:calibration}; see Fig.~\ref{fig:calibration_k_filter} specifically to see how it affects the clustering of matter in cubic Galileon gravity. 

\subsection{Density field smoothing} \label{ssec:dens_smoothing}

The density in a particular grid-cell of a cosmological simulation is computed using the volume of that grid-cell, and therefore the resolution of the simulation. As we increase the resolution the volume of each grid-cell decreases, and thus the density in cells that contain a simulation particle becomes large.
Since our screening coefficient $S$ is dependent on density, this means our screened fifth force is very sensitive to the simulation resolution, as mentioned by \cite{2015PhRvD..91l3507W}.

To alleviate this issue, we apply a Gaussian smoothing to the density field before computing the screened fifth forces. Since a convolution in real-space is equivalent to a product in Fourier-space, the density smoothing is applied as
\begin{align}
\delta_{\rm smooth}(k,R_{\rm smooth}) = \delta(k) \exp\left({-\frac{k^2 R_{\rm smooth}^2}{2}}\right)\,, \label{eq:dens_smooth}
\end{align}
where $R_{\rm smooth}$ is the smoothing radius to be calibrated. Alternative smoothing filters, such as a top-hat or a sharp cut-off, could be used instead.

We discuss how to calibrate the value of $R_{\rm smooth}$ in Appendix~\ref{sapp:calibration}; see Fig.~\ref{fig:calibration_R_smooth} specifically to see how it affects the clustering of matter in cubic Galileon gravity.


\section{\HiCOLA{} implementation} \label{sec:COLA_implemenation}

The approach for estimating screened fifth forces in reduced Horndeski gravity that we outline in \S\ref{sec:screened_fifth_force} could be implemented in any \nbody{} simulation code. In this work, we have chosen to implement it on top of the COLA solver contained within \fml{}\footnote{Information about \fml{} can be found at \url{https://fml.wintherscoming.no/} and the COLA solver is specifically available at \url{https://github.com/HAWinther/FML/tree/master/FML/COLASolver}.}. We call the resulting modified code \HiCOLA{}\footnote{Can be found at \url{https://github.com/Hi-COLACode/Hi-COLA}, along with documentation to guide those who wish to use \HiCOLA{}.} (Horndeski-in-COLA).

In the COLA approach to simulating structure formation \cite{OriginalCOLA}, instead of solving for the full particle trajectories, we solve for the deviations of the full trajectories relative to the trajectories predicted by 2$^{\rm nd}$ order Lagrangian perturbation theory (2LPT). The evolution of the particles on large scales will be very close to that predicted by 2LPT. Thus in this approach, we can decrease the number of simulation timesteps to trade accuracy at small scales for overall simulation speed while maintaining good accuracy at large scales.

The COLA solver in \fml{} has been modified in three key areas to create \HiCOLA{}: the background expansion, 2LPT, and particle-mesh (PM) sections of the code. Conveniently, it already contains many of the necessary components we have discussed above in \S\ref{sec:screened_fifth_force}, including density-dependent screening (via the method described in \cite{2015PhRvD..91l3507W}, although for DGP gravity only), linear-screened force interpolation, and density field smoothing.

One entirely new addition in \HiCOLA{} is the Python module component which takes a Horndeski Lagrangian as input, solves the background for that model following the approach described in \S\ref{sec:Horndeski_cosmo}, then outputs $\beta$ and $\chi/\delta_{\rm m}$ as computed by Eqs.~(\ref{eq:coupling}) \& (\ref{eq:chioverdelta}). Various quantities from these pre-computations are then given to the COLA solver as additional input and used as required (for example to compute the screened fifth force).

In the remainder of this section, we describe the various components of \HiCOLA{} in more detail.

\subsection{Background computation} \label{ssec:HiCOLA_precomp}

The \HiCOLA{} front-end module directly receives as inputs the functions $K(\phi,X)$, $G_3(\phi,X)$ and $G_4(\phi)$ appearing in the Horndeski Lagrangian, subject only to a few manipulations to factor out the dimensions of these objects (described in Appendix \ref{app:mass}). This immediate connection between action and outputs is designed to enable rapid investigation of theoretical models and their observational consequences for large-scale structure. We note this is something often obscured by parameterisations of MG at the field equation level, which usually cannot be linked to terms in the gravitational action in a straightforwards manner.

The front-end module is responsible for the pre-computation of cosmological background quantities. These are $E$ and $E^{\prime}$ (the Hubble rate and its derivative), the scalar field trajectory $\phi$, and the energy densities $\Omega_{\rm m}$, $\Omega_{\rm r}$, $\Omega_\Lambda$ and $\Omega_\phi$. From the scalar field and Hubble solutions the dimensionless coefficients in Eq.~(\ref{eq:alpha_def}) are computed, followed by the combinations $\cal B$ and $\cal C$ in Eq.~(\ref{eq:BCmu}). These enter the coupling function $\beta$ (Eq.~(\ref{eq:coupling})) and the time-dependent object $\chi/\delta_{\rm m}$ (Eq.~(\ref{eq:chioverdelta})), the latter of which is needed for the screening coefficient $S$ (Eq.~(\ref{eq:screen_coeff})). $\beta$ and $\chi/\delta_{\rm m}$, together with $E$ and $E^\prime$, are passed into the COLA solver to compute modified forces as described in \S\ref{sec:screened_fifth_force}.

Pragmatically, the first step is for the user to specify forms for the functions $K$, $G_3$ and $G_4$. Our front-end module then forms the necessary derivatives of these objects and constructs the equations of motion shown in \S\ref{sec:Horndeski_cosmo}, utilising the Python symbolic computation library \texttt{SymPy}\footnote{\url{https://www.sympy.org/en/index.html}} to do so. The equations of motion are manipulated so that they provide the derivatives $E^{\prime}$ and $\phi^{\prime\prime}$ in a solvable format. Note that although the Horndeski equations are generally quite complicated, they remain linear in $E^\prime$ and $\phi^{\prime\prime}$. Once constructed symbolically, the resulting expressions are then rendered useable by other Python functions and solved numerically.

A number of support modules are available to visualise the cosmological background solutions, and also to perform further tasks. In particular, one module performs scans over sets of model parameters and initial conditions, locating those that result in background expansion histories consistent with given criteria. This procedure is described in more detail in \S\ref{ssec:cuGal_viability} and \S\ref{ssec:ESS_viability}\footnote{We highlight that it would be inconsistent to use \fml{}'s capabilities to consider more complicated extensions such as curvature or non-zero neutrino masses in our \HiCOLA{} simulations, because these are not currently implemented in the front-end background solver.}.

\subsection{2LPT} \label{ssec:HiCOLA_2LPT}

Computing the particle trajectories predicted by 2LPT is a vital component of the COLA method. The LPT equations for the first and second order growth factors implemented in \fml{} are
\begin{align}
D_1^{\prime\prime} &= \frac{3}{2}\frac{\Omega_{\rm m0}}{E^2a^3} \mu^{(1)}D_1 - \left( 2+\frac{E^{\prime}}{E} \right) D_1^{\prime}\,,
\label{eq:lin_growth}
\\
D_2^{\prime\prime} &= \frac{3}{2}\frac{\Omega_{\rm m0}}{E^2a^3} \mu^{(2)}\left[D_2-D_1^2\right] - \left( 2+\frac{E^{\prime}}{E} \right) D_2^{\prime}\,,
\end{align}
where $\mu^{(1)}$ \& $\mu^{(2)}$ parameterise the effect of modified gravity in the Poisson equation at first and second order respectively. For a more detailed overview of LPT in modified gravity theories, see \cite{Winther2017, Wrightetal2017, Aviles:2017aor, Aviles:2018qot}.

In the parameterisation above, $\mu^{(1)}$ is simply
\begin{align}
\mu^{(1)}=1+\beta\,,
\end{align}
where $\beta$ is the coupling from Eq.~(\ref{eq:MGforce_simple}). Note that $\beta$ depends only on redshift, meaning that scale-independent growth is maintained. When the form of $\mu^{(2)}$ in this parameterisation is derived for modified gravity theories (for example, see Appendix~B of \cite{BoseKoyama2016} for the derivation in Horndeski gravity), we expect $\mu^{(2)}\neq\mu^{(1)}$ as $\mu^{(2)}$ should include the leading order non-linear screening behaviour. However, for simplicity in \HiCOLA{}, we choose to make the approximation $\mu^{(2)}=\mu^{(1)}$. We have tested this does not affect the $P(k)$ output by our simulations above the $1\%$ level for the cubic Galileon theory; we expect this to hold for ESS gravity, given the similarity of the two theories. Our use of this 2LPT approximation is further supported by the fact that, in the COLA method, the 2LPT prediction is essentially only used as a first guess that is then corrected by the PM part of the code. While these PM corrections dominate on small scales, the 2LPT does normally impact the simulation at large scales due to the low number of PM timesteps typical of COLA simulations. However, on these large scales there is no non-linear screening and thus our approximation is valid as $\mu^{(2)}\rightarrow\mu^{(1)}$. See \cite{Valogiannis:2016ane} for another work where the role of $\mu$ in LPT is simplified.

We acknowledge that the above approximation may be less appropriate for other theories within Horndeski gravity, such as those that demonstrate screening on quasi-linear scales, and may introduce larger errors for quantities other than the real-space matter power spectrum, such as bispectra or halo statistics. Our 2LPT treatment could be improved by extending the approach described in \cite{Song:2021msd} for cubic Galileon gravity to general Vainshtein-screened Horndeski theories.

\subsection{Forces between particles} \label{ssec:HiCOLA_forces}

To include the effect of the screened fifth force in the PM part of \HiCOLA{}, we implement Eq.~(\ref{eq:MGforce_simple}). Fortunately, the \fml{} code already contains an implementation of density-dependent screening for DGP gravity following the method described in \cite{2015PhRvD..91l3507W}, which we build upon to implement Eq.~(\ref{eq:MGforce_simple}). We spline the two time-dependent quantities $\beta$ and $\chi/\delta_{\rm m}$ produced as described in \S\ref{ssec:HiCOLA_precomp} for the full redshift range of the simulations, such that they can be quickly read and given to Eq.~(\ref{eq:MGforce_simple}) whenever the simulation needs to compute the screened fifth force.

Additionally, the existing \fml{} code already allows for interpolation between the linear and screened force solutions at large scales using Eq.~(\ref{eq:force_interp}), and smoothing of the density field $\delta_{\rm m}$ before it is passed to Eq.~(\ref{eq:MGforce_simple}) using Eq.~(\ref{eq:dens_smooth}). We discussed the importance of these two steps in \S\ref{ssec:force_interpolation} and \S\ref{ssec:dens_smoothing} respectively. These two steps introduce two new free parameters: $k_{\rm filter}$ \& $R_{\rm smooth}$.

To calibrate these two free parameters, we vary them to maximise the agreement between the output of \HiCOLA{} and the equivalent from \nbody{} codes that compute the screened fifth force directly via slowly solving the Klein--Gordon equation. 
Ideally, this calibration should be done for each modified gravity theory for which we intend to use our approximate method. However, this presents a problem, as \nbody{} codes that directly solve the Klein--Gordon equation only exist for a handful of individual modified gravity theories, whereas we want our approximate method to be applicable across a wide section of modified gravity theory space.

For now, we carry out the calibration using \nbody{} simulations for cubic Galileon gravity only, and then, using these calibrated parameters, validate \HiCOLA{} against the same cubic Galileon \nbody{} simulations. This validation yields an understanding of what regimes our fast, approximate \HiCOLA{} code produces accurate results for. We present this validation in Appendix~\ref{app:valid}, and the calibration specifically in Appendix~\ref{sapp:calibration}.


\section{Cubic Galileon simulations with \HiCOLA{}} \label{sec:cuGalsims}

As noted earlier, although \HiCOLA{} is capable of creating simulations for any theory within the Horndeski Lagrangian, in this work we focus in-depth on two specific models. 
We first explore the parameter space of the well-studied cubic Galileon model that we introduced in \S\ref{ssec:cuGal_theory}.

We note that throughout this section, we use the same base \lcdm{} cosmological parameters as \cite{Barreira:2013eea}; we recap the key values in Table~\ref{tab:base_cosmo_params}. However, we do not use the same cubic Galileon parameters. We will describe the cubic Galileon parameters we use instead below. In this section, we focus on the behaviour that is essential for understanding the non-linear matter power spectrum result. We investigate the cubic Galileon phenomenology in greater detail in Appendix~\ref{app:cuGal_results}.

\begin{table}
\begin{center}
\begin{tabular}{|c|c|c|c|}
\hline
$\Omega_{\rm r0}h^2$   & $\Omega_{\rm c0}h^2$ & $\Omega_{\rm b0}h^2$ & $h$    \\ \hline
$4.28\times10^{-5}$ & 0.1274            & 0.02196           & 0.7307 \\ \hline
\end{tabular}
\caption[Base cosmological parameters]{Key base cosmological parameters used throughout; they match those used for the simulations in \cite{Barreira:2013eea}.}
\label{tab:base_cosmo_params}
\end{center}
\end{table}

\subsection{Viability of expansion history} \label{ssec:cuGal_viability}

In \S\ref{ssec:cuGal_theory} we explained that after exploiting the rescaling symmetry of cubic Galileon gravity, we are left with two free parameters for the model, $E_{\textrm{dS}}$ and $f_{\phi}$\footnote{As a brief reminder, the dark energy sector in our simulations can be made up of both a scalar field and cosmological constant, in proportions controlled by $f_\phi$ (Eq.~(\ref{eq:fphi})); $f_{\phi}=1$ implies no cosmological constant contribution. $E_{\rm dS}$ is the value of the rescaled Hubble parameter $E=H/H_0$ in a future de Sitter limit. }. Additionally we have the initial conditions for $\{\phi^{\prime}, \Omega_{\rm m}, \Omega_{\rm r}, \Omega_{\Lambda}\}$ that need to be specified. 
For $\{\Omega_{\rm m}, \Omega_{\rm r}\}$ we use the values in Table~\ref{tab:base_cosmo_params} which match those of \cite{Barreira:2013eea}, which were obtained from fitting cubic Galileon gravity to WMAP 9yr results, SNLS supernovae and BAO measurements from SDSS. The initial value of $\Omega_\Lambda$ is then fixed by ensuring $\Omega_\Lambda+\Omega_\phi=\Omega_{\Lambda}^{\rm LCDM}$ for a given value of $f_{\phi}$. The initial condition for $\phi^{\prime}$ is determined through enforcement of the closure equation at $z=0$. 

\begin{figure*}
    \centering
    \includegraphics[width=1.\textwidth]{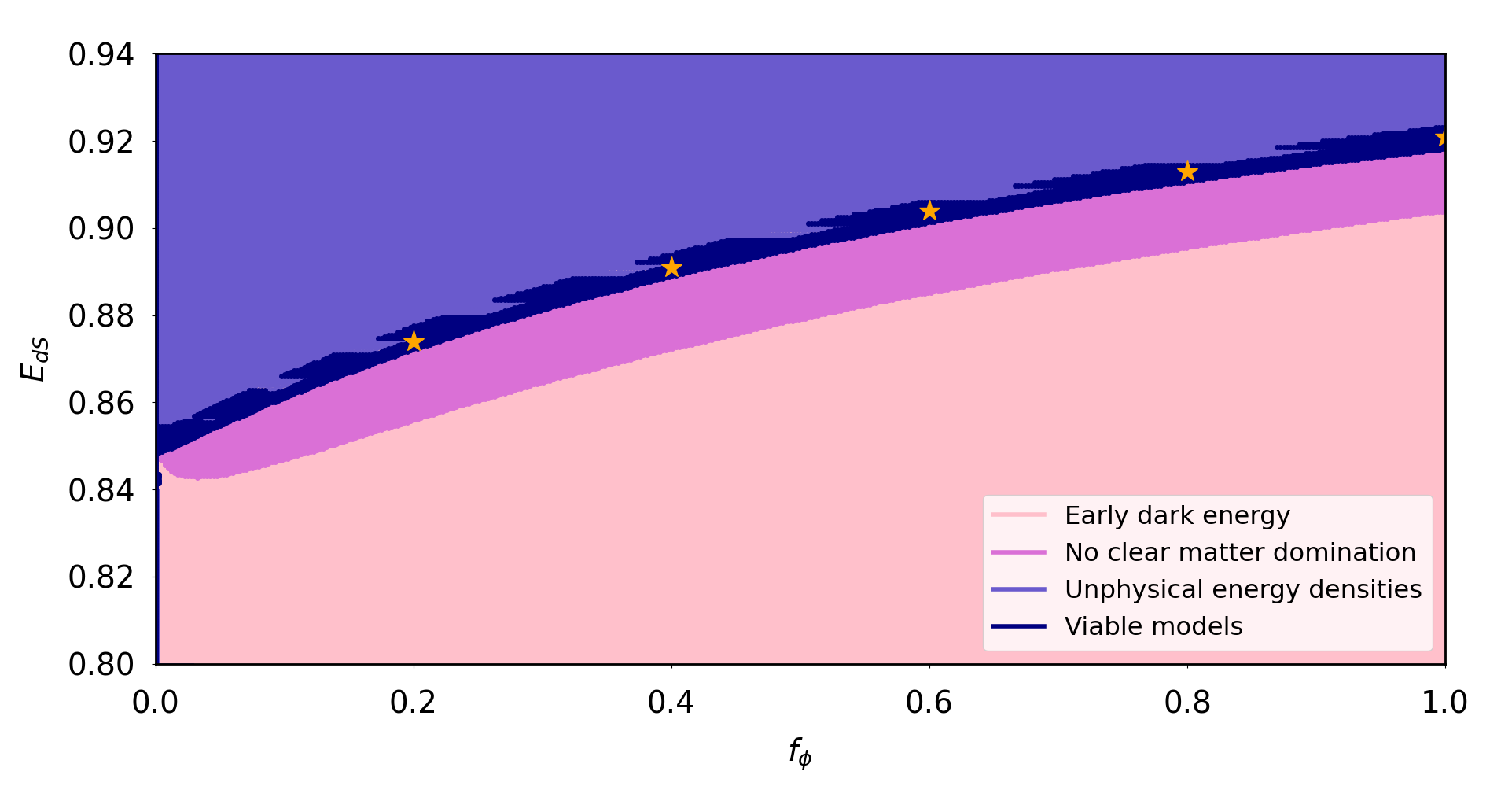}
    \caption{A plot showing the categorisation of $(f_{\phi}, E_{\rm dS})$ space according to the viability criteria. The purple-blue \crule[unphysical]{0.3cm}{0.3cm} region at the top corresponds to cubic Galileon models where $\Omega_i \notin [0-\epsilon,1+\epsilon]$, where $\epsilon=10^{-6}$ is a numerical tolerance value. The bottom-most light pink \crule[earlyde]{0.3cm}{0.3cm} region corresponds to cubic Galileon models with early dark energy. The dark pink \crule[nomatterdom]{0.3cm}{0.3cm} region corresponds to cubic Galileon models where the matter energy density fails to reach a critical value, meant to represent matter energy domination. Finally, the dark blue region \crule[viable]{0.3cm}{0.3cm} shows cubic Galileon models that pass all criteria, with the orange \crule[star]{0.3cm}{0.3cm} stars representing the five cubic Galileon models featured in this paper. The `fins' in the dark blue region are numerical artefacts and depend on the value of $\epsilon$ (larger values have the effect of increasing the fin sizes). For simplicity, models that satisfy multiple criteria have been coloured so as to keep the coloured regions in the plot as continuous as possible.
    }
    \label{fig:cG_viability}
\end{figure*}

Thus, the question that remains is how to select appropriate values of $E_{\textrm{dS}}$ and $f_{\phi}$. We scan over the joint parameter space of $E_{\textrm{dS}}$ and $f_{\phi}$ in order to find cubic Galileon models with expansion histories that are broadly similar to \lcdm{}, since this is what observations currently support. In practice, this is achieved by subjecting the background solutions to a set of criteria that reject models which are either: i) qualitatively different from \lcdm{}, or ii) have unphysical solutions. These rejection criteria are as follows, and are checked in the order presented: 
\begin{enumerate}
     \item $\Omega_i < - \epsilon$ or $\Omega_i>1+\epsilon$, or $\Omega_i = \texttt{NaN}$ at any point, where $i$ is one of $\{m,r,\Lambda,\phi\}$. This identifies when energy densities become unphysical. This criterion is implemented to within a small tolerance of $\epsilon = 10^{-6}$ to allow for numerical errors. In Fig.~\ref{fig:cG_viability} these models are coloured purple-blue  \crule[unphysical]{0.3cm}{0.3cm}. 
    \item $\max\left[ \Omega_{\textrm{DE}}(z>1) \right] > \Omega_{\textrm{DE}}(z=0)$. When satisfied, this indicates that there exists a value in the dark energy density at $z>1$ that is greater than its value at $z=0$. This criterion identifies when a model is exhibiting early dark energy, though it allows for dark energy density to exceed values today in the redshift range $0 \leq z < 1$. In Fig.~\ref{fig:cG_viability} these models are coloured light pink \crule[earlyde]{0.3cm}{0.3cm}. 
    \item $\max\left[ \Omega_{\rm m} \right] < \Omega_{\rm m}^{\textrm{crit}}$. This indicates that the maximum value for the matter energy density fails to equal or exceed a critical value, $\Omega_{\rm m}^{\textrm{crit}}$. This criterion identifies when a model fails to produce a matter-dominated era in the universe. For the results in this paper, $\Omega_{\rm m}^{\textrm{crit}} = 0.99$. In Fig.~\ref{fig:cG_viability} these models are coloured dark pink \crule[nomatterdom]{0.3cm}{0.3cm}.
    \item If all above criteria are passed, then the model's expansion history is considered viable. In Fig.~\ref{fig:cG_viability} these are coloured dark blue \crule[viable]{0.3cm}{0.3cm}.
\end{enumerate}

\noindent For cubic Galileon gravity, we show the model space in terms of the criteria listed above in Fig.~\ref{fig:cG_viability}. We can see that for cubic Galileon gravity, there is a dark blue band of fully viable, LCDM-like models in $(f_{\phi}, E_{\rm dS})$ parameter space. The `fins' in this band are numerical artefacts that depend on $\epsilon$. Larger values of $\epsilon$ tend to increase the sizes of these fins. The models that we study in this paper are picked from this viable band, and the model parameter values are listed in Table~\ref{tab:cuGal_params} and represented by orange stars in Fig.~\ref{fig:cG_viability}. They all possess the usual traits of a Hubble rate that decreases with time, a radiation-dominated phase followed by a period of matter domination, followed by a dark energy-dominated phase at late times.
A further investigation of why cubic Galileon gravity has a particular region of viability, along with analogous discussions for the ESS model, will be studied in a future work. As a reminder, one should note that in the context of this work, models are called ``viable" after passing \textit{background-level} criteria; constraints coming from large-scale structure data are not currently considered in the model-finding process.

Given the relative narrowness of the viable band in Fig.~\ref{fig:cG_viability} we can consider an effective one-parameter parameterisation of cubic Galileon gravity.
Excluding the fin artefacts, one can fit a function to the upper edge of the remainder of the band, and use it to determine the corresponding maximum valid value of $E_{\rm dS}$ for a chosen value of $f_\phi$. In the next section we show \HiCOLA{} simulations for the five $\{f_{\phi}, E_{\rm dS}(f_{\phi})\}$ cases listed in Table~\ref{tab:cuGal_params}, determined in this way.

We note that one could also use priors like positivity bounds \cite{Melville:2019sky, Melville:2022ykg} to reduce the parameter space being considered, particularly before a scan is performed. While it appears the sign of $k_{1\textrm{dS}}$ is consistent with the bounds detailed in \cite{Melville:2022ykg}, as the authors explain, these bounds are flat-space calculations. It would therefore be of interest to utilise similar bounds derived in curved spacetimes like de Sitter space as priors in future parameter scans. Such an approach may be of significant aid in reducing the time taken to complete scans in the higher-dimensional ESS parameter space discussed in \S\ref{ssec:ESS_viability}.

\begin{table}
\begin{center}
\begin{tabu}{|[2pt] c | c |[2pt] c | c |   }
\tabucline[2pt]{1-2} \hline
$f_{\phi}$   & $E_{\rm dS}$ & $k_{1\textrm{dS}}$ & $g_{31\textrm{dS}}$ \\ \hline
0.2 & 0.874 & -1.475 & 0.492 \\
0.4 & 0.891 & -2.734 & 0.911 \\
0.6 & 0.904 & -3.885 & 1.295  \\
0.8 & 0.913 & -4.963 & 1.654 \\[-2pt]
1.0 & 0.921 & -6.000 & 2.000 \\[-2pt] \tabucline[2pt]{1-2} \hline
\end{tabu}
\caption[de Sitter-cubic Galileon parameters]{Parameters for the five cubic Galileon models considered in \S\ref{sec:cuGalsims}. $f_{\phi}$ and $E_{\rm dS}$ are highlighted in bold as these are the independent parameters. $k_{1\textrm{dS}}$ and $g_{31\textrm{dS}}$ are fixed by these values as per Eq.~(\ref{eq:CuGal_dS}).}\label{tab:cuGal_params}
\end{center}
\end{table}

\subsection{Simulation setup}\label{ssec:cuGal_sim_setup}

\subsubsection{Initial conditions}\label{sssec:cuGal_ICs}

A common approach for generating initial conditions (ICs) in \nbody{} simulations is to use backscaling (referred to as `Newtonian backscaling' in \cite{Fidler:2017ebh}). The backscaling approach is designed to account for the flaw of \nbody{} simulations not fully including the effects of radiation during their forward evolution. This flaw manifests when such an \nbody{} code is used with ICs generated directly at the initial simulation redshift.

To generate backscaled ICs, the linear power spectrum at a target redshift (often $z=0$), generated by a Einstein--Boltzmann code including the full effects of radiation, is evolved backwards to the initial simulation redshift using the flawed linear growth factors computed by the simulation code (which do not include the full effects of radiation). This ensures that, when the simulation evolves forwards to the target redshift (effectively using those same flawed linear growth factors on linear scales), it recovers the correct linear power spectrum containing the full effects of radiation as computed by the Einstein--Boltzmann code.

In this work, we do not have access to a Einstein--Boltzmann code that has been modified to account for either of our two modified gravity cosmologies. This prevents us from not only creating appropriately backscaled ICs, but also from generating modified gravity ICs directly at the initial simulation redshift. What we chose to use instead is to use \lcdm{} ICs generated directly at the initial simulation redshift, which in our case is $z=49$. Specifically, to set our initial conditions (ICs), we use \fml{}'s existing functionality to read directly from the particle data in the $z=49$ snapshot of \cite{Barreira:2013eea}, such that our ICs are identical to theirs.

With our approach, in addition to the error caused by the lack of backscaling, any deviation from \lcdm{} at $z=49$ will introduce error via the ICs. We can estimate the size of this error from the ratio of modified gravity and \lcdm{} growth factors at $z=49$, a quantity which we plot in Fig.~\ref{fig:cuGal_lin_growth} for our strongest cubic Galileon case where $(f, E_{\rm dS})=(1, 0.921)$.
We find the ratio of cubic Galileon and \lcdm{} growth factors at $z=49$ is $1.00225$, or a $0.225\%$ deviation from \lcdm{}. Ignoring the effects of radiation, we can approximate the ratio of cubic Galileon and \lcdm{} linear power spectra at $z=49$ to be $1.00225^2\approx1.00451$, or a $0.451\%$ deviation from \lcdm{}. This small error in the linear power spectra at $z=49$ will propagate to give a similarly-sized offset at large, linear scales in the non-linear power spectra measured from the simulation output at late times, and may also introduce transient effects at non-linear scales.
We discuss the linear growth factor in cubic Galileon gravity further in Appendix~\ref{sapp:cuGal_lin_growth}.

\subsubsection{Other simulation specifics}\label{sssec:cuGal_other_sim_setup}

We use a boxsize of $L=400~{\rm Mpc}/h$ with $N_{\rm part}=512^3$ particles per dimension and a force grid of size $N_{\rm mesh}=3(N_{\rm part})^{1/3}=1536$. We use $40$ timesteps linearly spaced in $a$ between $a_{\rm ini}=0.02$ and $a=1$. This choice aims to balance accuracy at small scales against computational cost, and is based on both our previous experience with COLA \cite{Valogiannis:2016ane,Winther2017,Wrightetal2017,Ramachandra:2020lue,Fiorini:2021dzs} and the thorough study of the optimal number of timesteps in COLA simulations from Section~4.1 of \cite{2016MNRAS.459.2327I}.
The base cosmological parameters used are listed in Table~\ref{tab:base_cosmo_params}, again matching the simulations from \cite{Barreira:2013eea}.\footnote{Although in this section we don't use the same cubic Galileon parameters as \cite{Barreira:2013eea} or compare our simulations against theirs, it was still convenient to use the same ICs and base cosmological parameters as we use for the simulations in Appendix~\ref{app:valid} where we do directly compare against \cite{Barreira:2013eea}.} 
As described in \S\ref{sec:COLA_implemenation}, our \HiCOLA{} simulation takes the time-dependent $E$, $E^{\prime}/E$, $\beta$, and $\chi/\delta_{\rm m}$ functions produced by the \HiCOLA{} front-end module as input.

In addition to the reference \lcdm{} simulation and the full cubic Galileon gravity simulations where the expansion is modified relative to \lcdm{} and the screened fifth forces are computed, we run a simulation that is a hybrid between \lcdm{} and full cubic Galileon gravity for each of the five cases listed in Table~\ref{tab:cuGal_params}.
This type of hybrid, where the expansion of the simulation matches that of the full cubic Galileon simulation but the forces are purely GR, is referred to as a QCDM simulation in \cite{Barreira:2013eea}\footnote{The error introduced via the ICs discussed in \S\ref{sssec:cuGal_ICs} will be marginally different in the QCDM simulations, as the absent modification to gravity has a very small impact on the linear growth factor at $z=49$.}.
By taking the ratio between the outputs of the QCDM and \lcdm{} simulations, we can isolate the effect of cubic Galileon gravity's modified expansion history. By taking the ratio between the outputs of the full cubic Galileon and QCDM simulations, we can isolate the effect of cubic Galileon gravity's screened fifth force. These two isolated effects are shown for the real-space matter power spectrum in the lower panels of Figs.~\ref{fig:cuGal_background_Pkbg} and \ref{fig:cuGal_coupl_screenfac} respectively.
In Appendix~\ref{sapp:cuGal_Pk_breakdown}, we use these hybrid simulations and others to further investigate the isolated impacts of various components that are modified in cubic Galileon gravity relative to \lcdm{}.

Thus the final requirement is two simulations for each of the five cubic Galileon cases described in Table~\ref{tab:cuGal_params}, plus a single \lcdm{} reference simulation; a total of 11 simulations.

The real-space non-linear matter power spectra $P(k)$ are measured from the simulation snapshots using \fml{}'s built-in on-the-fly power spectrum measurement option.
To ensure we can trust our results, we have validated \HiCOLA{} against traditional \nbody{} simulations of cubic Galileon gravity \cite{Barreira:2013eea,Barreira:2015vra} in Appendix~\ref{app:valid}. This validation indicates that the simulation setup described above allows our results for boost factors (ratios of real-space matter power spectra, here specifically between different cosmologies, for example cubic Galileon gravity vs \lcdm{})
\footnote{Accurate boost factors are still very useful, albeit in a less direct way than accurate absolute power spectra. For example, a beyond-\lcdm{} vs \lcdm{} `cosmology boost factor' computed by fast, approximate code (such as \HiCOLA{}) can be multiplied by the \lcdm{} power spectrum computed by a more accurate method (such as traditional \nbody{}) to yield an accurate beyond-\lcdm{} power spectrum. Other kinds of boost factor can also be used to shift cosmological parameters from a reference set (a `parameter boost factor'), or even just evolve a power spectrum from a reference redshift to a target redshift (a `redshift boost factor'). One could even design a boost factor to do multiple tasks at once, for example to transform a \lcdm{} power spectrum with base cosmological parameters `A' at $z=49$ to a beyond-\lcdm{} power spectrum with base cosmological parameters `B' at $z=0$. See \cite{Brando:2022gvg} for more information, although note that in their terminology our boost factors are referred to as `response functions'.} 
to be trusted within $2.5\%$ up to $k_{\rm max}=1.2~h/{\rm Mpc}$.

At scales smaller than $k_{\rm max}$ our force computation becomes too inaccurate to successfully reproduce the results of traditional \nbody{} simulations to the level we desire. This is due to the combination of the low number of timesteps (a defining feature of COLA simulations), the force resolution of our simulations (set by the ratio $N_{\rm mesh}/N^{1/3}_{\rm part}$), and the approximations involved in our specific approach to estimating the screened fifth force.

\subsection{Simulation results} \label{ssec:cuGal_sim_res}

\begin{figure*}
\includegraphics[width=1.\textwidth]{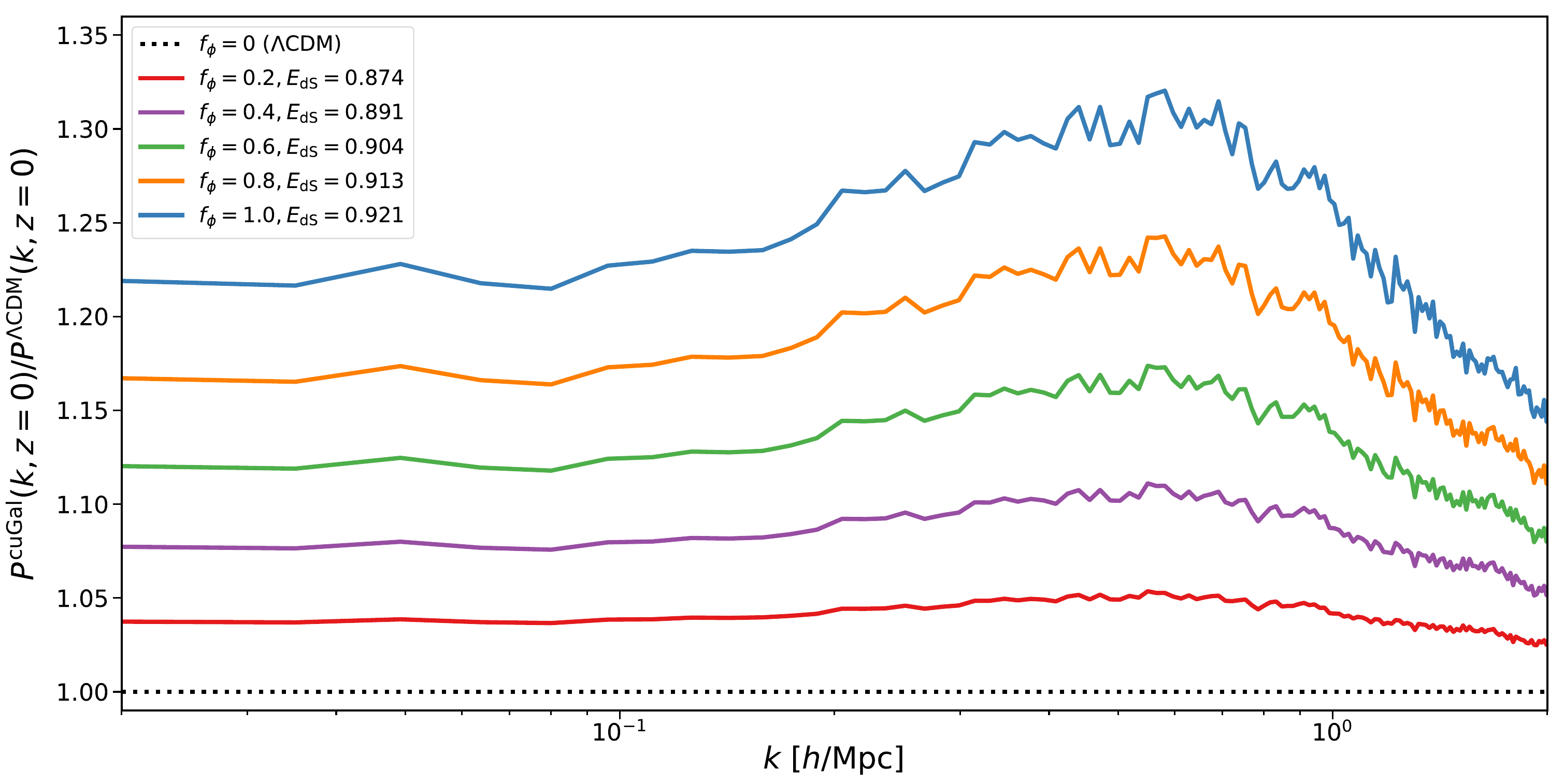}
\caption[Ratios of real-space non-linear matter power spectra at $z=0$ in cubic Galileon gravity and \lcdm{}.]{\small Ratios of real-space non-linear matter power spectra at $z=0$ in cubic Galileon gravity and \lcdm{} for a variety of $\left(f_{\phi}, E_{\rm dS}(f_{\phi})\right)$ values.
The equivalent of this figure for ESS gravity is Fig.~\ref{fig:ESS_Pk_GR_ratio}.
}
\label{fig:cuGal_Pk_GR_ratio}
\end{figure*}

We study the effect of cubic Galileon gravity on structure formation by taking the ratio between the $P(k)$ from the full cubic Galileon gravity simulation and the \lcdm{} simulation at $z=0$, which we plot in Fig.~\ref{fig:cuGal_Pk_GR_ratio}.

Firstly, we see that the impacts of cubic Galileon gravity are stronger for increasing values of $f_{\phi}$, the fraction of effective DE contained in the scalar field. This is reassuring, given that we know we should recover \lcdm{} for $f\rightarrow0$.

Secondly, we see the $P(k)$ have an approximately scale-independent enhancement in cubic Galileon gravity relative to \lcdm{} on large scales, and that this enhancement increases slightly with $k$ at intermediate scales. However, this enhancement is then reduced significantly at small scales. But what specifically is responsible for this behaviour?

\begin{figure*}
\includegraphics[width=1.\textwidth]{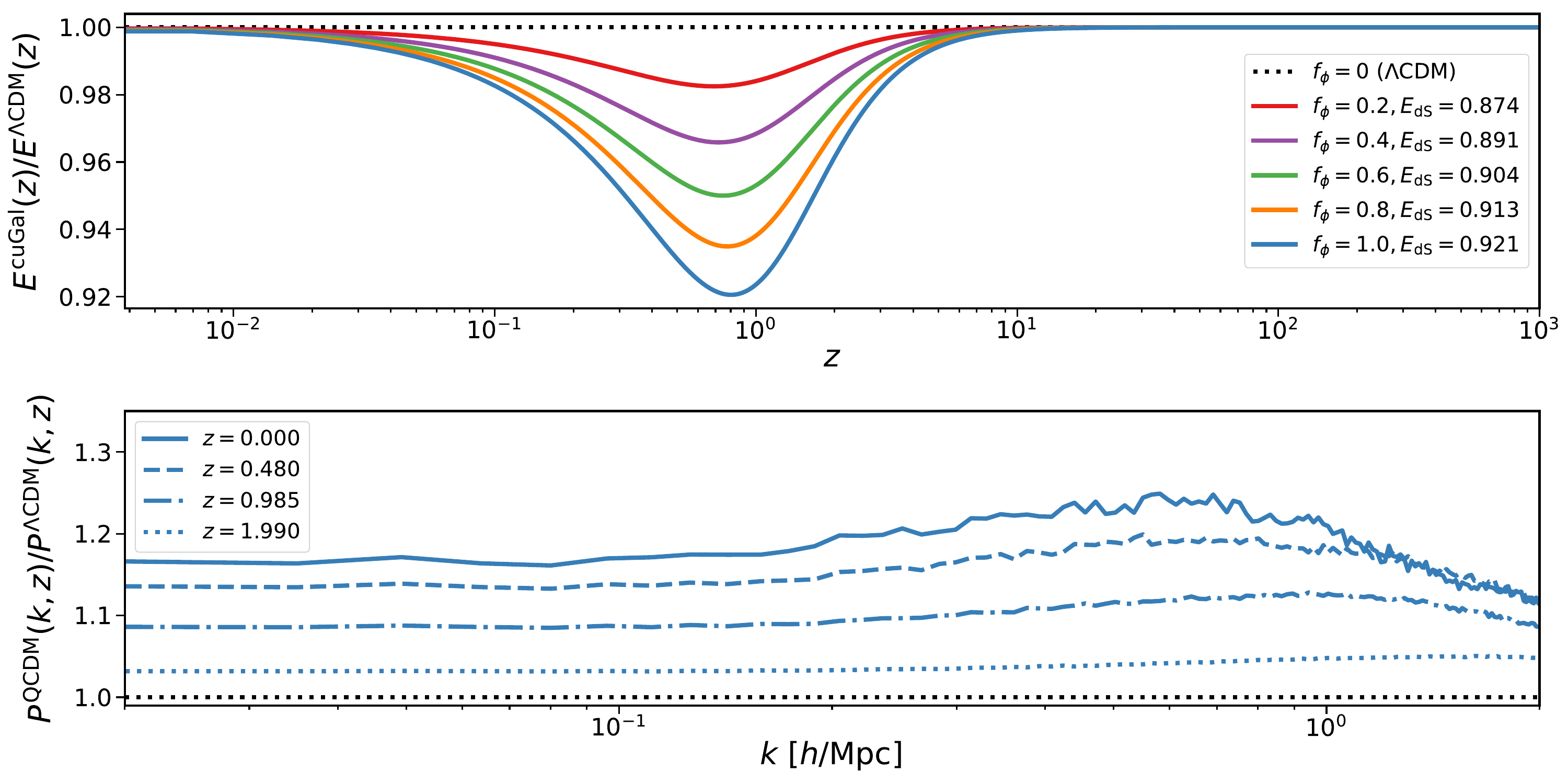}
\caption[Expansion history in cubic Galileon gravity and its effect on structure formation.]{\small \textit{Upper panel}: The ratio of expansion rate $E$ in cubic Galileon gravity and \lcdm{} for the five $\left(f_{\phi}, E_{\rm dS}(f_{\phi})\right)$ cases.
\textit{Lower panel}: The ratio of power spectra in QCDM and \lcdm{} at several redshifts for the $(f_{\phi}=1, E_{\rm dS}=0.921)$ case, which demonstrates the isolated effect of the modified expansion history in cubic Galileon gravity.
The equivalent of this figure for ESS gravity is Fig.~\ref{fig:ESS_background_Pkbg}.
}
\label{fig:cuGal_background_Pkbg}
\end{figure*}

The first key feature of cubic Galileon gravity is that there is a period of slower-than-LCDM expansion around $z\approx1$, as can be seen in the upper panel of Fig.~\ref{fig:cuGal_background_Pkbg}. Cosmic expansion suppresses structure formation; thus the slower-than-\lcdm{} expansion rate reduces this suppression effect, leading to a net enhancement of structure formation relative to \lcdm{}. However, cosmic expansion will have less of a suppressing effect on small scales where structures are locally gravitationally bound. So the impact of the different expansion history in cubic Galileon gravity is to produce an enhancement of the $z=0$ power spectrum on large scales relative to \lcdm{} that falls away towards small scales. This effect can be seen in isolation via the ratio of power spectra in QCDM and \lcdm{} in the lower panel of Fig.~\ref{fig:cuGal_background_Pkbg}. Note that at high redshifts the enhancement is almost scale-independent since the linear regime extends to larger $k$ at early times.

\begin{figure*}
\includegraphics[width=1.\textwidth]{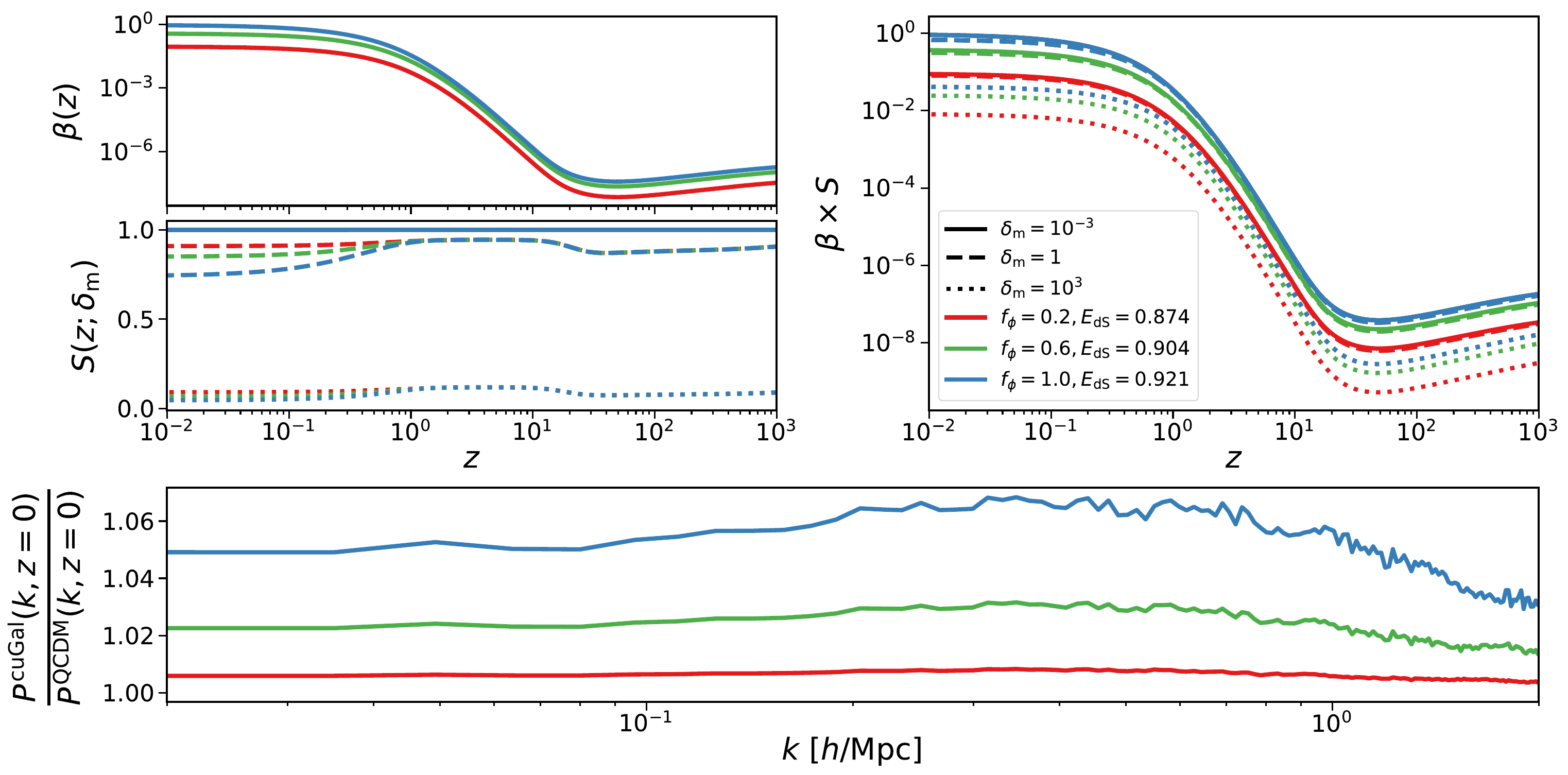}
\caption[Coupling and screening coefficient in cubic Galileon gravity.]{\small The effect of varying $\left(f_{\phi}, E_{\rm dS}(f_{\phi})\right)$ in cubic Galileon gravity on the approximate screened fifth force and its component quantities. \textit{Upper left} shows the coupling (as defined in Eq.~(\ref{eq:coupling})) as a function of redshift. \textit{Middle left} shows the screening coefficient (as defined in Eq.~(\ref{eq:screen_coeff})) as a function of redshift for three different densities. \textit{Upper right} shows the product of the coupling and screening coefficient as a function of redshift for three different densities; this quantity is the approximate screened fifth force. The \textit{lower panel} shows the ratio of power spectra from the full cubic Galileon and QCDM simulations at $z=0$, which demonstrates the isolated impact of the screened fifth force.
We have omitted the $\left(f_{\phi}, E_{\rm dS}(f_{\phi})\right)=(0.8, 0.913)$ and $(0.4, 0.874)$ cases for clarity.
The equivalent figure for ESS gravity is Fig.~\ref{fig:ESS_coupl_screenfac}.
}
\label{fig:cuGal_coupl_screenfac}
\end{figure*}

The second key feature of cubic Galileon gravity is the screened fifth force. In Fig.~\ref{fig:cuGal_coupl_screenfac}, we study the two components of the screened fifth force in cubic Galileon gravity: the coupling $\beta$ and the screening coefficient $S$, which we introduced in \S\ref{ssec:Vainshtein_spherical_symmetry}. The coupling is obtained from the background solutions via Eqs.~(\ref{eq:coupling}). To compute the screening coefficient, we first compute the Vainshtein radius ratio $\chi/\delta_{\rm m}$ via Eq.~(\ref{eq:chioverdelta}). We can then use $\chi/\delta_{\rm m}$ to compute the screening coefficient $S$ for various test values of the overdensity $\delta_{\rm m}$ using Eq.~(\ref{eq:screen_coeff}). Since $G_{G_4}=G_{\rm N}$ for cubic Galileon gravity, the product $\beta S$ is then the strength of the screened fifth force at each test value of $\delta_{\rm m}$. In Fig.~\ref{fig:cuGal_coupl_screenfac} we plot the evolution of $\beta$ with redshift for each of the five cubic Galileon cases described above, as well as the evolution of $S$ and the product $\beta S$ for three test values of $\delta_{\rm m}=\{ 0.1, 1, 10 \}$. Finally, we want to understand how the screened fifth force, which is a function of redshift and density, maps onto the power spectrum, which is a function of redshift and scale. To do so, we also plot the ratio of power spectra from the full cubic Galileon simulation and that of the QCDM simulation in the lower panel. This ratio isolates the effect of the screened fifth force, as both simulations feature the cubic Galileon expansion history.

Firstly, the upper left panel of Fig.~\ref{fig:cuGal_coupl_screenfac} shows that the coupling in cubic Galileon gravity is very small at early times but grows rapidly around $z=1$, leading to a significant fifth force activating at late times. This is consistent with our criteria in \S\ref{ssec:cuGal_viability}, which reject models where the Horndeski scalar field is significant at early times.
Secondly, the lower left panel of Fig.~\ref{fig:cuGal_coupl_screenfac} shows the screening coefficient decreases with increasing density. For low densities, $S\approx1$ meaning that screening is inactive and the full fifth force operates. For high densities, $S\ll1$ indicating that screening is active.
The upper right panel displays the product of the coupling and screening coefficient, and we see, unsurprisingly, that the fifth force activates at late times but is suppressed at high densities due to screening.
Finally, the lowest panel shows the effect of this screened fifth force on the power spectrum is to produce an approximately scale-dependent enhancement to the $z=0$ power spectrum at large scales, which increases slightly with $k$ at intermediate scales, before falling off at small scales as screening activates.

For the interested reader, we study and discuss the phenomenology of cubic Galileon gravity in greater detail in Appendix~\ref{app:cuGal_results}.
But in summary, Fig.~\ref{fig:cuGal_Pk_GR_ratio} shows that the combined impact of the period of slower-than-LCDM expansion and the screened fifth force is to enhance $P(k)$ on large linear scales, but that this enhancement falls off at small non-linear scales due to the screening and the cosmic expansion not affecting structures that are locally gravitationally bound. We highlight that including the modified expansion history of cubic Galileon gravity has a significant impact on the resulting non-linear power spectrum.

This study of cubic Galileon gravity has demonstrated the power of \HiCOLA{} to take us, consistently, from a Lagrangian all the way through to simulating non-linear clustering. While a \lcdm{}-like background is sometimes assumed in MG studies (and particularly in MG simulations), we emphasise that our method does not make this assumption. We find that, for the models considered here, the modified expansion history has an effect that is perhaps surprising in magnitude. In the next section we repeat this demonstration for a previously un-simulated theory.


\section{Extended shift-symmetric (ESS) simulations with \HiCOLA{}} \label{sec:ESSsims}
In this section we will present results from the first simulations of non-linear structure in the extended shift-symmetric (ESS) gravity models described in \S\ref{ssec:ESS_theory}. An extended discussion of power spectrum features can be found in Appendix~\ref{app:ESS_results}.

\subsection{Viability of expansion history} \label{ssec:ESS_viability}

The procedure for finding viable ESS models is similar to the process carried out for cubic Galileon gravity; the initial conditions for the cosmological density parameters are fixed in the same way. A difference is that there are now four model parameters to fix: $E_{\textrm{dS}}, f_\phi, k_{1\textrm{dS}}$ and $g_{31\textrm{dS}}$. This time, we scan over a four-dimensional space, subjecting the background solutions to the criteria listed in \S\ref{ssec:cuGal_viability}. With increased dimensionality, we face the difficulties of a much larger space to scan and a less intuitive landscape to visualise. To address the former, we first perform broad and coarse scans over the model space, which we then use to inform narrower and finer scans once regions of viability are identified. For the latter, we can only look at 2D slices and 3D volumes of the overall 4D space to understand the placement of the viable models.
The three ESS models chosen for analysis in this paper are listed in Table~\ref{tab:ESS_params}. Unlike for cubic Galileon gravity, it does not appear that viable ESS models are located in a clearly identifiable and continuous band. A discussion of the structure of the viable ESS model regions will be studied further in a future work. 

\begin{table}[]
\begin{center}
\begin{tabu}{|[2pt]c|c|c|c|c|[2pt]c|c|}
\tabucline[2pt]{1-5} \hline
Name & $f_{\phi}$   & $E_{\rm dS}$ & $k_{1\textrm{dS}}$ & $g_{31\textrm{dS}}$ & $k_{2\textrm{dS}}$ & $g_{32\textrm{dS}}$ \\ \hline
A & 0.812 & 0.125 & -0.114 & -1.945 & -0.293 & 1.040  \\
B & 0.751 & 0.368 & -0.871 & -8.106 & -0.567 & 4.290  \\[-2pt]
C & 0.713 & 0.421 & -0.344 & -8.290 & -1.460 & 4.450  \\[-2pt] \tabucline[2pt]{1-5} \hline
\end{tabu}
\caption[de Sitter-ESS parameters]{Parameters for the three ESS models considered in \S\ref{sec:ESSsims}. Like in Table~\ref{tab:cuGal_params}, the independent variables are highlighted in bold, with the others being determined by the values of the independent variables as per Eq.~(\ref{eq:ESS_dS}).}
\label{tab:ESS_params}
\end{center}
\end{table}

\subsection{Simulation setup} \label{ssec:ESS_sim_setup}

We use the same simulation setup as described in \S\ref{ssec:cuGal_sim_setup}. 
This includes \lcdm{} initial conditions generated directly at $z=49$ (i.e. without backscaling). As discussed in \S\ref{sssec:cuGal_ICs}, with this approach, in addition to the error due to the lack of backscaling, any deviation from \lcdm{} at $z=49$ will introduce error via the ICs. We can estimate the size of this error from the ratio of ESS and \lcdm{} growth factors at $z=49$, a quantity which we plot in Fig.~\ref{fig:ESS_lin_growth} for our strongest ESS case (ESS-C).
We find the ratio of ESS and \lcdm{} growth factors at $z=49$ is approximately $0.991$, or a $0.9\%$ deviation from \lcdm{}. Ignoring the effects of radiation, we can approximate the ratio of ESS and \lcdm{} linear power spectra at $z=49$ to be $0.991^2\approx0.982$, or a $1.8\%$ deviation from \lcdm{}. Again, this error in the linear power spectra at $z=49$ will propagate to give a similarly-sized offset at large, linear scales in the non-linear power spectra measured from the simulation output at late times, and may also introduce transient effects at non-linear scales.
We discuss the linear growth factor in ESS gravity further in Appendix~\ref{sapp:ESS_lin_growth}.

As in \S\ref{sec:cuGalsims}, in addition to the full ESS simulation with both the expansion modified relative to \lcdm{} and screened fifth forces between particles, we run QCDM hybrid simulations. These QCDM simulations feature the ESS expansion history but GR forces\footnote{The error introduced via the ICs discussed above will be marginally different in the QCDM simulations, as the absent modification to gravity has a very small impact on the linear growth factor at $z=49$.}.
In Appendix~\ref{sapp:ESS_Pk_breakdown}, we use these hybrid simulations and others to isolate the impact of the different expansion history from the impact of the screened fifth force.
Thus we require two simulations for each of the three ESS cases from Table~\ref{tab:ESS_params}, which results in a total of 6 simulations (since we re-use the reference \lcdm{} simulation produced for \S\ref{sec:cuGalsims}).

\subsection{Simulation results} \label{ssec:ESS_sim_res}

\begin{figure*}
\includegraphics[width=1.\textwidth]{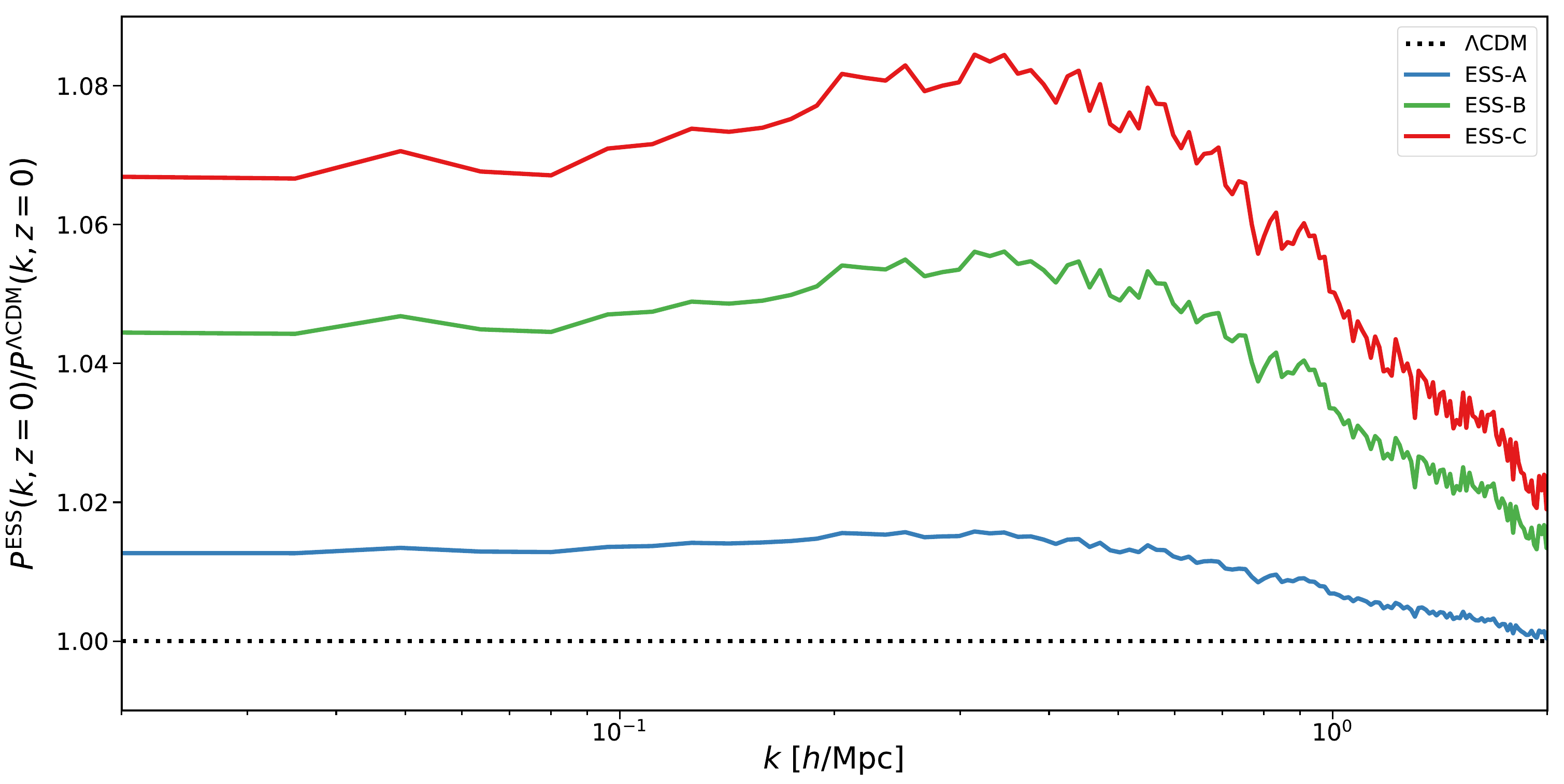}
\caption[Ratios of real-space non-linear matter power spectra at $z=0$ in ESS gravity and \lcdm{}.]{\small Ratios of real-space non-linear matter power spectra at $z=0$ in ESS gravity and \lcdm{} for a variety of ESS parameter values.
The equivalent of this figure for cubic Galileon gravity is Fig.~\ref{fig:cuGal_Pk_GR_ratio}.
}
\label{fig:ESS_Pk_GR_ratio}
\end{figure*}

We study the effect of ESS gravity on structure formation by taking the ratio between the $P(k)$ from the full ESS gravity simulation and the \lcdm{} simulation at $z=0$, which we plot in Fig.~\ref{fig:ESS_Pk_GR_ratio}.
The behaviour of the ESS power spectrum is broadly similar to the cubic Galileon power spectrum discussed in \S\ref{ssec:cuGal_sim_res}.
Instead of repeating ourselves, we will therefore focus on highlighting the differences between ESS gravity and cubic Galileon gravity in this section.

\begin{figure*}
\includegraphics[width=1.\textwidth]{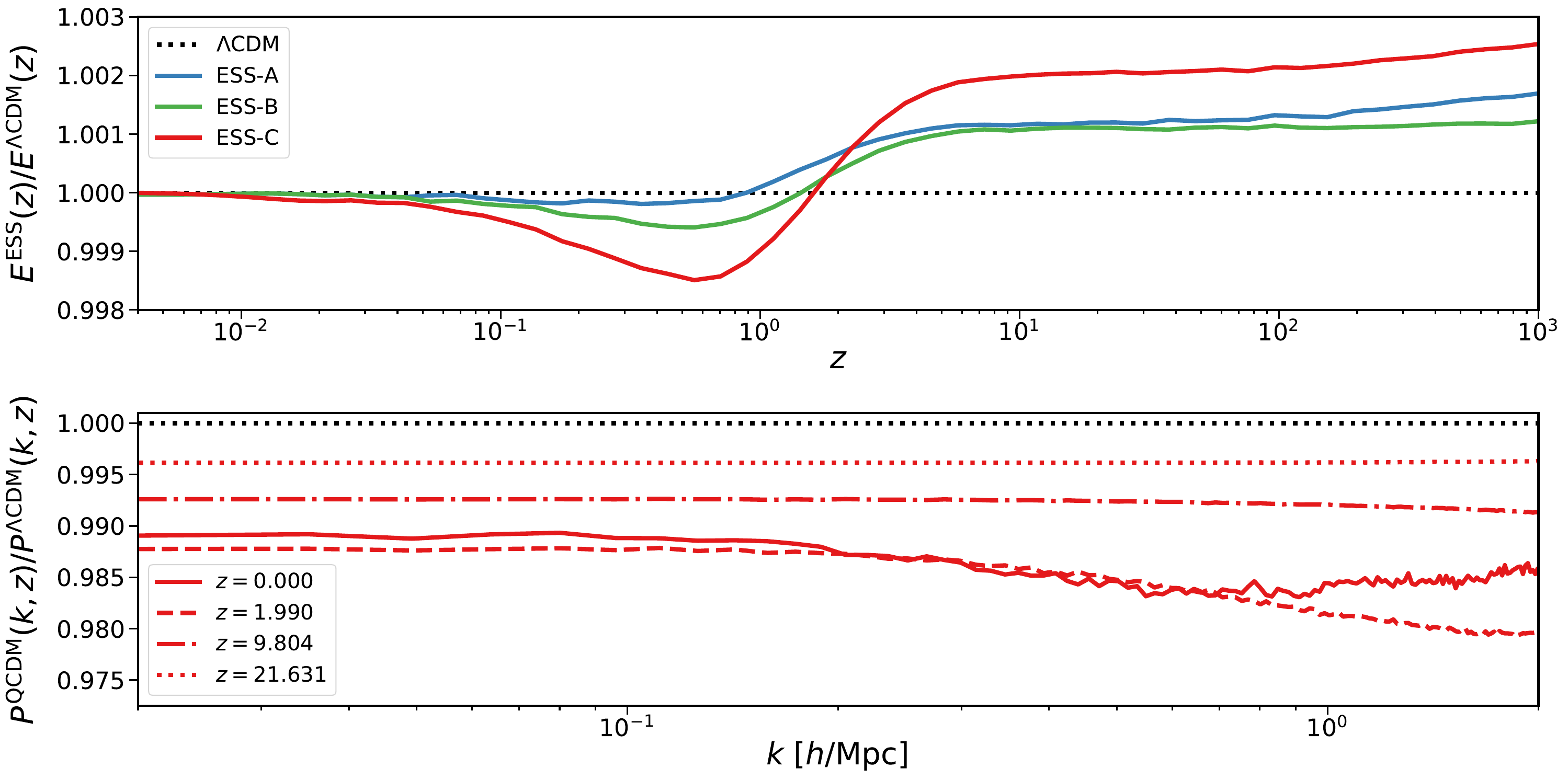}
\caption[Expansion history in ESS gravity and its effect on structure formation.]{\small \textit{Upper panel}: The ratio of expansion rate $E$ in ESS gravity and \lcdm{} for the three different sets of model parameters.
\textit{Lower panel}: The ratio of power spectra in QCDM and \lcdm{} at several redshifts for the ESS-C case, which demonstrates the isolated effect of the modified expansion history in ESS gravity.
The equivalent of this figure for cubic Galileon gravity is Fig.~\ref{fig:cuGal_background_Pkbg}.
}
\label{fig:ESS_background_Pkbg}
\end{figure*}

The first key difference in ESS gravity is related to the expansion history, displayed in Fig.~\ref{fig:ESS_background_Pkbg}. Firstly, the upper panel shows the size of the deviations from \lcdm{} are around $0.2\%$ compared to $8\%$ in cubic Galileon gravity, so they will have a smaller effect. Secondly, in ESS gravity there is a phase where expansion is faster than in \lcdm{} for $z\gtrsim2$, in addition to the phase where expansion is slower than in \lcdm{} for $0.1\lesssim z\lesssim2$ which is qualitatively similar to cubic Galileon gravity.
Cosmic expansion suppresses structure formation, thus faster-than-\lcdm{} expansion increases this suppression, whereas slower-than-\lcdm{} expansion reduces this suppression leading to a net enhancement of structure formation relative to \lcdm{}.

This effect can be seen in isolation in the lower panel of Fig.~\ref{fig:ESS_background_Pkbg} via the ratios of power spectra in QCDM and \lcdm{}. We see that the suppression of clustering increases from $z=21.631$ to $z=9.804$ to $z=1.990$, but then weakens between $z=1.990$ and $z=0.000$; this reflects the transition from faster-than-\lcdm{} expansion to slower-than-\lcdm{} expansion that occurs around $z\sim2$ as seen in the upper panel of Fig.~\ref{fig:ESS_background_Pkbg}.
The combined result of the different expansion history in ESS gravity is to produce a small net suppression of the $z=0$ power spectrum on large scales relative to \lcdm{} that becomes stronger at intermediate scales before weakening towards small scales where structures tend to be locally gravitationally bound and thus unaffected by cosmic expansion. 
This suppression of clustering is in contrast to the enhancement of the power spectrum caused by the different expansion history in cubic Galileon gravity seen in Fig.~\ref{fig:cuGal_background_Pkbg}.

\begin{figure*}
\includegraphics[width=1.\textwidth]{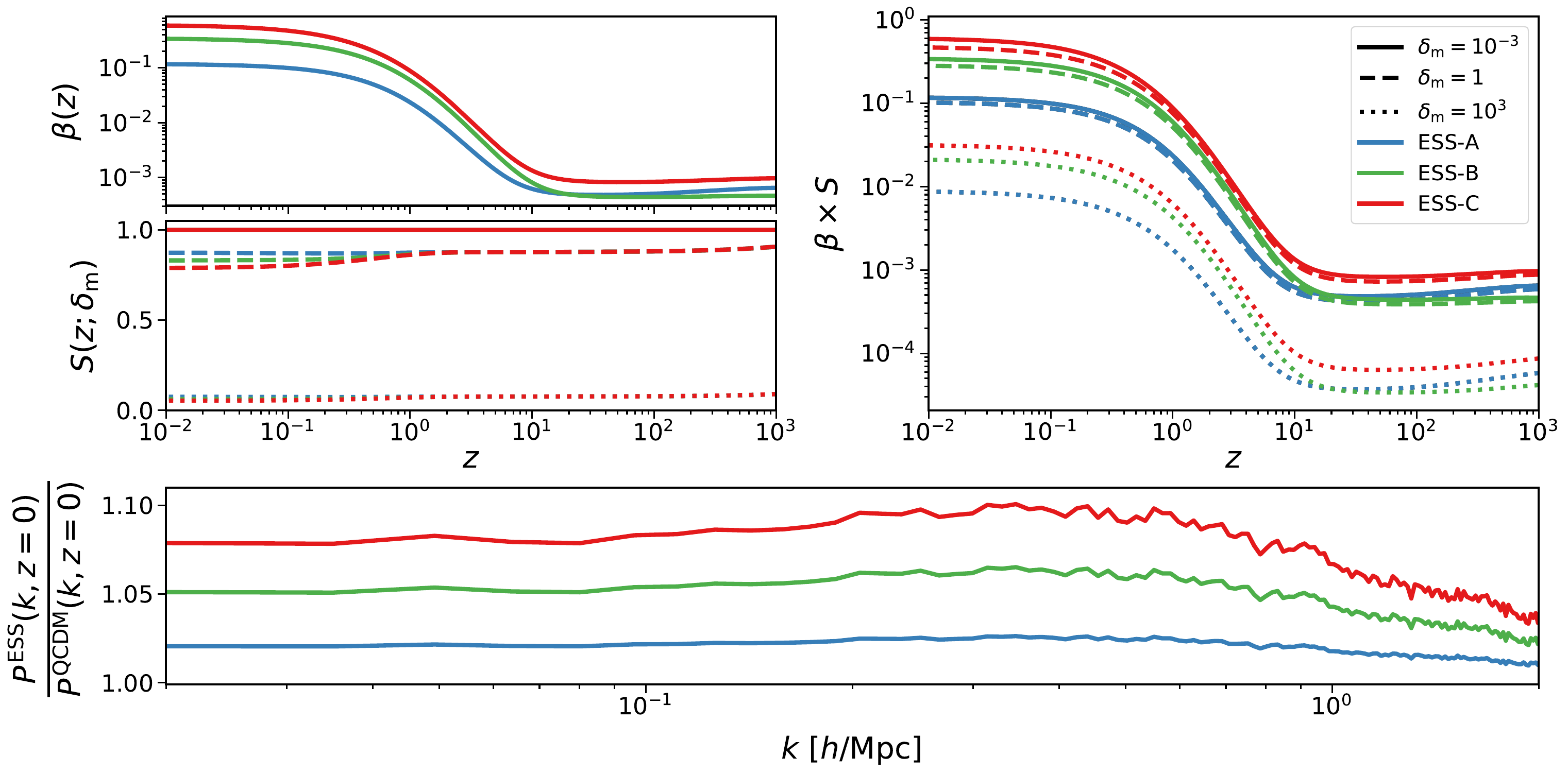}
\caption[Coupling and screening coefficient in ESS gravity.]{The effect of varying ESS parameters on the approximate screened fifth force and its component quantities. 
\textit{Upper left} shows the coupling (as defined in Eq.~(\ref{eq:coupling})) as a function of redshift. 
\textit{Middle left} shows the screening coefficient (as defined in Eq.~(\ref{eq:screen_coeff})) as a function of redshift for three different densities. 
\textit{Upper right} shows the product of the coupling and screening coefficient as a function of redshift for three different densities; this quantity is the approximate screened fifth force.
The \textit{lower panel} shows the ratio of power spectra from the full ESS and QCDM simulations at $z=0$; therefore it shows the isolated impact of the ESS screened fifth force.
The equivalent of this figure for cubic Galileon gravity is Fig.~\ref{fig:cuGal_coupl_screenfac}.
}
\label{fig:ESS_coupl_screenfac}
\end{figure*}

Looking at Fig.~\ref{fig:ESS_coupl_screenfac}, we see that the second key difference in ESS gravity is related to the screened fifth force. While both the screening and the late time strength of the coupling are very similar in ESS and cubic Galileon gravity, the early time coupling in ESS gravity is a few orders of magnitude stronger than that of cubic Galileon gravity; this reflects the corresponding behaviours for the respective scalar fields, with the gradient of the scalar field $y$ being larger at early times in ESS gravity than in cubic Galileon gravity. As a result, the impact of the screened fifth force on the $z=0$ power spectrum in ESS gravity is slightly larger than in cubic Galileon gravity.

For the interested reader, we study and discuss the phenomenology of ESS gravity in greater detail in Appendix~\ref{app:ESS_results}.
But in summary, the small amount of suppression of clustering due to the modified expansion history combines with the larger amount of enhancement due to the screened fifth force. As seen in Fig.~\ref{fig:ESS_Pk_GR_ratio}, this produces an overall moderate enhancement of clustering on large scales that briefly increases with $k$ at intermediate scales before screening activates to eliminate the enhancement at small scales.

Comparing the different ESS cases, we see that ESS-A produces the smallest deviations from \lcdm{} and ESS-C the largest. The ESS-B case is very similar to ESS-A at early times, but at late times is more similar in strength to ESS-C. We intend to elaborate on the parameter dependence of ESS phenomenology in future work.

While we have focused on the differences between cubic Galileon gravity and ESS gravity in this section, the phenomenology of these two theories is broadly similar. This is not surprising, given that they share the same screening mechanism, and the rather conservative criteria we applied in \S\ref{ssec:cuGal_viability} and \S\ref{ssec:ESS_viability} ensure fairly similar background solutions. We leave investigations of more complicated reduced Horndeski theories, such as those that go beyond shift symmetry or feature different screening mechanisms, for future work.


\section{Conclusions}\label{sec:conc}

In this work, we have implemented reduced Horndeski gravity in a fast, approximate COLA simulation code called \HiCOLA{}\footnote{Can be found at \url{https://github.com/Hi-COLACode/Hi-COLA}.}. We believe \HiCOLA{} is the first implementation of reduced Horndeski gravity in an \nbody{} code, and one of only a small number of codes that have the flexibility to simulate many different gravity theories. Thus \HiCOLA{} has excellent potential to be used by modified gravity model builders to explore the quasi-non-linear regime up to scales of $k\sim 1\, {\mathrm{h} \,\mathrm{Mpc}}^{-1}$, in the same way that the modified Einstein--Boltzmann codes \hiclass{} and EFTCAMB \cite{Hu:2013twa,Raveri:2014cka,2015PhRvD..91f3524H, Zumalacarregui:2016pph,Bellini:2017avd,Bellini:2019syt} have enabled general exploration of the linear cosmological regime\footnote{We also note the existence of tools based on the halo model that allow the computation of the real-space power spectrum in modified gravity theories to be extended beyond the linear regime such as MGHalofit \cite{Zhao:2013dza}, HMCode \cite{Mead:2016zqy}, and ReACT \cite{Bose:2020wch}.}.

\HiCOLA{} computes screening effects on non-linear scales using a coupling function and a screening coefficient, both of which are determined by the cosmological background of the model. A consistent solution of both cosmological background and screening quantities with the same model parameters is automatically implemented in \HiCOLA{}. The approach we adopt makes the mathematical connection between background and non-linear scales easy to follow.

We have used \HiCOLA{} to compute non-linear structure formation in two example theories belonging to the Horndeski family -- cubic Galileon gravity and the extended shift-symmetric theories of \cite{Traykova:2021hbr} -- validating the code extensively against traditional \nbody{} simulations of the former. Our work to date has centred on shift-symmetric theories for convenience; however, we plan to next turn attention to theories without this property. This will enable the functionality needed for, e.g. classic quintessence models. Likewise, we chose to first focus on theories that screen via the Vainshtein mechanism; further density-dependent functionality needed for chameleon screening will be implemented in the future.

The key limit to \HiCOLA{}'s accuracy beyond quasi-non-linear scales is the use of a spherically symmetric approximation when deriving the screening coefficient in \S\ref{ssec:Vainshtein_spherical_symmetry}. We have shown that, provided steps are taken to alleviate errors (\S\ref{ssec:force_interpolation} \& \S\ref{ssec:dens_smoothing}), this approximation is acceptable and is worthwhile given the rapid simulations it enables. However, in order to calibrate the free parameters in these mitigation steps, and to validate \HiCOLA{} more generally, we must still compare to the results of traditional \nbody{} codes. These currently only exist for a handful of specific MG theories; we will continue to validate \HiCOLA{} against these for available theories. 
We intend to investigate methods to improve the accuracy of \HiCOLA{}, for example by finding faster ways to solve the full non-linear Klein--Gordon equation where necessary. This would remove the need to calibrate parameters in our current spherically symmetric approach.

There are many directions and applications to be explored with \HiCOLA{}, and we intend to make the code publicly available in due course. Whilst we have validated \HiCOLA{}'s ability to produce dark matter power spectra, there are many other quantities that \nbody{} codes are relied upon to produce, including halo and lensing quantities. Integration of \HiCOLA{} into such pipelines will ultimately allow for the broad span of Horndeski theories to be tested with data from upcoming galaxy and lensing surveys. In this way, we hope to open up the landscape of modified gravity theory space to non-linear constraints and continue to progress our understanding of what does, and what does not, constitute a viable theory of gravity.


\section*{Contribution Statements}

B.S.W. modified the \fml{} code, ran the \HiCOLA{} simulations, carried out the validation, led the analysis of simulation results, and led the manuscript writing.
A.S.G. wrote the code for the front-end module, assisted with analysing the simulation results, and contributed to the manuscript.
T.B. led the development of the project concept, led the analytical calculations, assisted with analysing the simulation results, contributed to the manuscript, and managed the project.
G.V. helped develop an early version of the simulation code and guided the project within LSST-DESC. B.F. assisted with the analysis and validation of results for the final version of this paper.

\section*{Acknowledgements}

We are very grateful to Alex Barreira and Baojiu Li for generously providing us with the cubic Galileon \nbody{} data used in our validation. We are also grateful for useful discussions with Cristhian Garcia-Quintero, Wojciech Hellwing, Kazuya Koyama, Mustapha Ishak, Johannes Noller, Shankar Srinivasan, Dan Thomas and Hans Winther.

This paper has undergone internal review in the LSST Dark Energy Science Collaboration. 
We are grateful to the internal reviewers, who were Alejandro Aviles, Matteo Cataneo, and Kazuya Koyama. 

We acknowledge use of the {\tt NumPy} \cite{NumPy}, {\tt SciPy} \cite{SciPy}, {\tt SymPy} \cite{SymPy}, {\tt matplotlib} \cite{matplotlib}, and {\tt Pylians} Python libraries; the \fml{} C++ template library; as well as the colourblind-friendly {\tt PyPlot} colour scheme by GitHub user `thriveth' (\url{https://gist.github.com/thriveth/8560036}) and the Coblis colourblindness simulator. Some of the numerical computations were done on the Sciama High Performance Compute (HPC) cluster which is supported by the ICG, SEPNet, and the University of Portsmouth. This research utilised Queen Mary's Apocrita HPC facility \cite{Apocrita}, supported by QMUL Research-IT.

B.S.W. is supported by a Royal Society Enhancement Award (grant no.~RGF$\backslash$EA$\backslash$181023). A.S.G. is supported by a STFC PhD studentship. T.B. is supported by ERC Starting Grant \textit{SHADE} (grant no.~StG 949572) and a Royal Society University Research Fellowship (grant no.~URF$\backslash$R1$\backslash$180009). G.V. recognises partial support by NSF grant AST-1813694. B.F. is supported by a Royal Society Enhancement Award (grant no. RF$\backslash$ERE$\backslash$210304 ).

The DESC acknowledges ongoing support from the Institut National de 
Physique Nucl\'eaire et de Physique des Particules in France; the 
Science \& Technology Facilities Council in the United Kingdom; and the
Department of Energy, the National Science Foundation, and the LSST 
Corporation in the United States.  DESC uses resources of the IN2P3 
Computing Center (CC-IN2P3--Lyon/Villeurbanne - France) funded by the 
Centre National de la Recherche Scientifique; the National Energy 
Research Scientific Computing Center, a DOE Office of Science User 
Facility supported by the Office of Science of the U.S.\ Department of
Energy under Contract No.\ DE-AC02-05CH11231; STFC DiRAC HPC Facilities, 
funded by UK BEIS National E-infrastructure capital grants; and the UK 
particle physics grid, supported by the GridPP Collaboration.  This 
work was performed in part under DOE Contract DE-AC02-76SF00515.


\appendix

\section{Mass scales}\label{app:mass}

In order to avoid our numerical solvers handling very small or large numbers, \HiCOLA{} has been constructed to use dimensionless analogues of the quantities shown in Eq.~(\ref{eq:Horn_action}), and dimensionless equations of motion. In this appendix we describe first the quantities, and then the equations implemented in the code. Since we work in geometric units, where $c=1$, we typically express the dimensionality as a mass dimension. We remind the reader of our convention where tilded quantities have dimensions and untilded quantities are dimensionless. 

\subsection{Dimensionless quantities}

As an example, let us consider $\tphi$, which has a mass dimension of $1$. We can straightforwardly define a dimensionless analogue of the scalar field $\phi$ as
\begin{align}
\label{eq:dimless_phi}
    \tphi &\equiv M_{\rm s} \phi \,,
\end{align}
where $M_{\rm s}$ is the mass scale associated with the scalar field. Using $\phi$, we can then construct a dimensionless analogue of $\tX$, which we call $X$. This involves an extra step of also performing a change of variables so that we use $x = \ln(a)$ as a dimensionless coordinate for time in place of the scale factor, $a$. We denote derivatives with respect to time with a dot, and to $x$ with a prime. Substituting $\phi$ and performing the change of variables (implicitly we are evaluating $\tilde{X}$ on the background FRW metric) yields
\begin{align}
     \tX &= \frac{M_{\rm s}^2}{2} {\dot\phi}^2 \equiv \frac{M_{\rm s}^2 H^2}{2} {\phi^{\prime}}^2 \equiv {M_{\rm s}^2 H_0^2} X\,.
\end{align}
Our dimensionless kinetic variable is therefore
\begin{align}
\label{eq:dimless_X}
    X &\coloneqq \frac{H^2 {\phi^{\prime}}^2}{2H^2_0} \equiv \frac{E^2 {\phi^{\prime}}^2}{2}\,,
\end{align}
where we note that $E = H/H_0$.

We can then follow through with these newly defined objects to construct dimensionless analogues of the Horndeski functions and their derivatives. For example, given that $\tgthree$ has dimensions of mass, we can define the dimensionless analogue $G_3$ through
\begin{equation}
    \tgthree = M_{\tgthree} G_3\,,
\end{equation}
where $M_{\tgthree}$ is the mass scale associated to $\tgthree$. Then derivatives of $\tgthree$ can be expressed using the dimensionless analogues as follows:
\begin{align}
    \tilde{G}_{3\tX} &\coloneqq \partial_{\tX} \tgthree = \frac{\partial X}{\partial \tX} \partial_X \tgthree =  \frac{\partial X}{\partial \tX} \partial_X \left( M_{\tgthree} G_3 \right)\\
    &= M^{-2}_s H^{-2}_0 M_{\tgthree} G_{3X},
\end{align}
where in the last line $G_{3X}$ is the dimensionless analogue of $\tilde{G}_{3\tilde{X}}$. We construct dimensionless analogues of all the remaining Horndeski functions and their derivatives in a manner similar to this example.

\subsection{Dimensionless equations of motion}
Having pulled out the mass scales of all quantities used, our equations of motion will consist of dimensionless variables multiplied by ratios of masses. These are generally (or generally expected to be) within a few magnitudes of order unity, at least for MG theories which have effects on cosmological scales. As an example, below is the equation for $\Omega_{\phi}$ with the mass ratios left unspecified:
\begin{align}\label{eq:EOM_OmPhi}
    \Omega_{\phi} =\ &(\Omega_{\rm r} + \Omega_{\rm m} + \Omega_{\Lambda})\bigg(\frac{M^2_{pG_4}}{2G_4} -1 \bigg) + \frac{1}{3G_4} \bigg[ \frac{M^2_{KG_4}}{E^2}  X K_X - \frac{M^2_{KG_4}}{2E^2} K + 3{M_{G_3{\rm s}} M_{{\rm s}G_4}} X \phi^{\prime} G_{3X} \nonumber\\
    &- \frac{ M_{G_3G_4} M_{{\rm s}G_4}}{E^2} X G_{3\phi} -  3\phi^{\prime} G_{4\phi} \bigg]\,.
\end{align}
Terms of the form $M_{ab}$ are the mass ratios, $M_{ab} \coloneqq M_a / M_b$, where $M_a$ and $M_b$ are the mass scales for quantities $a$ and $b$ respectively. Here $a,b$ stand for subscripts ${\rm P}, {\rm s}, X, K, G_3, G_4$, indicating the Planck mass and mass scales for the scalar field, $X$, $K$, $G_3$, and $G_4$ respectively.

For the results in this paper, it is assumed that all mass ratios are $1$. This implies that the Horndeski functions $\tilde{K}$, $\tilde{G}_3$, and $\tilde{G}_4$ have the same order of magnitude, and $M_{{\rm P}G_4}=1$ fixes this scale to be approximately $M_{\rm P}$.

\section{Code Validation \& Calibration}\label{app:valid}

To understand the regime of validity for \HiCOLA{} and how this depends on the simulation setup, we compare against cubic Galileon \nbody{} simulations described in \cite{Barreira:2013eea}. Unlike the cubic Galileon model considered elsewhere in this paper, the model in \cite{Barreira:2013eea} is the original cubic Galileon model that does not include a cosmological constant. The screened fifth force in these simulations was computed using the slow but accurate method of solving the full non-linear scalar field equation, rather than the approximate approach we outlined in \S\ref{ssec:Vainshtein_spherical_symmetry}.

In addition to running simulations using the full cubic Galileon theory, \cite{Barreira:2013eea} also produced simulations for linear cubic Galileon theory (i.e. without screening), as well as the case where the forces in the simulation are purely GR but the background expansion history is that of the cubic Galileon model (referred to as the QCDM case).

We use \HiCOLA{} to produce simulations for all three cosmological cases: full cubic Galileon theory, linear cubic Galileon theory, and QCDM. We use \fml{}'s existing functionality to read directly from the $z=49$ \nbody{} snapshot of \cite{Barreira:2013eea}, such that our ICs are identical. This allows us to run only a single simulation per cosmology while eliminating realisation noise at large scales. We replicate the simulation setup used to generate the \nbody{} data as closely as possible, using a boxsize of $L=400~{\rm Mpc}/h$ and $N_{\rm part}=512^3$ particles. However, since the \nbody{} data was produced using {\tt ECOSMOG}, which is an adaptive mesh refinement (AMR) code, we cannot replicate its force computation approach exactly with \fml{}'s particle-mesh (PM) approach; instead we choose a force mesh of size $N_{\rm mesh}=3(N_{\rm part})^{1/3}=1536$. Also, since the key advantage of the COLA method is to allow us to use a smaller number of timesteps without losing accuracy at large scales, we again use only $40$ timesteps linearly spaced in scale factor $a$ between $z_{\rm ini}=49$ and $z=0$, which should still yield good accuracy on small scales.

To ensure the power spectra are computed consistently from both the \HiCOLA{} and \nbody{} simulations, we use the {\tt Pylians} tool\footnote{Available at \url{https://github.com/franciscovillaescusa/Pylians3}.} for measuring power spectra from the simulation snapshots at $a=\{0.2, 0.6, 1.0\}$.

\subsection{Validation}
For the validation, we compare power spectra for individual cosmologies and ratios of power spectra for different cosmologies which we refer to as boost factors. COLA simulations are well known to perform better at reproducing boost factors (ratios of power spectra in two cosmologies) than power spectra themselves \cite{Winther2017,Wrightetal2017,Winther:2019mus,Fiorini:2021dzs}. As discussed in \S\ref{sssec:cuGal_other_sim_setup}, accurate boost factors are still very useful. The boost factors (denoted $B$) we focus on for this validation are the ratio of power spectra between 1) the full cubic Galileon and QCDM cases; and 2) the full and linear cubic Galileon cases. The first of these negates the effect of the modified expansion history to isolate the effects of modified forces, while the latter isolates further the effect of screening. We reproduce the output at three redshifts corresponding to $a=\{0.6, 0.8, 1\}$.

As described in \S\ref{sec:screened_fifth_force}, we have some flexibility in our screening approach with two parameters to calibrate: $k_{\rm filter}$ \& $R_{\rm smooth}$. As part of our validation, we compared our \HiCOLA{} simulations to the \nbody{} simulations for several different values of these two parameters. We direct interested readers to Appendix~\ref{sapp:calibration} for the details of that calibration, but the summary is that we determine the best parameter values in terms of agreement with the \nbody{} sims are $k_{\rm filter}=0.2~h/{\rm Mpc}$ \& $R_{\rm smooth}=3~{\rm Mpc}/h$. We use these parameters for our headline validation below.

\begin{figure*}
\includegraphics[width=1.\textwidth]{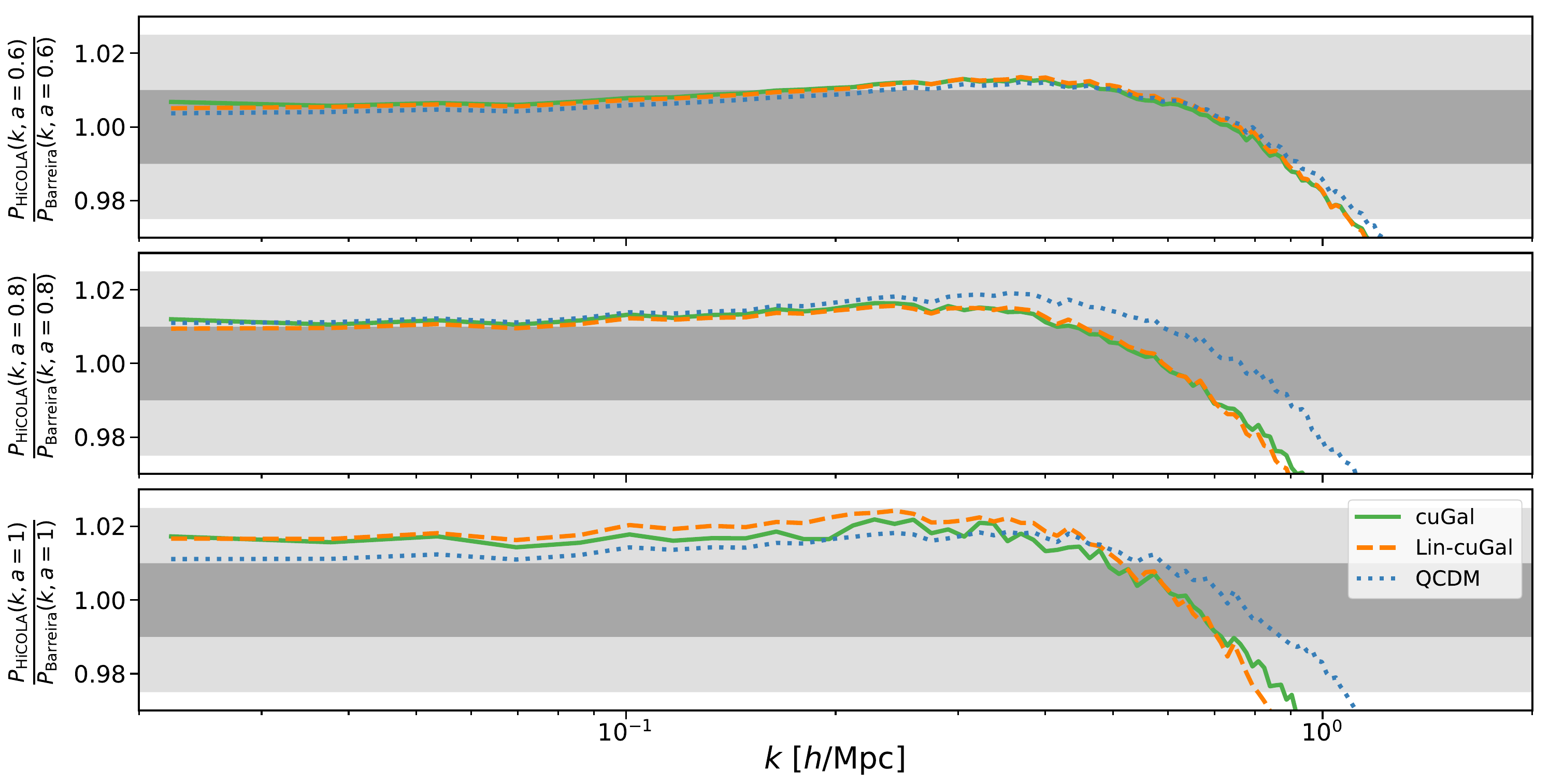}
\caption[Power spectrum comparison between \HiCOLA{} and \nbody{}.]{
Power spectrum comparison between \HiCOLA{} and \nbody{} results from \cite{Barreira:2013eea}. We consider three cosmologies: the full cubic Galileon case, the linear cubic Galileon case (i.e. without screening), and the QCDM case with GR forces but a cubic Galileon expansion history. The top, middle, and bottom panels correspond to $a=0.6$, $a=0.8$, and $a=1$ respectively. Dark and light grey bands indicate the $\pm 1\%$ and $\pm 2.5\%$  regions respectively.
}
\label{fig:validation_headline_Pk}
\end{figure*}

Figure~\ref{fig:validation_headline_Pk} shows the power spectrum comparison between \HiCOLA{} and \nbody{} results from \cite{Barreira:2013eea}.
We note that there is some offset at large scales in the power spectrum comparison for all cosmologies and redshifts. We strongly believe this is due to a small disagreement in our respective expansion histories; however, we have been unable to identify the exact source. This makes it difficult to draw reliable conditions from the power spectrum comparison. Fortunately, this expansion history disagreement does not propagate to our boost factors, giving further reason to focus on these for our validation.

\begin{figure*}
\includegraphics[width=1.\textwidth]{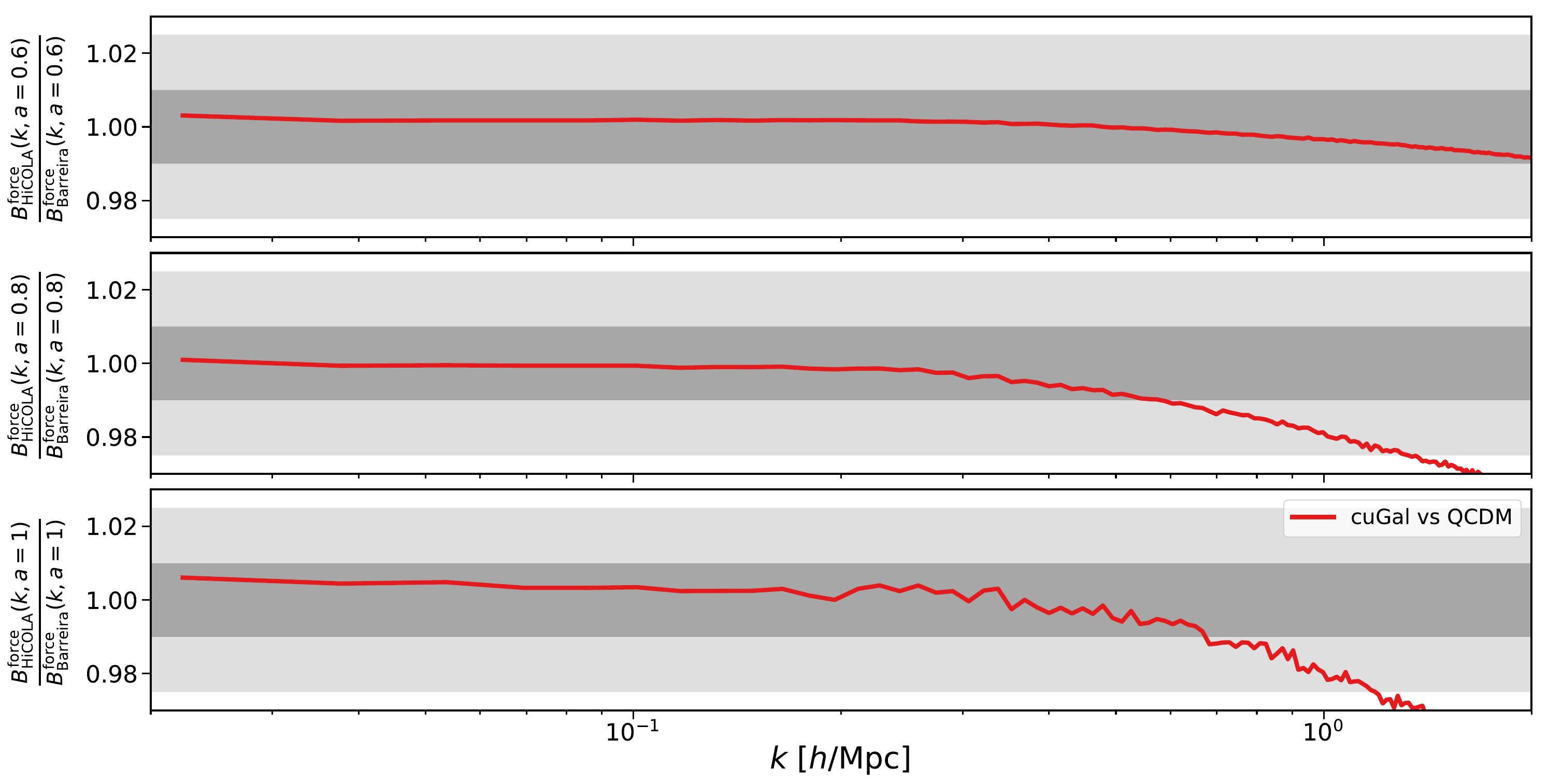}
\caption[Force boost factor comparison between \HiCOLA{} and \nbody{}.]{
Boost factor comparison between \HiCOLA{} and \nbody{} results from \cite{Barreira:2013eea}. The boost factor in this plot is the ratio of power spectra in the full cubic Galileon and QCDM cosmologies, which negates the effect of the modified expansion history to isolate the effects of modified forces. The top, middle, and bottom panels correspond to $a=0.6$, $a=0.8$, and $a=1$ respectively. Dark and light grey bands indicate the $\pm 1\%$ and $\pm 2.5\%$  regions respectively.
}
\label{fig:validation_headline_boost_force}
\end{figure*}

\begin{figure*}
\includegraphics[width=1.\textwidth]{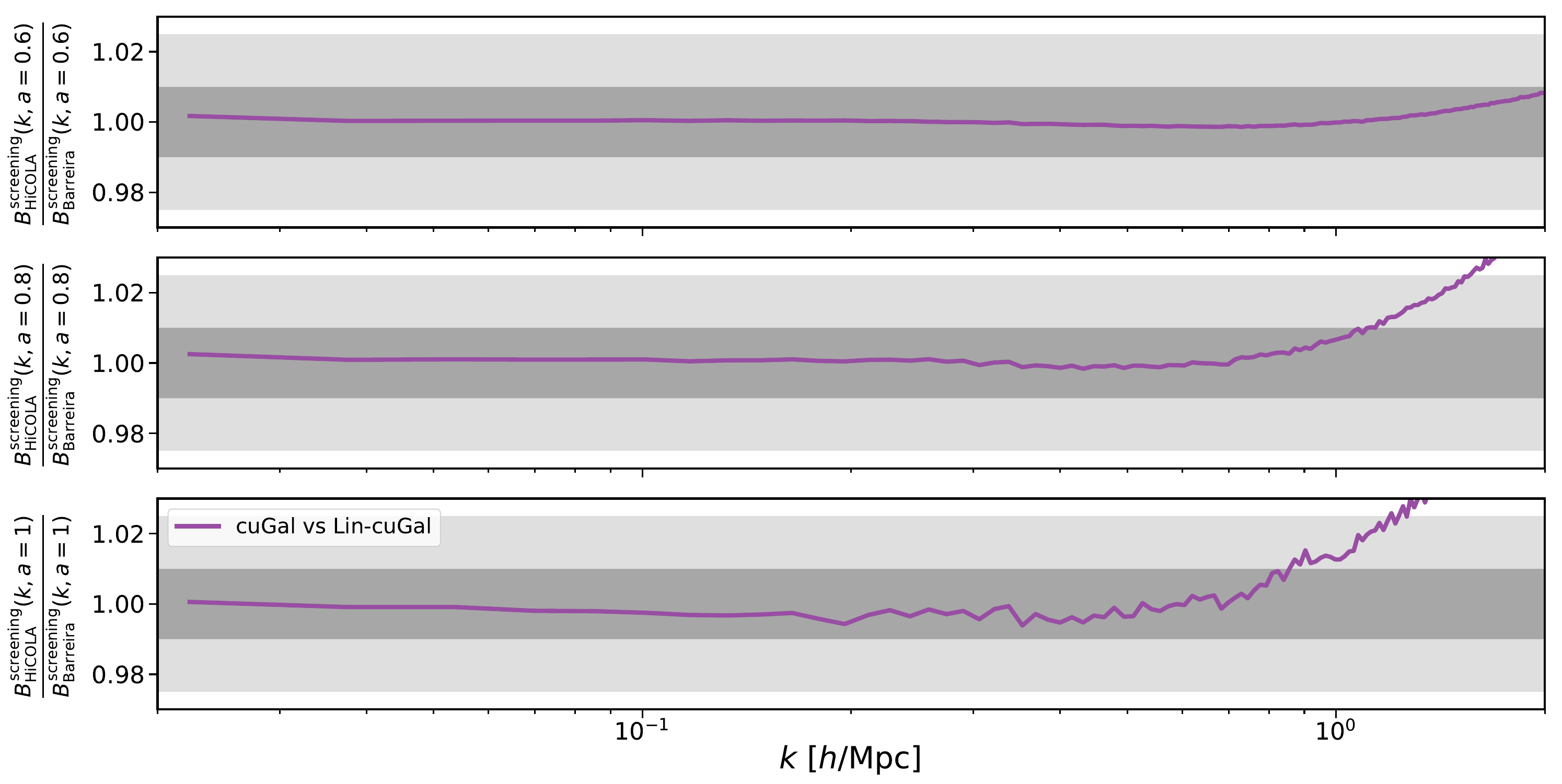}
\caption[Screening boost factor comparison between \HiCOLA{} and \nbody{}.]{
Boost factor comparison between \HiCOLA{} and \nbody{} results from \cite{Barreira:2013eea}. The boost factor in this plot is the ratio of power spectra in the full and linear cubic Galileon cosmologies, which isolates the effects of screening. The top, middle, and bottom panels correspond to $a=0.6$, $a=0.8$, and $a=1$ respectively. Dark and light grey bands indicate the $\pm 1\%$ and $\pm 2.5\%$  regions respectively.
}
\label{fig:validation_headline_boost_screen}
\end{figure*}

Figures~\ref{fig:validation_headline_boost_force} \& \ref{fig:validation_headline_boost_screen} show the two boost factor comparisons between \HiCOLA{} and \nbody{} results from \cite{Barreira:2013eea}.
Figure~\ref{fig:validation_headline_boost_force} is for the force boost factor, while Figure~\ref{fig:validation_headline_boost_screen} is for the screening boost factor.
Table~\ref{tab:valid_boost_accuracy} records the $k_{\rm max}$ values at which the boost factors are accurate at the $1\%$ \& $2.5\%$ levels at each redshift. Therefore, we determine \HiCOLA{} is valid at the $2.5\%$ level down to $k_{\rm max}=1.2~h/{\rm Mpc}$ at $z=0$. This is a measure of accuracy determined with respect to the Cubic Galileon, and strictly speaking, comparisons should be made on a per-model basis where possible. However, as detailed in \S\ref{sapp:calibration}, making use of \HiCOLA{}'s calibration parameters can help ensure this accuracy is maintained across models in the reduced Horndeski framework.

\begin{table}
\begin{center}
\begin{tabular}{cc|cc|}
\cline{3-4}
                                                 &       & \multicolumn{2}{c|}{$k_{\rm max}~[h/{\rm Mpc}]$}      \\ \cline{3-4}
                                                 &       & \multicolumn{1}{c|}{$\phantom{.5}1\%$} & $2.5\%$ \\ \hline
\multicolumn{1}{|c|}{\multirow{3}{*}{Force}}     & $a=0.6$ & 2.4                      & 5.4 \\ \cline{2-2}
\multicolumn{1}{|c|}{}                           & $a=0.8$ & 0.58                     & 1.3 \\ \cline{2-2}
\multicolumn{1}{|c|}{}                           & $a=1$   & 0.68                     & 1.2 \\ \hline
\multicolumn{1}{|c|}{\multirow{3}{*}{Screening}} & $a=0.6$ & 2.2                      & 3.4 \\ \cline{2-2}
\multicolumn{1}{|c|}{}                           & $a=0.8$ & 1.1                      & 1.6 \\ \cline{2-2}
\multicolumn{1}{|c|}{}                           & $a=1$   & 0.85                     & 1.2 \\ \hline
\end{tabular}
\caption[Boost factor $k_{\rm max}$ values.]{The $k_{\rm max}$ values at which the force and screening boost factors are accurate at the $1\%$ \& $2.5\%$ levels at each redshift, taken from Figures~\ref{fig:validation_headline_boost_force} \& \ref{fig:validation_headline_boost_screen}.}
\label{tab:valid_boost_accuracy}
\end{center}
\end{table}

\subsection{Calibration} \label{sapp:calibration}

We have two calibration parameters as described in \S\ref{sec:screened_fifth_force}. The first, $k_{\rm filter}$, controls the scale at which we interpolate between the linear and screened solutions for the fifth force. The second, $R_{\rm smooth}$, controls the scale at which we smooth the density field before computing the screened fifth force.

To decide on a `best' set of parameters, we focus on the second boost factor described above, the one that isolates the effect of the screening, since it is the screening that these two parameters affect and that we want to match in the more accurate \nbody{} simulations. We also focus on $a=1$, although we expect the best parameters to be relatively independent of redshift.

\begin{figure*}
\includegraphics[width=1.\textwidth]{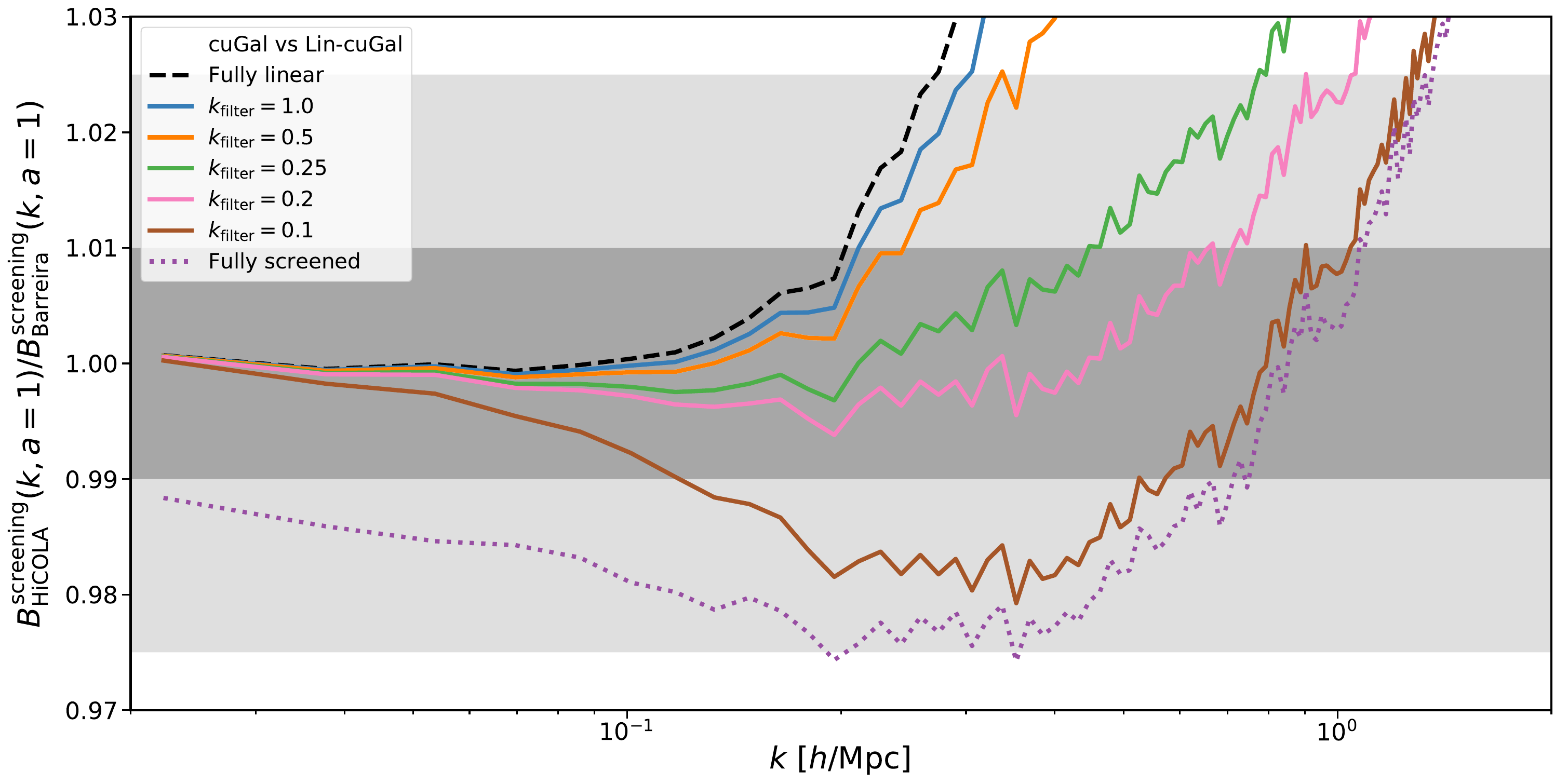}
\caption[$k_{\rm filter}$ calibration.]{
Effect of $k_{\rm filter}$ on the boost factor comparison between \HiCOLA{} and \nbody{} results from \cite{Barreira:2013eea} at $a=1$. The boost factor in this plot is the ratio of power spectra in the full and linear cubic Galileon cosmologies, which isolates the effects of screening. We have fixed $R_{\rm smooth}=2~{\rm Mpc}/h$ where applicable. We also plot the fully linear ($k_{\rm filter}\rightarrow \infty$) and fully screened ($k_{\rm filter}\rightarrow 0$) results to demonstrate what varying $k_{\rm filter}$ interpolates between.
Dark and light grey bands indicate the $\pm 1\%$ and $\pm 2.5\%$  regions respectively.
}
\label{fig:calibration_k_filter}
\end{figure*}

We first investigate the effect of varying $k_{\rm filter}$ while fixing $R_{\rm smooth}=2~{\rm Mpc}/h$, which is seen in Fig.~\ref{fig:calibration_k_filter}. We also compute the results for the cases with fully linear and fully screened forces, which represent the limits $k_{\rm filter}\rightarrow \infty$ and $k_{\rm filter}\rightarrow 0$ respectively. Simply examining the figure, we determine that $k_{\rm filter}=0.2~h/{\rm Mpc}$ yields the best results.

\begin{figure*}
\includegraphics[width=1.\textwidth]{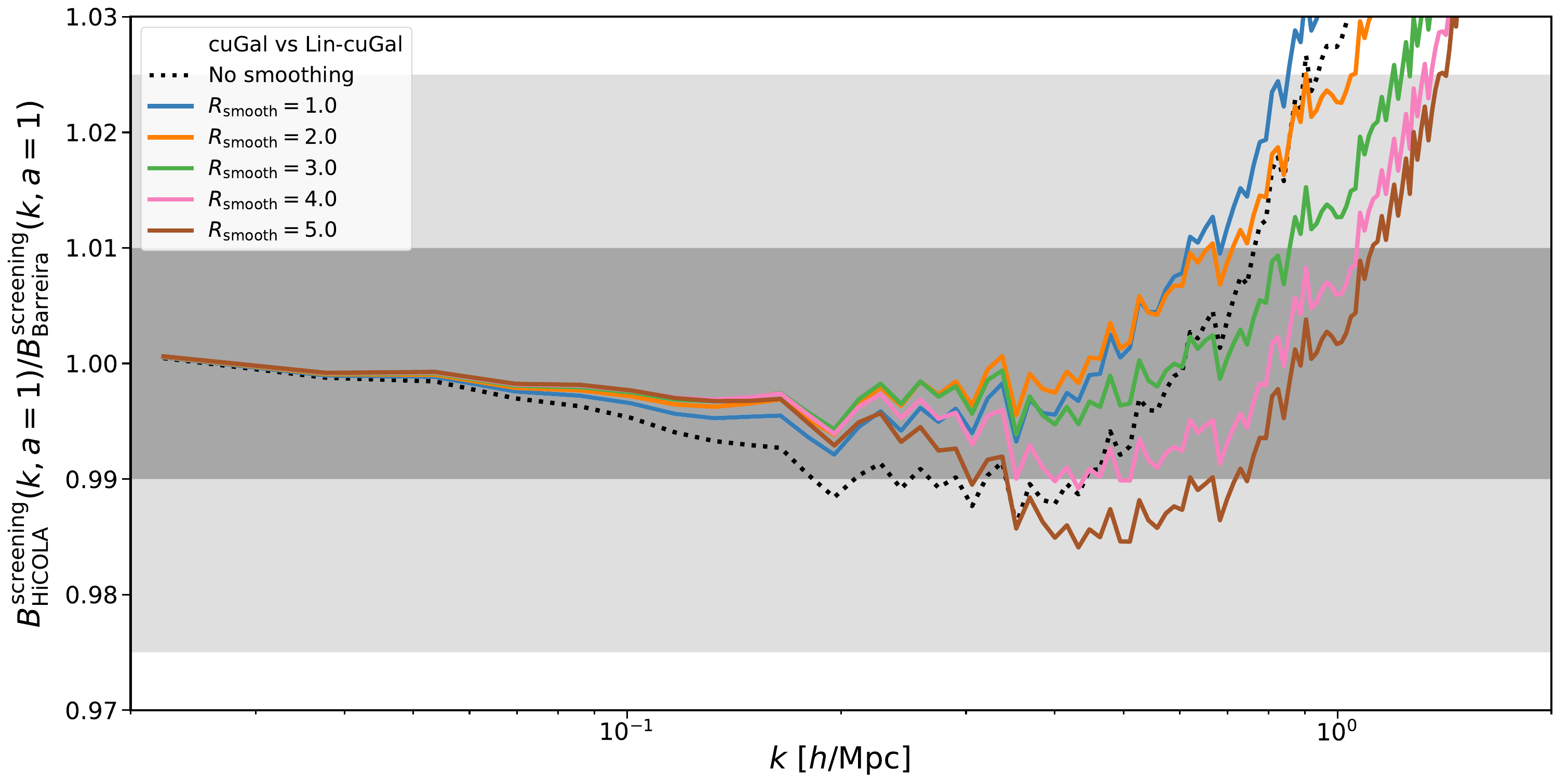}
\caption[$R_{\rm smooth}$ calibration.]{
Effect of $R_{\rm smooth}$ on the boost factor comparison between \HiCOLA{} and \nbody{} results from \cite{Barreira:2013eea} at $a=1$. The boost factor in this plot is the ratio of power spectra in the full and linear cubic Galileon cosmologies, which isolates the effects of screening. We have fixed $k_{\rm filter}=0.2~h/{\rm Mpc}$.
Dark and light grey bands indicate the $\pm 1\%$ and $\pm 2.5\%$  regions respectively.
}
\label{fig:calibration_R_smooth}
\end{figure*}

We next investigate the effect of varying $R_{\rm smooth}$ while fixing $k_{\rm filter}=0.2~h/{\rm Mpc}$, which is seen in Fig.~\ref{fig:calibration_R_smooth}. Again by simply examining the figure, we determine that $R_{\rm smooth}=3~{\rm Mpc}/h$ yields the best results.

\begin{figure*}
\includegraphics[width=1.\textwidth]{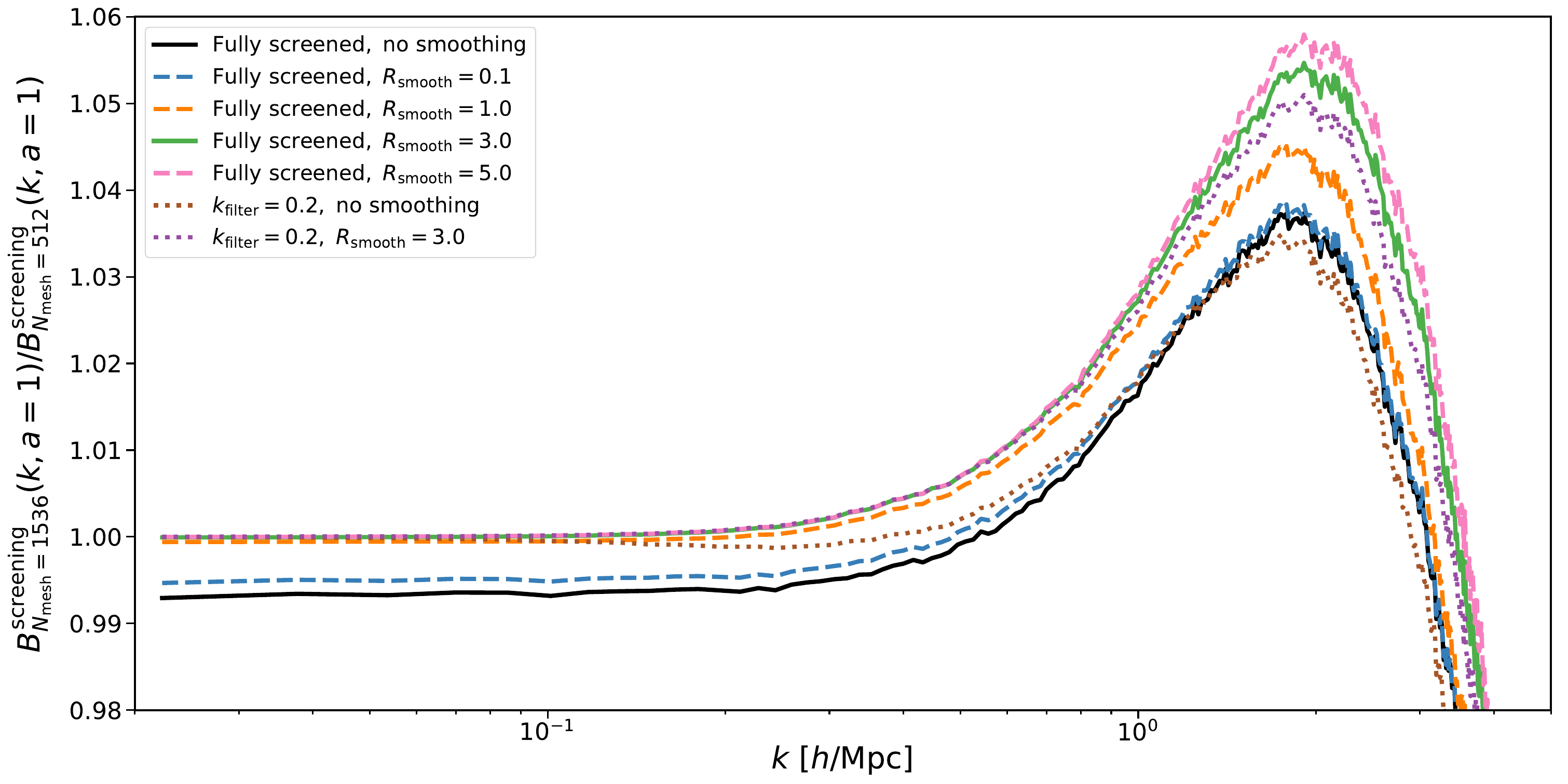}
\caption[Effect of varying resolution on screening boost factor.]{
Effect of varying resolution of our \HiCOLA{} simulations on the screening boost factor at $a=1$ for various screening setups.
}
\label{fig:calibration_resolution_ratios}
\end{figure*}

We also need to confirm that our screening setup prevents simulations from being spuriously affected by changing the force resolution, as discussed in \S\ref{ssec:dens_smoothing}. To do so, we run a set of lower resolution \HiCOLA{} simulations with $N_{\rm mesh}=512$ instead of our standard $N_{\rm mesh}=1536$ for various values of $k_{\rm filter}$ and $R_{\rm smooth}$. We compute the screening boost factor for this lower resolution set and then compare the results against our standard higher resolution set. We plot this as a ratio in Fig.~\ref{fig:calibration_resolution_ratios}.

Firstly, we note that changing the force resolution of a simulation will have a natural effect on clustering statistics at small scales, as can be seen for $k\gtrsim0.2~h/{\rm Mpc}$ in Fig.~\ref{fig:calibration_resolution_ratios}. It is not this effect we are trying to avoid with our screening setup, but any `unnatural' effects, such as those on large scales.
The figure shows that if we neither interpolate between the linear and screened force solutions at large scales (i.e. we use the screened force solution at all scales) nor apply any density field smoothing before computing the screened force solution, then there is an offset between the high- and low-resolution simulations at large and intermediate scales.
The figure shows that $R_{\rm smooth}\gtrsim1~{\rm Mpc}/h$ is necessary to mitigate this effect.
It also shows that, even without any density field smoothing, interpolating between the linear and screened force solutions at large scales is sufficient to eliminate the offset at large scales. However, using interpolation alone without smoothing may still lead to spurious resolution effects at intermediate scales, and Fig.~\ref{fig:calibration_R_smooth} demonstrated that smoothing improves our agreement with \nbody{} regardless.

Thus based on the `screening' boost factor, we determine by examination that the approximate values of our \HiCOLA{} parameters that give the `best' agreement with \nbody{} are $k_{\rm filter}=0.2~h/{\rm Mpc}$ \& $R_{\rm smooth}=3~{\rm Mpc}/h$, and we use these values elsewhere in the paper.

\section{Detailed cubic Galileon results} \label{app:cuGal_results}

In this Appendix, we discuss the results of our cubic Galileon simulations in greater detail. We aim to walk the interested reader through some of the phenomena that result in the figures of \S\ref{ssec:cuGal_sim_res}. We break this down by first discussing the background solutions and screening coefficient, then the linear growth rate, and finally the power spectra.

\subsection{Background evolution} \label{sapp:cuGal_bg}

\begin{figure*}
\includegraphics[width=1.\textwidth]{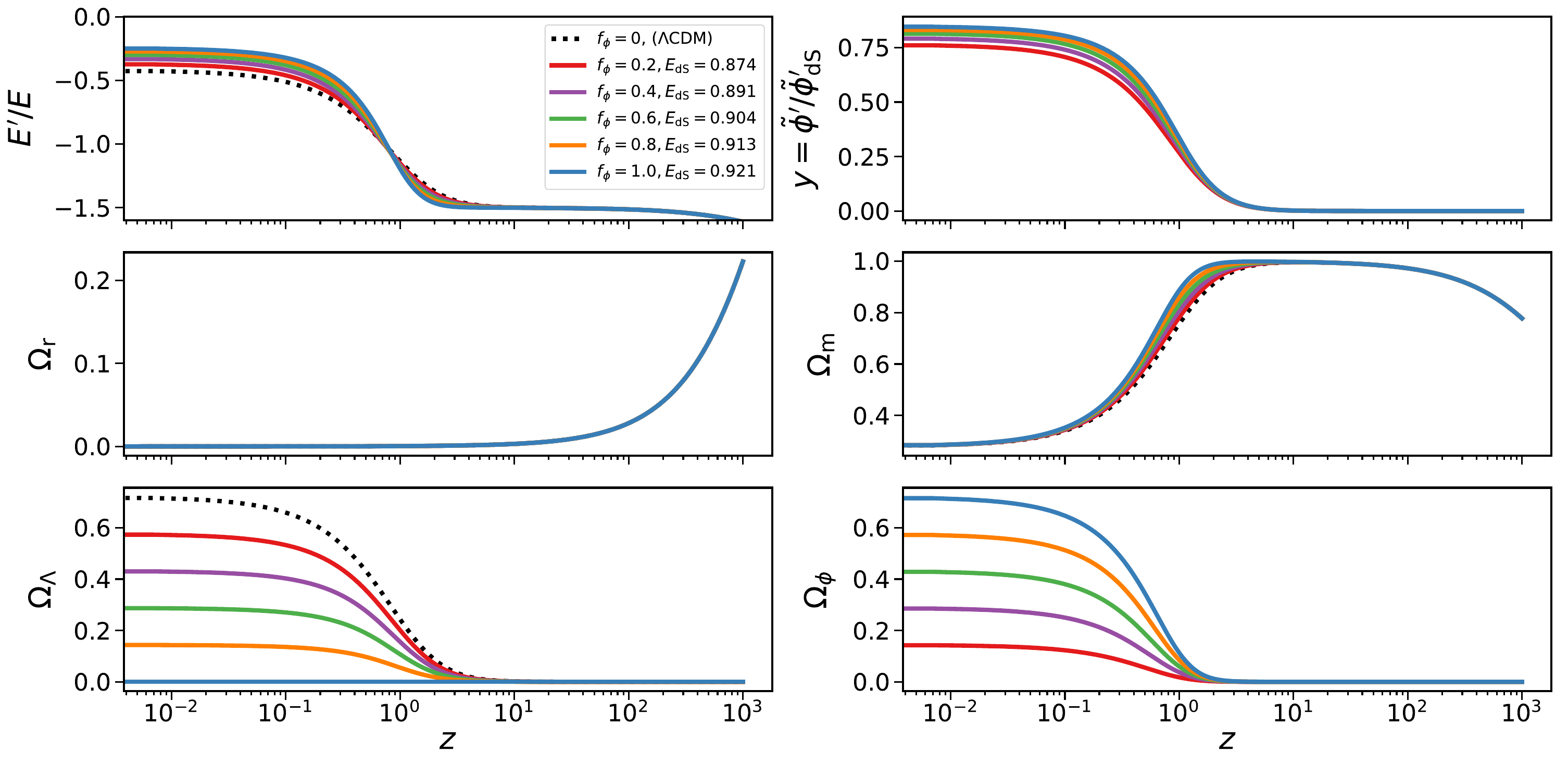}
\caption[Background quantities in cubic Galileon gravity as a function of redshift.]{\small The effect of varying $\{f_{\phi}, E_{\rm dS}(f_{\phi})\}$ in cubic Galileon gravity on various background quantities as a function of redshift: $E^{\prime}/E$ (\textit{upper left}), our scalar field variable $y=\tilde{\phi}^{\prime}/\tilde{\phi}^{\prime}_{\rm dS}$ (\textit{upper right)}, $\Omega_{\rm r}$ (\textit{middle left}), $\Omega_{\rm m}$ (\textit{middle right}), $\Omega_{\Lambda}$ (\textit{lower left}), and $\Omega_{\phi}$ (\textit{lower right}). We also show \lcdm{} predictions for comparison.
The equivalent figure for ESS gravity is Fig.~\ref{fig:ESS_background}.
}
\label{fig:cuGal_background}
\end{figure*}

We start by assessing the background solutions we obtain from solving the equations in \S\ref{ssec:cuGal_theory} using the \HiCOLA{} front-end module described in \S\ref{ssec:HiCOLA_precomp}. In Fig.~\ref{fig:cuGal_background}, we plot the evolution of various background quantities with redshift in each of the five models listed in Table~\ref{tab:cuGal_params}, and also in the corresponding \lcdm{} model, which is reached in the limit $f_{\phi}\rightarrow0$. To highlight the departures from \lcdm{}, we also plot the ratio between the $f_{\phi}=1$ cubic Galileon gravity case and \lcdm{} for $E$ \& $E^{\prime}/E$ in the upper panel of Fig.~\ref{fig:cuGal_lin_growth}. As we will describe below, these two quantities are important in determining the rate of linear growth, as seen in Eq.~(\ref{eq:lin_growth}).

We can see that the main effect of the cubic Galileon scalar field is to produce a brief epoch around $z=1$ where the expansion is temporarily slower relative to the \lcdm{} case. This coincides with the period of rapid evolution of the scalar field derivative (top right panel). Unsurprisingly, we see that as $f_{\phi}$, and thus the proportion of effective DE coming from the cubic Galileon scalar field, increases the deviation from \lcdm{} also increases. In Fig.~\ref{fig:cuGal_background} we see that $\Omega_{\rm r}$ and $\Omega_{\rm m}$ are marginally affected, whilst the sum of the two lowest panels approximately reproduces the behaviour of $\Omega_{\Lambda}$ in \lcdm{} (dotted line in lower left panel).

As we discussed in \S\ref{ssec:cuGal_sim_res}, once we have these background solutions, we then use them to compute the quantities we will need for estimating the screened fifth force, the coupling $\beta$ and Vainshtein radius $\chi/\delta_{\rm m}$, using Eqs.~(\ref{eq:coupling}) \& (\ref{eq:chioverdelta}). These two time-dependent functions along with $E$ and $E^{\prime}/E$ make up the set needed as input for our \HiCOLA{} simulations.

\subsection{Linear growth rate} \label{sapp:cuGal_lin_growth}

\begin{figure*}
\includegraphics[width=1.\textwidth]{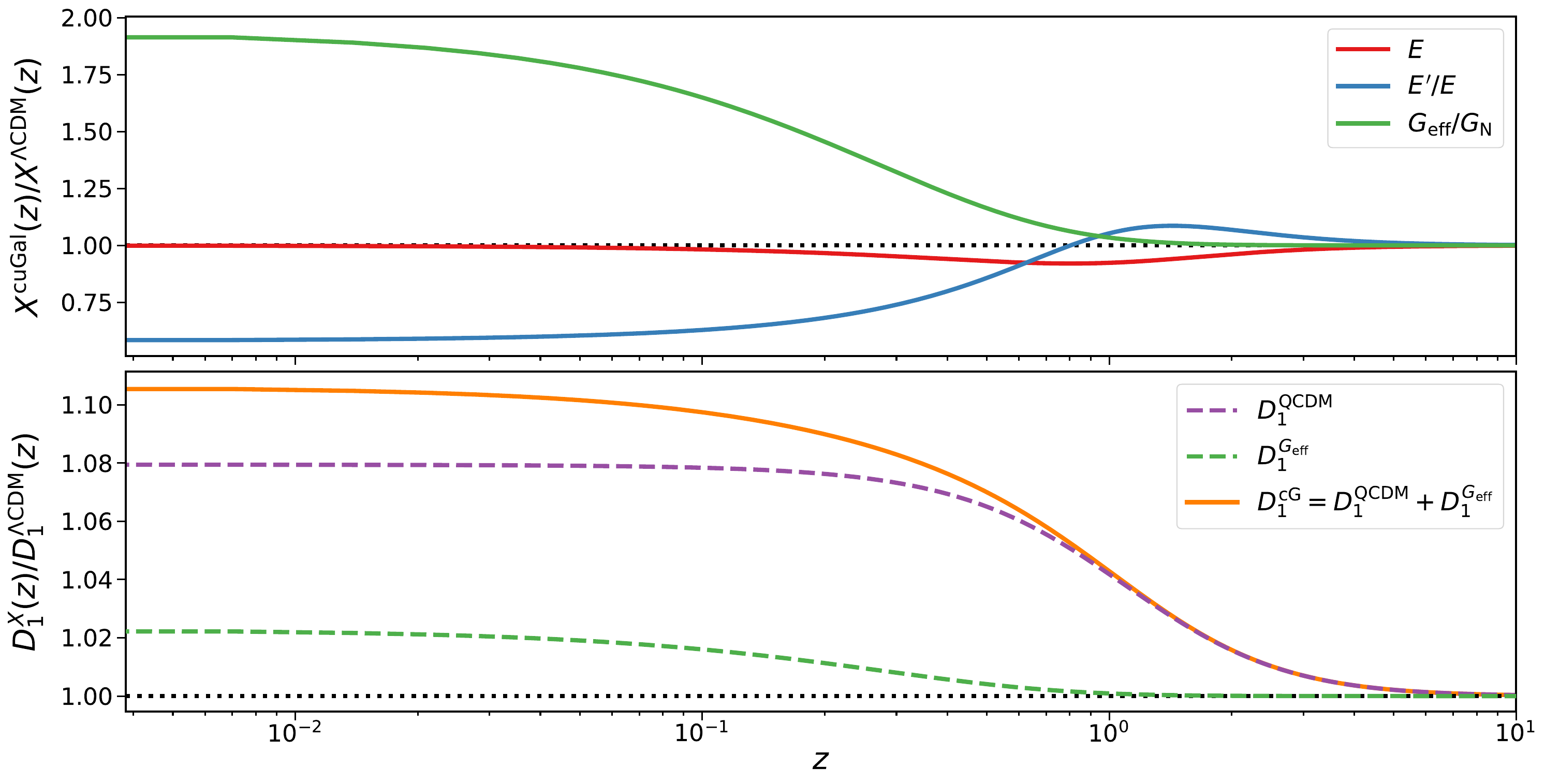}
\caption[Linear growth in cubic Galileon gravity.]{\small The effect of the $(f_{\phi}=1, E_{\rm dS}=0.921)$ cubic Galileon case on linear growth relative to \lcdm{}. {\it Upper panel} shows the impact on the three key quantities that appear in the linear growth equation (\ref{eq:lin_growth}): $E, E^{\prime}/E, G_{\rm eff}/G_{\rm N}$. {\it Lower panel} shows the effect on the linear growth factor $D_1$ relative to \lcdm{} for the two background quantities $E$ \& $E^{\prime}/E$ ($D_1^{\rm QCDM}$) and the linear modification to gravity $G_{\rm eff}/G_{\rm N}$ ($D_1^{G_{\rm eff}}$) separately, then for their combination ($D_1^{\rm cG}=D_1^{\rm QCDM}+D_1^{G_{\rm eff}}$).
The equivalent of this figure for ESS gravity is Fig.~\ref{fig:ESS_lin_growth}.
}
\label{fig:cuGal_lin_growth}
\end{figure*}

We can also compute the linear growth in these cubic Galileon cases using Eq.~(\ref{eq:lin_growth}), which shows that linear growth is dependent on the expansion history through $E$ \& $E^{\prime}/E$ and the linear modification to the strength of gravity $\mu^{(1)}=G_{\rm eff}/G_{\rm N}=1+\beta$. In Fig.~\ref{fig:cuGal_lin_growth}, we first display the redshift evolution of these three quantities relative to their values in \lcdm{} in the upper panel for the $f_\phi=1$ cases. Then in the lower panel, we present the impact of cubic Galileon gravity on linear growth relative to \lcdm{} for a) only the two expansion history quantities ($E$ \& $E^{\prime}/E$) contributing i.e. what we refer to in the main text as the QCDM scenario, b) only $G_{\rm eff}/G_{\rm N}$ contributing, c) the full scenario with all three quantities contributing.

From the magenta dashed curve we see that the expansion in these cubic Galileon models goes through a period around $z=1$ where it is slower than in \lcdm{}, which causes a reduced suppression of structure formation, and thus a boost to linear growth relative to \lcdm{}.
We see that the effective strength of gravity is stronger than in GR (green dashed curve), and this further increases the linear growth relative to \lcdm{}, although the impact on growth is smaller than that of the change to the expansion history.

\subsection{$P(k)$ breakdown} \label{sapp:cuGal_Pk_breakdown}

In \S\ref{ssec:cuGal_sim_setup} we mentioned that we ran hybrid QCDM simulations in addition to the full cubic Galileon and \lcdm{} simulations.
Here, we run an additional second type of hybrid simulation, which we refer to as `Lin-cG', for each of the five cases listed in Table~\ref{tab:cuGal_params}. These Lin-cG simulations are identical to the full cubic Galileon simulations, except that the screening has been artificially turned off, resulting in linear modifications to GR on all scales, or equivalently unscreened fifth forces between particles in all density environments.
For convenience, the features of the two types of hybrid simulation are also summarised in Table~\ref{tab:sim_expl}, along with \lcdm{} and the full cubic Galileon gravity case.

\begin{table}
\begin{center}
\begin{tabular}{c|ccc}
        & Modified Background & Modified Linear Forces & Screening \\ \hline
\lcdm{}      & No                  & No                     & No        \\
QCDM    & Yes                 & No                     & No        \\
Lin-cuGal  & Yes                 & Yes                    & No        \\
cuGal & Yes                 & Yes                    & Yes
\end{tabular}
\caption[Simulation label explanation.]{A table explaining which aspects of \lcdm{} are modified in each simulation discussed in Appendix~\ref{sapp:cuGal_Pk_breakdown}.}
\label{tab:sim_expl}
\end{center}
\end{table}

\begin{figure*}
\includegraphics[width=1.\textwidth]{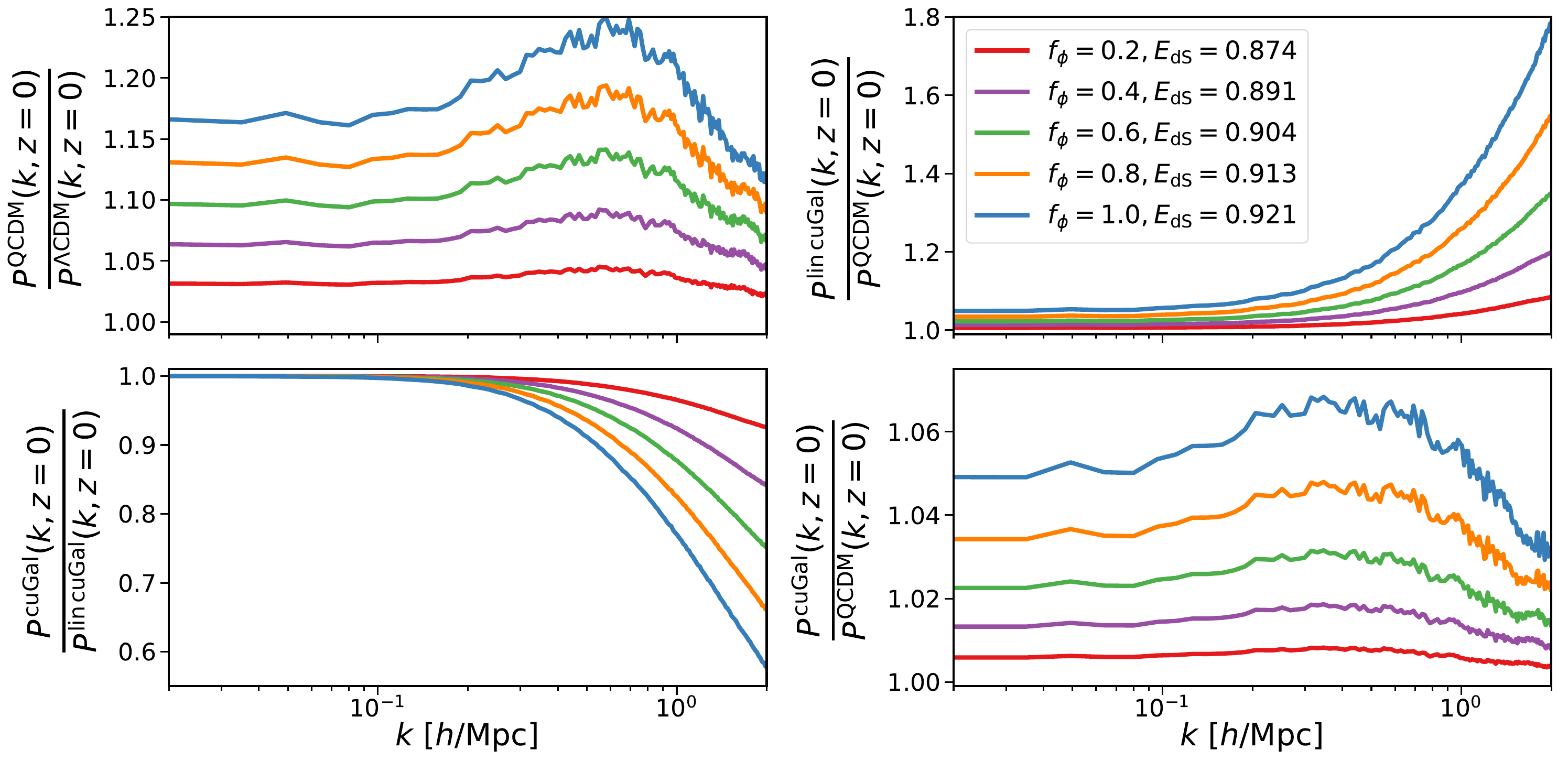}
\caption[The isolated impacts of various aspects of cubic Galileon gravity.]{The impacts of various aspects of cubic Galileon gravity at $z=0$ for a variety of $\left(f_{\phi}, E_{\rm dS}(f_{\phi})\right)$ values.
\textit{Upper left} shows the first boost factor from the list in Appendix~\ref{sapp:cuGal_Pk_breakdown}, \textit{upper right} shows the second, \textit{lower left} shows the third, and \textit{lower right} shows the fourth.
The equivalent of this figure for ESS gravity is Fig.~\ref{fig:ESS_Pk_effects_breakdown}.
}
\label{fig:cuGal_Pk_effects_breakdown}
\end{figure*}

By taking boost factors or $P(k)$ ratios between the various hybrid simulations described in Table~\ref{tab:sim_expl}, we can isolate the impact of different aspects of cubic Galileon gravity. Specifically, in Fig.~\ref{fig:cuGal_Pk_effects_breakdown} we look at four boost factors:
\begin{enumerate}
    \item QCDM vs \lcdm{} -- this isolates the impact of the modified expansion history.
    \item Lin-cuGal vs QCDM -- this isolates the impact of the unscreened fifth force.
    \item cuGal vs Lin-cuGal -- this isolates the impact of the screening.
    \item cuGal vs QCDM -- this isolates the impact of the full fifth force, including screening.
\end{enumerate}

The upper left panel of Fig.~\ref{fig:cuGal_Pk_effects_breakdown} demonstrates that the isolated effect of the modified expansion history in cubic Galileon gravity is to produce an approximately scale-independent enhancement of clustering on large linear scales, which then briefly increases with $k$ at intermediate scales before disappearing at small non-linear scales. We saw in the upper panel of Fig.~\ref{fig:cuGal_lin_growth} that in cubic Galileon gravity there is a brief period where the expansion of the universe is slower than in \lcdm{}. Physically, this reduction in the expansion will produce less of a suppression (i.e. a net enhancement) of structure formation on large linear scales relative to \lcdm{}. This enhancement of linear growth at large scales produces an increasing enhancement on quasi-linear intermediate scales as the physics of structure formation begins to become non-linear\footnote{Interestingly, \cite{Srinivasan:2021gib} found that the quasi-linear growth at $z=0$ is mostly insensitive to the timing of the source of modified growth (in our case, the upper panel of Fig.~\ref{fig:cuGal_lin_growth} shows the slower-than-\lcdm{} expansion is most significant around $z=1$), but that the same is not true of growth at non-linear scales.}. However, the change in expansion rate will have little impact on structure formation on small non-linear scales since these structures are locally gravitationally bound and thus not affected by the expansion of the universe; this explains why the enhancement of clustering decreases at small non-linear scales in the plot.

The upper right panel of Fig.~\ref{fig:cuGal_Pk_effects_breakdown} demonstrates that the isolated effect of the \textit{unscreened} fifth force in cubic Galileon gravity is to produce an approximately scale-independent enhancement of clustering on large linear scales, which then begins to rapidly increase with $k$ towards small non-linear scales. Although the effect on large linear scales is similar to the impact of the modified expansion history, the behaviour on small non-linear scales is different. Physically, the unscreened fifth force in cubic Galileon gravity produces a stronger attraction between particles relative to \lcdm{}. Unlike the impact of the slower expansion, the impact of the stronger attraction between particles is just as relevant on small scales as it is on large scales. Thus the enhancement of clustering does not vanish on small scales and in fact increases rapidly due to the non-linear nature of structure formation.

The lower left panel of Fig.~\ref{fig:cuGal_Pk_effects_breakdown} demonstrates that screening in cubic Galileon gravity has no effect on clustering on large linear scales (as expected), but suppresses structure formation with increasing effect towards small non-linear scales. Physically, this behaviour can be understood by recognising that density increases on small non-linear scales, given that we saw in Fig.~\ref{fig:cuGal_coupl_screenfac} that screening is stronger in high-density environments.

The lower right panel of Fig.~\ref{fig:cuGal_Pk_effects_breakdown} demonstrates the effect of the total fifth force in cubic Galileon gravity, effectively combining the isolated impacts of the unscreened fifth force and the screening. The result is to produce an approximately scale-independent enhancement of clustering on large scales where the screening is not active, which then tends to increase with $k$ at intermediate scales due to growing non-linearity before the screening activates on small scales and suppresses it.

The combination of the enhancement of clustering from the modified expansion history and the enhancement due to the screened fifth force gives the final result we presented in Fig.~\ref{fig:cuGal_Pk_GR_ratio}.


\section{Detailed ESS results} \label{app:ESS_results}

Analogously to Appendix~\ref{app:cuGal_results}, here we discuss the results of our ESS simulations in greater detail than \S\ref{ssec:ESS_sim_res}. However, as in \S\ref{ssec:ESS_sim_res}, we will focus our discussion on the points of difference between ESS gravity and cubic Galileon gravity to minimise repetition.

\subsection{Background evolution} \label{sapp:ESS_bg}

\begin{figure*}
\includegraphics[width=1.\textwidth]{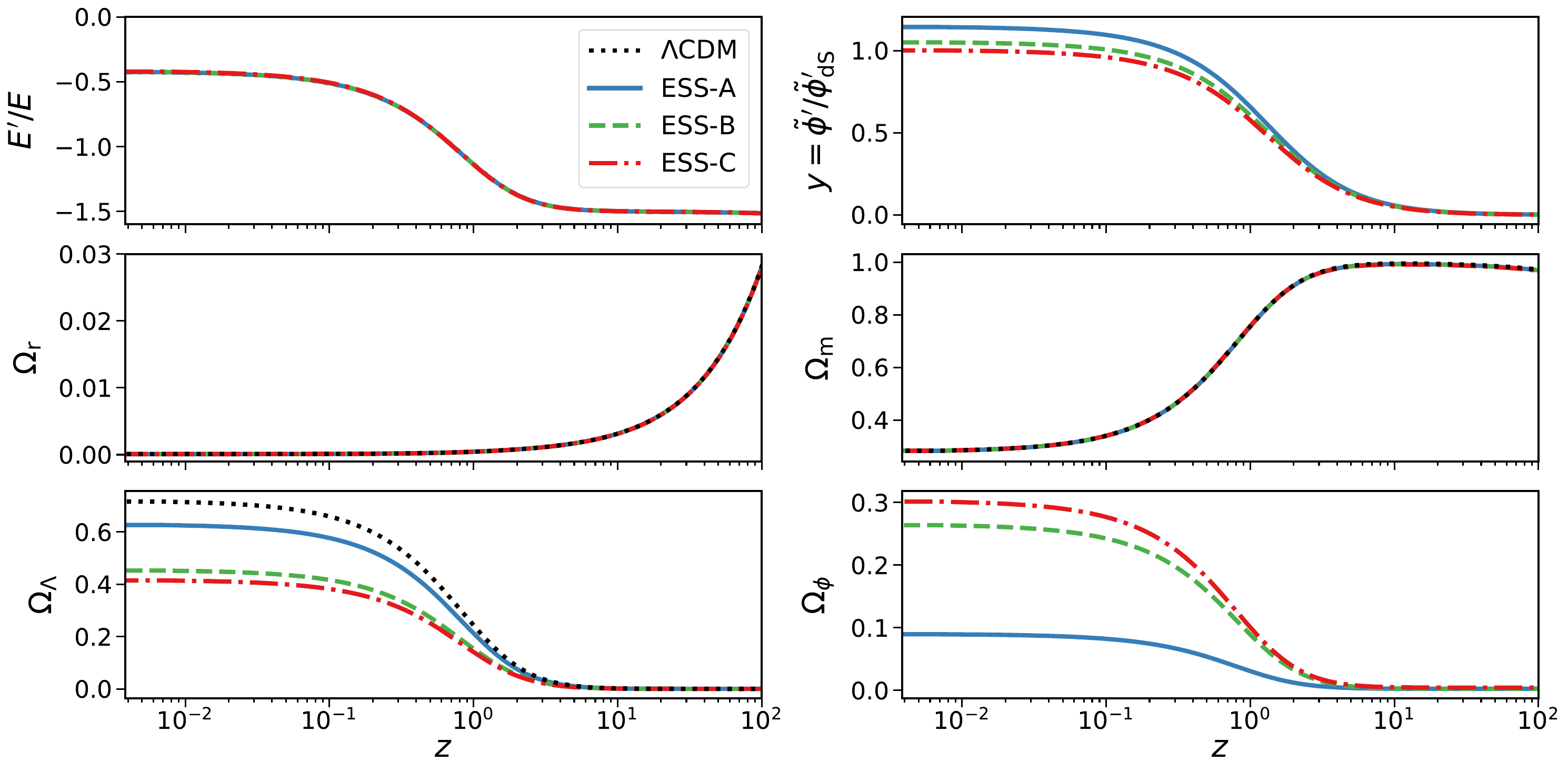}
\caption[Background quantities in ESS gravity as a function of redshift.]{\small The evolution of various background quantities as a function of redshift in three different cases of ESS gravity. We show $E^{\prime}/E$ (\textit{upper left}), our scalar field variable $y=\tilde{\phi}^{\prime}/\tilde{\phi}^{\prime}_{\rm dS}$ (\textit{upper right)}, $\Omega_{\rm r}$ (\textit{middle left}), $\Omega_{\rm m}$ (\textit{middle right}), $\Omega_{\Lambda}$ (\textit{lower left}), and $\Omega_{\phi}$ (\textit{lower right}). We also show \lcdm{} predictions for comparison. The model parameters are found in Table~\ref{tab:cuGal_params}. 
The equivalent of this figure for cubic Galileon gravity is Fig.~\ref{fig:cuGal_background}.
}
\label{fig:ESS_background}
\end{figure*}

As in Appendix~\ref{sapp:cuGal_bg}, we begin our investigation by studying the background solutions we obtain from solving the ESS equations in \S\ref{ssec:ESS_theory}.
In Fig.~\ref{fig:ESS_background}, we plot the evolution of various background quantities with redshift for each of the three ESS models described Table~\ref{tab:ESS_params}, as well as for the corresponding \lcdm{} model (which is reached in the limit of $f_\phi\rightarrow 0$). We also plot the ratio between the ESS-C case and \lcdm{} for $E$ \& $E^{\prime}/E$ in the upper panel of Fig.~\ref{fig:ESS_lin_growth}. These two quantities are important in determining the rate of linear growth, as seen in Eq.~(\ref{eq:lin_growth}).

We find that for these ESS models, departures from \lcdm-like evolution in $E$, $\Omega_{\rm r}$ and $\Omega_{\rm m}$ are very small -- practically invisible on the scale of Fig.~\ref{fig:ESS_background}. However, deviations \textit{are} present, as can be seen in the inset for one of the models in Fig.~\ref{fig:ESS_lin_growth}. Cosmological expansion in these ESS cases is faster than in \lcdm{} at early times for $z>2$, then becomes slower than in \lcdm{} around $z=1$ before approaching the \lcdm{} rate as $z\rightarrow0$.

\subsection{Linear growth rate} \label{sapp:ESS_lin_growth}

\begin{figure*}
\includegraphics[width=1.\textwidth]{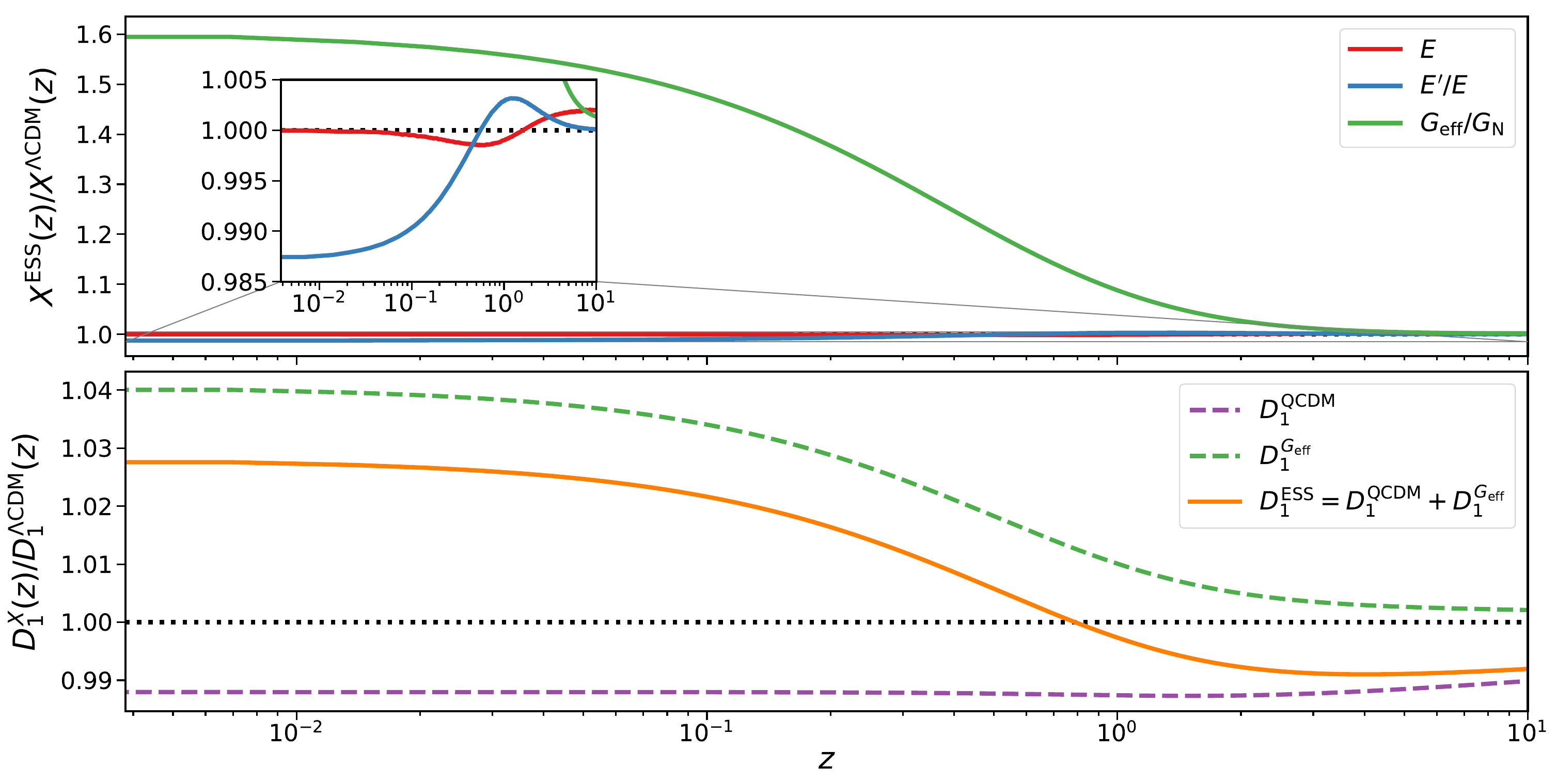}
\caption[Linear growth in ESS gravity.]{\small The effect of one of the ESS cases (ESS-C) on linear growth relative to \lcdm{}. {\it Upper panel} shows the impact on the three key quantities that appear in the linear growth equation: $E, E^{\prime}/E, G_{\rm eff}/G_{\rm N}$. {\it Lower panel} shows the effect on the linear growth factor $D_1$ relative to \lcdm{} for the two background quantities $E$ \& $E^{\prime}/E$ ($D_1^{\rm QCDM}$) and the linear modification to gravity $G_{\rm eff}/G_{\rm N}$ ($D_1^{G_{\rm eff}}$) separately, then for their combination ($D_1^{\rm ESS}=D_1^{\rm QCDM}+D_1^{G_{\rm eff}}$).
The equivalent of this figure for cubic Galileon gravity is Fig.~\ref{fig:cuGal_lin_growth}.
}
\label{fig:ESS_lin_growth}
\end{figure*}

In Fig.~\ref{fig:ESS_lin_growth}, we study the linear growth in ESS gravity and its dependence on expansion history and the linear modification to the strength of gravity.

We see that because the expansion in ESS gravity is faster than in \lcdm{} for $z\gtrsim2$, the suppression of structure formation due to expansion is increased which causes a reduction in linear growth relative to \lcdm{} (purple dashed line). We also see that the effective gravitational strength being stronger than in \lcdm{} causes an increase in linear growth relative to \lcdm{} (green dashed line). These two impacts compete, leading to linear growth in ESS gravity being suppressed for $z\gtrsim2$, but enhanced for $z\lesssim2$ (orange solid line). Note this is qualitatively different to the cubic Galileon case, where the modified expansion history was the dominant source of deviation from \lcdm{} in the growth rate, and both the modified expansion history and $G_{\rm eff}/G_{\rm N}$ acted to enhance growth.

\subsection{$P(k)$ breakdown} \label{sapp:ESS_Pk_breakdown}

\begin{figure*}
\includegraphics[width=1.\textwidth]{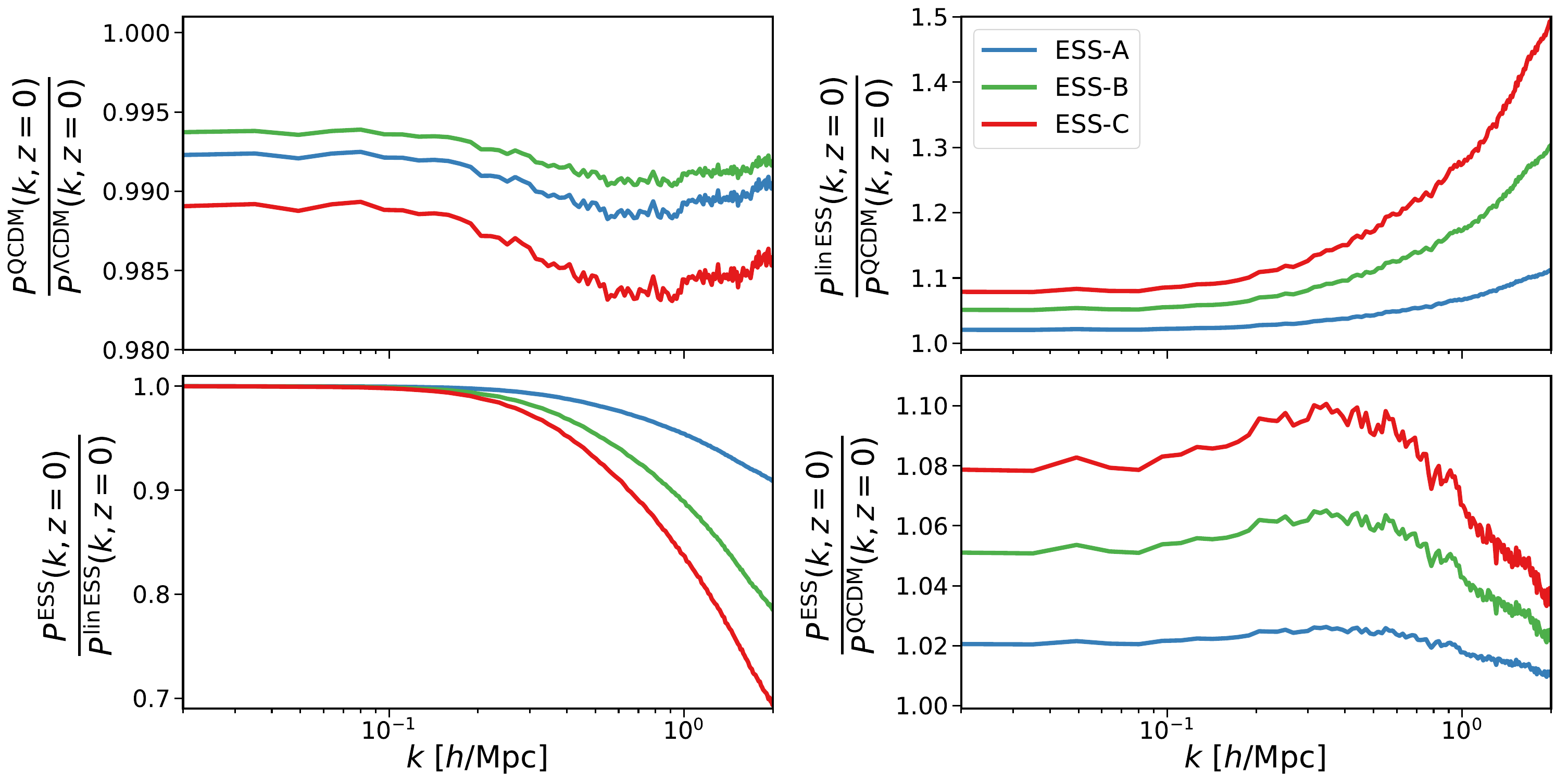}
\caption[The isolated impacts of various aspects of ESS gravity.]{The impacts of various aspects of ESS gravity at $z=0$ for a variety of ESS parameter values. \textit{Upper left} shows the first boost factor from the list in Appendix~\ref{sapp:cuGal_Pk_breakdown}, \textit{upper right} shows the second, \textit{lower left} shows the third, and \textit{lower right} shows the fourth.
The equivalent of this figure for cubic Galileon gravity is Fig.~\ref{fig:cuGal_Pk_effects_breakdown}.
}
\label{fig:ESS_Pk_effects_breakdown}
\end{figure*}

As in Appendix~\ref{sapp:cuGal_Pk_breakdown}, here we run an additional type of hybrid simulation, which we refer to as `Lin-ESS', for each of the three cases in Table~\ref{tab:ESS_params}. These Lin-ESS simulations are identical to the full ESS, except that the screening has been artificially turned off resulting in linear modifications to gravity on all scales, or equivalently unscreened ESS fifth forces in all density environments. As in Appendix~\ref{sapp:cuGal_Pk_breakdown}, we take ratios between $P(k)$ from the various simulations to isolate the impact of different aspects of the ESS model, which we then plot in Fig.~\ref{fig:ESS_Pk_effects_breakdown}.

The upper left panel of Fig.~\ref{fig:ESS_Pk_effects_breakdown} demonstrates that the effect of the modified expansion history in ESS gravity is to produce an approximately scale-independent suppression of clustering on large scales. This suppression then increases with $k$ at intermediate scales before beginning to reduce again at small scales. Physically, this result matches our expectation that the faster-than-\lcdm{} expansion we see for $z\gtrsim2$ in ESS gravity in the upper panel of Fig.~\ref{fig:ESS_lin_growth} will enhance the suppression of structure formation that occurs due to cosmic expansion. This suppression increases at quasi-linear scales as the physics of structure formation becomes non-linear. Again, cosmic expansion will have less of a suppressing effect on structure formation on scales where structures are locally gravitationally bound, which explains the decrease in the suppression of clustering we see on small scales.

The isolated impacts of the unscreened fifth force and the screening are shown in the upper right and lower left panels of Fig.~\ref{fig:ESS_Pk_effects_breakdown} respectively. Unlike the impact of the modified expansion history, the impacts of these two quantities in ESS gravity are qualitatively similar to the corresponding impacts in cubic Galileon gravity. In quantitative terms, the enhancement of clustering due to the unscreened fifth force is larger at large scales in our ESS cases, but smaller at small scales. However, the suppression of clustering at small scales due to the screening is also reduced in our ESS cases. As a result, we actually find the net enhancement to clustering from the combination of these two effects is greater for our ESS cases at all scales.

The suppression of clustering from the modified expansion history and the enhancement due to the screened fifth force combine to give the final result we presented in Fig.~\ref{fig:ESS_Pk_GR_ratio}.


\section{Core count scaling} \label{sapp:corescaling}

\begin{figure*}[h!]
    \centering
    \includegraphics[width=0.7\textwidth]{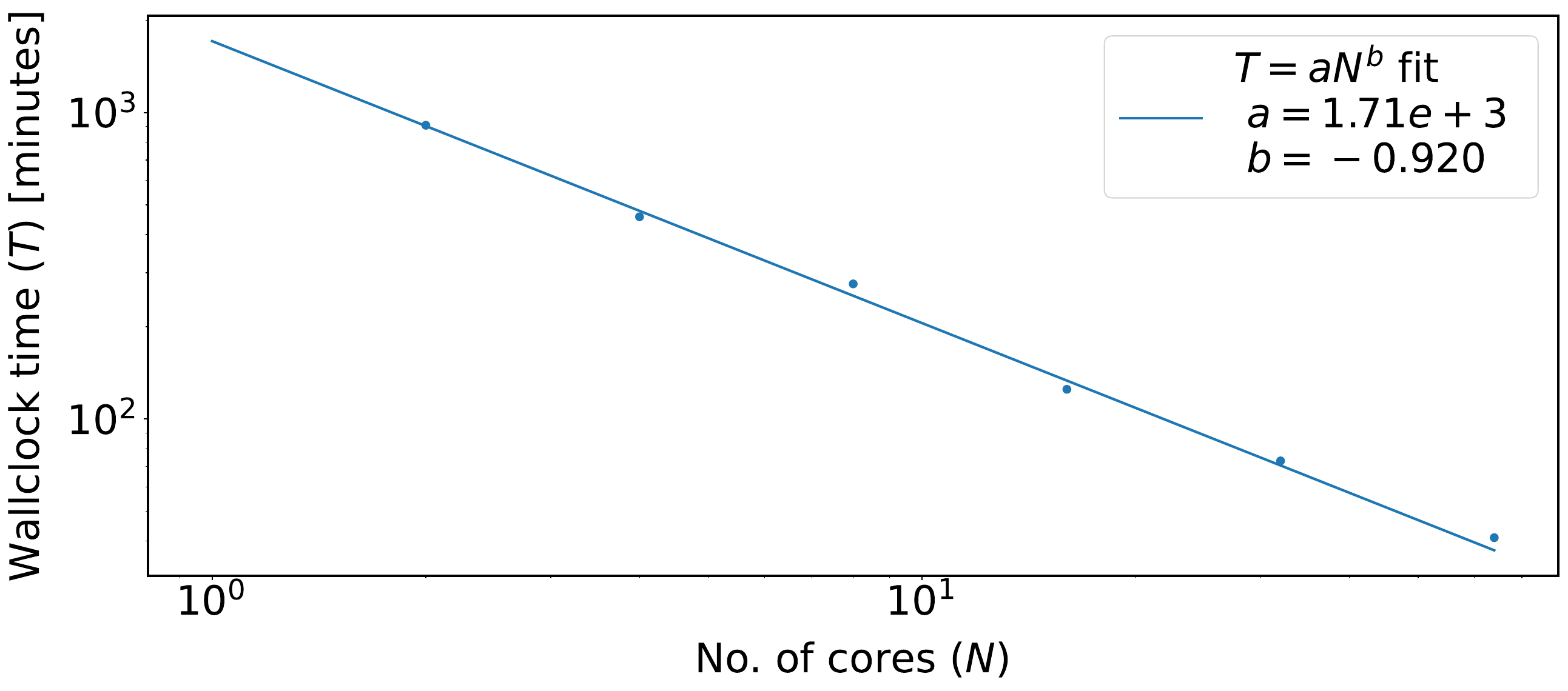}
    \caption{Relationship between the wallclock time for simulations of ESS-C and the numbers of cores used. We find that in logarithmic space, the relationship between wallclock time and core count is well described by a linear fit, in particular, $T = 1710N^{-0.92} \, {\rm minutes}$. In other words, \HiCOLA{} scales well and makes effective use of the resources given to it.}
    \label{fig:corescaling}
\end{figure*}

Being a relatively fast simulation code, naturally one may want to know how \HiCOLA{}'s run time scales with the number of cores used. This is shown in Fig.~\ref{fig:corescaling}, where multiple runs of ESS-C (with the same setup as described in the main text, see \S\ref{ssec:cuGal_sim_setup} and \S\ref{ssec:ESS_sim_setup}) were conducted, only differing by the number of cores given for the simulation. We find that the relationship between the wallclock time and core count scales approximately like $T \propto N^{-1}$, showing that \HiCOLA{} has good scaling and is making efficient use of resources given to it, with no obvious signs of running into overheads.


\bibliographystyle{JHEP}
\bibliography{References}

\end{document}